\documentclass[letterpaper,twocolumn,10pt]{article}
\usepackage{usenix-2025-08-28}

\usepackage{tikz}

\usepackage{filecontents}
\usepackage{booktabs}
\usepackage{tabularx}
\usepackage{xspace}

\usepackage{enumitem}
\usepackage{listings}
\usepackage{color}
\usepackage{caption}
\usepackage{subcaption}
\usepackage{bm}
\usepackage{bbm}
\usepackage{tabularray}
\usepackage{multirow}
\usepackage{arydshln}
\usepackage{pifont}
\usepackage{enumitem}
\usepackage[export]{adjustbox}
\usepackage{threeparttable}
\UseTblrLibrary{diagbox}
\usepackage{algorithm}
\usepackage{algorithmic}
\usepackage{amsfonts} 
\usepackage{url}
\usepackage{amssymb}
\usepackage[available]{usenixbadges}

\definecolor{keywordsColor}{rgb}{0, 0, 255}
\definecolor{methodsColor}{rgb}{255, 0, 255}
\definecolor{codeColor}{rgb}{0, 0, 0}

\lstdefinestyle{mystyle}{ %
  language={Python},
  frame=b,
  backgroundcolor=\color{white},   
  basicstyle=\rmfamily\fontsize{6}{7}\selectfont,        
  breaklines=true,                 
  captionpos=b,                    
  escapeinside={\%*}{*)},          
  keywordstyle=\color{keywordsColor}\bfseries,    
  morekeywords={None},
  deletekeywords=[2]{buffer,len,reversed,range,type},
  moredelim=**[is][\color{methodsColor}]{@}{@}
}
\newcommand{\cmark}{\ding{52}}
\newcommand{\xmark}{\ding{56}}

\newcommand{\watermark}{\textsc{Themis}\xspace}

\newcommand{\plain}[1]{\textcolor{black}{#1}}

\begin{document}

\date{}

\title{THEMIS: Towards Practical Intellectual Property Protection for Post-Deployment On-Device Deep Learning Models}

\author{
\rm Yujin Huang\textsuperscript{1}\ \ \
Zhi Zhang\textsuperscript{2\dag}\ \ \
Qingchuan Zhao\textsuperscript{3}\ \ \
Xingliang Yuan\textsuperscript{1}\ \ \
Chunyang Chen\textsuperscript{4}\ \
\\
\textsuperscript{1}\textit{The University of Melbourne} \ \ 
\textsuperscript{2}\textit{The University of
Western Australia} \ \ \ \\
\textsuperscript{3}\textit{City University of Hong Kong} \ \ \
\textsuperscript{4}\textit{Technical University of Munich}
}

\maketitle
{
\renewcommand{\thefootnote}{\fnsymbol{footnote}}
\footnotetext[2]{Corresponding author.}
}
\pagestyle{empty}

\begin{abstract}
On-device deep learning (DL) has rapidly gained adoption in mobile apps, offering the benefits of offline model inference and user privacy preservation over cloud-based approaches.
However, it inevitably stores models on user devices, introducing new vulnerabilities, particularly model-stealing attacks and intellectual property infringement.
\plain{While system-level protections like Trusted Execution Environments (TEEs) provide a robust solution, practical challenges remain in achieving scalable on-device DL model protection, including complexities in supporting third-party models and limited adoption in current mobile solutions.
Advancements in TEE-enabled hardware, such as NVIDIA's GPU-based TEEs, may address these obstacles in the future.}
Currently, watermarking serves as a common defense against model theft but also faces challenges here as many mobile app developers lack corresponding machine learning expertise and the inherent read-only and inference-only nature of on-device DL models prevents third parties like app stores from implementing existing watermarking techniques in post-deployment models.

To protect the intellectual property of on-device DL models, in this paper, we propose \watermark, an automatic tool that lifts the read-only restriction of on-device DL models by reconstructing their writable counterparts and leverages the untrainable nature of on-device DL models to solve watermark parameters and protect the model owner’s intellectual property.
Extensive experimental results across various datasets and model structures show the superiority of \watermark in terms of different metrics.
Further, an empirical investigation of 403 real-world DL mobile apps from Google Play is performed with a success rate of 81.14\%, showing the practicality of \watermark.

\end{abstract}
\section{Introduction}

With the proliferation of smartphones, mobile apps have permeated and greatly enhanced convenience in people's lives, facilitating activities such as medical diagnosis and driving assistance.
To fully unlock the potential of mobile apps, mobile developers have embraced on-device deep learning (DL) to integrate artificial intelligence features like image classification into their apps~\cite{xu2019first}.
Compared to offloading deep learning
from mobile devices to the cloud, on-device DL provides several distinct advantages.
For example, it eliminates the need to transmit sensitive user data to the cloud, resulting in bandwidth saving, accelerated inference, enhanced privacy preservation~\cite{huang2021robustness}, and even allows
apps 
function 
without requiring internet connectivity~\cite{sun2021mind}.

\begin{figure}[t]
    \centering
    \includegraphics[width=0.925\linewidth]{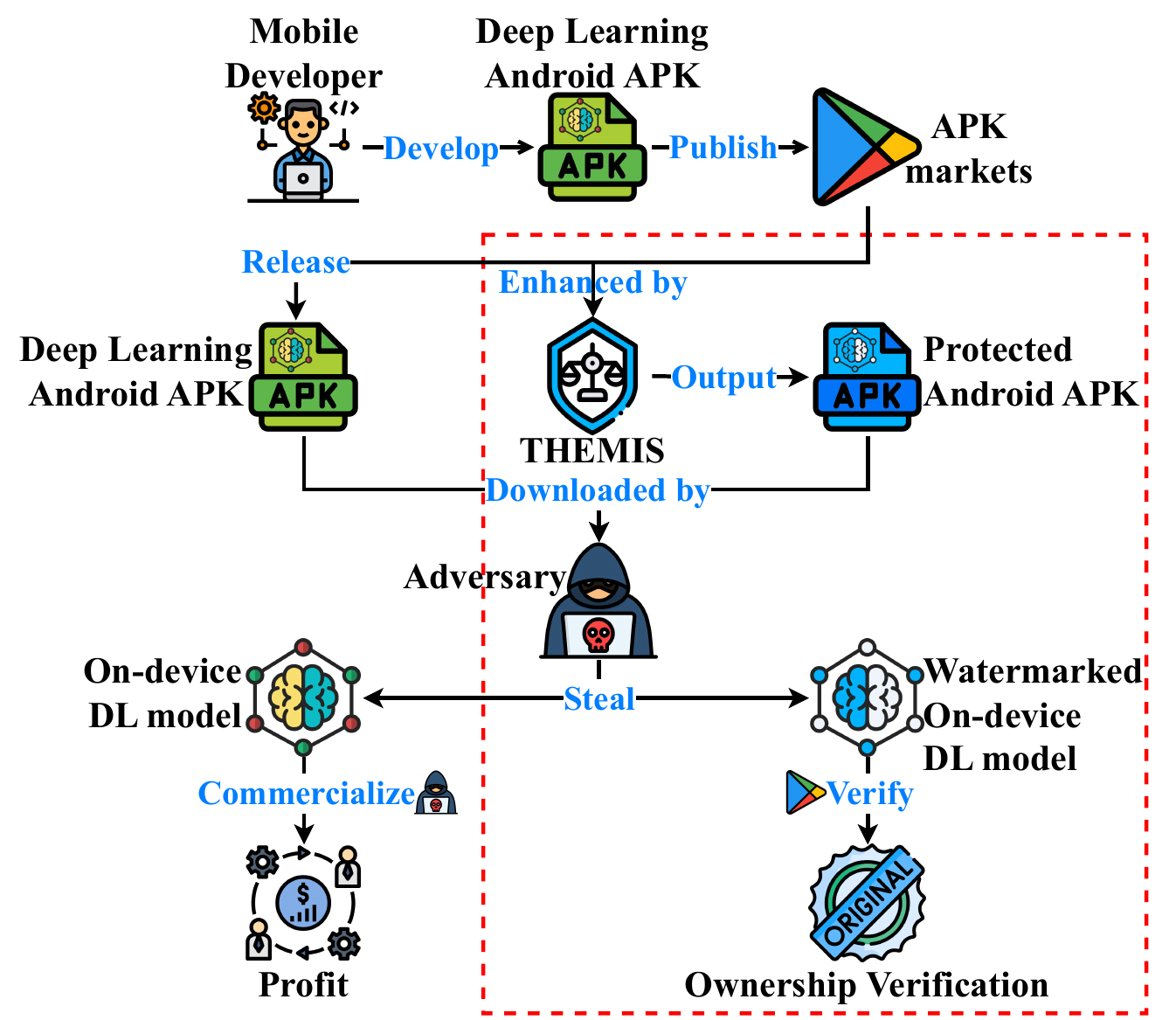}
    \caption{The common scenario of model stealing. The red block indicates that \watermark enables on-device DL model protection like watermarking.}
    \label{fig:model_steal_scenario}
\end{figure}

Unfortunately, on-device DL necessitates the local storage of DL models on user devices, which enables adversaries to disassemble deep learning mobile apps (DL apps) to steal the models and thus jeopardize the intellectual property (IP) of model owners, as shown in Figure~\ref{fig:model_steal_scenario}.
Such an issue is theoretically unsolvable unless with system-level protections like Trusted Execution Environments.
\plain{However, these mechanisms encounter practical challenges in safeguarding on-device DL models at scale, primarily due to the complexity of supporting third-party models and limited adoption in mobile solutions to date.
Emerging advancements in TEE-enabled hardware, such as  NVIDIA’s GPU-based TEEs, suggest that these obstacles may be addressed in the future.}

Currently, watermarking is often used to protect the IP of machine learning models but also challenging to apply here as:
(1) Many mobile developers either lack awareness of model theft risks or do not possess the machine learning expertise necessary for implementing model watermarking~\cite{sun2021mind}.
(2) App stores or other parties may intend to protect the model IP of DL apps but the embedding of watermarks into post-deployment on-device models proves challenging due to the \textit{read-only} and \textit{inference-only} natures of the on-device model (See Section~\ref{sec:on_device_dl}).

Hence, in this paper, we propose \watermark, an automatic tool to demonstrate the feasibility of embedding watermarks into post-deployment on-device models from DL apps.
Specifically, \watermark takes the specific
characteristics (i.e., \textit{read-only and inference-only with backpropagation disabled}) of the on-device model into account and enables watermark embedding through four steps:
First, given a DL app, it extracts an on-device model inside by disassembling the app.
Then, it enables parameter modification of the on-device model as the successful watermark embedding relies on the ability to modify model parameters.
Next, it leverages existing training-free backdoor algorithms to solve the watermark parameters of the writable on-device model and rewrites these parameters back to the model accordingly.
Finally, it utilizes the watermarked model to rebuild the DL app, generating a protected counterpart.

There are three key challenges in \watermark.
First, certain post-deployment on-device models manifest in encrypted forms and thwart direct extraction.
While recent work~\cite{sun2021mind} has showcased the viability of extracting these models from memory, the circumscribed instrumentation strategies lead to fragmentary model extraction.
Thus, we advance their approach by introducing an execution tracing mechanism that monitors API calls within DL app operations, pinpoints those tied to encrypted models, captures corresponding data objects, and refines extracted models by integrating values retrieved from these objects if they resonate with model metadata.

The second challenge lies in enabling the writability of on-device models due to their inherent \textit{read-only} nature.
By analyzing the file format used for on-device models, we observe that all model information (e.g., operators and parameters) is serialized following a predefined model schema, rendering such information immutable.
To tackle this challenge, we propose Model Rooting, an on-device model reverse engineering technique that reconstructs a writable model from the original one by leveraging the model schema.
Particularly, we generate programming language classes aligned with the model schema's data structures to deserialize information from an on-device model.
Despite the model information can be retrieved, the generated classes derived from the model schema lack serialization support.
To address this issue, we enhance these classes with specially designed serialization functions that enable seamless reconstruction of a model.


Since post-deployment on-device models are \textit{inference-only with backpropagation disabled}, watermarking such models in a training-free manner is also challenging. 
Specifically, identifying appropriate parameters to modify and determining the optimal magnitude of changes are non-trivial without using iterative optimization.
Meanwhile, as some on-device models perform rare recognition tasks, collecting such data from the Internet is unrealistic, making the watermarking even more challenging.
To address these challenges, we present a data synthesis approach that can generate data for any on-device model by exploiting a public dataset.
\plain{Subsequently, we introduce Model Reweighting, a watermark embedding process that can leverage any training-free backdoor algorithm to solve the watermark parameters for on-device models.}

\plain{Furthermore, existing training-free backdoor algorithms, such as Targeted Weight Perturbations~\cite{dumford2020backdooring} and Handcrafted Perturbations~\cite{hong2021handcrafted}, heavily rely on substantial data for parameter search, limiting their effectiveness in data-constrained on-device scenarios like data-missing and data-scarce scenarios (see Section~\ref{sec:wat_effectiveness}).
To overcome these limitations, we propose Feed-Forward Knowledge Editing Watermarking (FFKEW), a training-free backdoor algorithm that embeds watermarks into on-device models by precisely editing model knowledge in a single feed-forward pass.}

We evaluate \watermark on three widely used on-device DL models including MobileNetV2~\cite{sandler2018mobilenetv2}, InceptionV3~\cite{szegedy2016rethinking}, and EfficientNetV2~\cite{tan2021efficientnetv2} with four computer vision datasets,
i.e., FMNIST~\cite{xiao2017fashion}, CIFAR10~\cite{krizhevsky2009learning}, GTSRB~\cite{stallkamp2011german}, and SVHN~\cite{netzer2011reading}.
Following the previous studies on watermarking~\cite{adi2018turning,jia2021entangled}, we use metrics that measure Watermark Success Rate and Accuracy.
\plain{Experimental results show that \watermark with FFKEW significantly outperforms all baselines across various on-device scenarios.
Specifically, FFKEW achieves over 80\% Watermark Success Rate with the lowest Accuracy drop in all evaluated cases.}
To further examine the feasibility of our tool on real-world DL apps, we perform end-to-end watermark embedding on 403 DL apps from Google Play. 
Of these apps, 81.14\% can be successfully watermarked, which includes popular ones in medicine, automation, and finance categories with critical usage scenarios such as skin cancer recognition, safety gear detection, cash recognition, etc.

\noindent
\textbf{Contributions.} We summarize the contributions as follows:
\begin{itemize}[leftmargin=*]
    \item  \textbf{New Problem:} This paper endeavors to tackle a novel research problem, i.e., how to successfully embed watermarks into post-deployment on-device models with their inherent \textit{read-only} and \textit{inference-only} natures.
    This issue has received little attention but becomes urgent with the surge in on-device DL adoption in mobile apps.
    \item \textbf{Automatic Tool:} We propose and implement a groundbreaking systematic tool, \watermark\footnote{\href{https://github.com/Jinxhy/THEMIS}{https://github.com/Jinxhy/THEMIS}}, to demonstrate the feasibility of watermarking on-device models in post-deployment stage and address challenges encountered in watermark embedding.
    Our tool introduces a series of novel mechanisms, including execution tracing for complementing partially extracted encrypted models, Model Rooting for constructing writable models, and Model Reweighting for solving watermark model parameters in a training-free manner.
    
    \item  \textbf{Comprehensive Evaluation:} We conduct an extensive evaluation to demonstrate the effectiveness of \watermark.
    The results indicate that it achieves watermark with a high success rate and reasonable utility drop in various watermark scenarios.
    In addition, an empirical investigation of 403 real-world DL apps, 327 of which successfully embedded watermarks, demonstrates the practicality of \watermark.
\end{itemize}

\section{Preliminaries}

\subsection{On-device DL}
\label{sec:on_device_dl}

On-device DL is commonly realized through DL frameworks such as Google TensorFlow~\cite{tensorflow} and TFLite~\cite{tensorflowlite}, Facebook PyTorch~\cite{pytorch} and Pytorch Mobile~\cite{pytorchmobile}, and Apple Core ML~\cite{coreml}.
Of these frameworks, TFLite is the most widely adopted framework for deploying DL models on mobile, microcontrollers, and edge devices, powering nearly 50\% of all DL apps in the past two years with a noticeable growth rate compared to other frameworks~\cite{huang2021robustness,xu2019first,huang2022smart,zhou2024model,zhou2024dynamo}.
This trend stems from TFLite's lightweight and efficient design, enabling DL deployment on resource-constrained mobile devices without compromising performance.
The development and deployment of a TFLite model unfolds as follows.
A mobile app developer first builds a TensorFlow model, either by training from scratch, fine-tuning a pre-trained model from TensorFlow Hub~\cite{tensorhub}, or directly using an existing pre-trained model.
The TensorFlow model is then converted into a TFLite model via the TFLite Converter, leveraging optimizations like quantization to reduce size and latency while maintaining accuracy.
Finally, the TFLite model is integrated into a mobile app and packaged as an Android Application Package (APK) for public release on app markets.

\textbf{TFLite Model Constraint.} It is worth noting that the generated TFLite model cannot be modified and converted back to the original TensorFlow model.
In other words, the TFLite model is \textit{strictly read-only} and \textit{exclusively used for inference purposes}.
This is due to the inherent storage of on-device DL models on user devices, which can be accessed by adversaries, leading Google to impose restrictions on the model's capabilities to prevent potential attacks.

\subsection{Watermarking in DL}
Watermarking has commonly been employed to safeguard the IP of DL models~\cite{adi2018turning,lukas2022sok}.
Rather than proactively preventing model theft, it enables ownership verification for suspected stolen models.
The process consists of two phases: watermark embedding and ownership verification.
In the embedding phase, the defender clandestinely injects watermarks into a to-be-protected DL model during its training phase, with these watermarks remaining confidential and known only to the defender.
During the verification phase, the defender scrutinizes a suspect DL model or queries it to verify the existence of watermarks.
Based on the defender's accessibility to a suspected model during verification, existing watermarking approaches can be categorized into two groups.

\textbf{White-box Watermarking.}
In the white-box setting, the defender is assumed to possess complete access to the internal of a suspect DL model~\cite{lv2023robustness}.
Such watermarking approaches~\cite{uchida2017embedding,darvish2019deepsigns,lv2023robustness} embed watermarks into model parameters or architecture, allowing the defender to extract them from a suspect model using its parameters or specific layer output representations in cases of IP infringement.
However, obtaining direct access to suspect models in real-world applications may not always be feasible due to the potential redeployment of stolen models via encryption, which makes the white-box watermarking impractical.

\textbf{Black-box Watermarking.}
In the black-box setting, the defender
can not access a suspect DL model (i.e., parameters and architecture) and only can perform queries on it~\cite{jia2021entangled}.
Typically, black-box watermarking approaches~\cite{adi2018turning,zhang2018protecting,li2019prove,namba2019robust,jia2021entangled} involve the insertion of backdoor triggers into a to-be-protected DL model and consider these triggers as the watermarks.
Therefore, the defender can verify the ownership of a suspect model by solely querying it with trigger-stamped inputs, such as a small square positioned in an unobtrusive area of an image.
Compared to the white-box setting, the black-box watermarking is deemed more realistic and practical as it enables watermark extraction without assuming access to a suspect model's parameters and architecture.

\section{Threat Model}
\label{sec:thr_model}
\begin{figure*}[hbt!]
\centering
\includegraphics[width=0.8\textwidth]{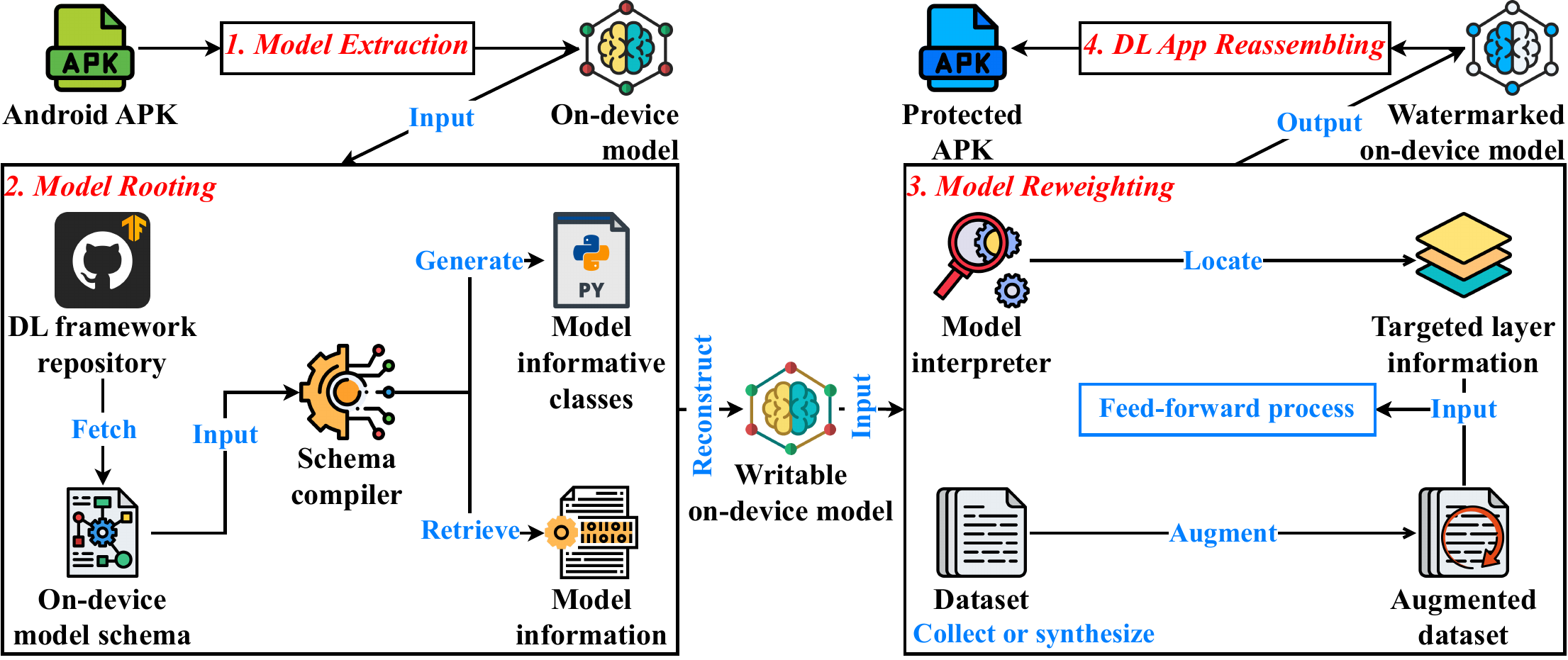}
\caption{The Overview of \watermark.}
\label{fig:watermark_overview}
\vspace{-1em}
\end{figure*}


In our threat model, we consider three parties: an Android app store offering mobile apps for widespread public access, a DL app developer creating apps with on-device DL, and an adversary aiming to illicitly steal on-device models for financial advantage.
In practice, most on-device models within DL apps are left unprotected due to app developers' scant knowledge of model theft and protection measures~\cite{xu2019first,sun2021mind}.
\plain{Even when developers are aware of model theft risks and consider solutions like watermarking, they may lack the expertise to implement them effectively.
Furthermore, the absence of readily available SDKs for on-device model watermarking leaves them without practical tools.
These lacks in awareness, expertise, and tools present opportunities for the adversary to steal on-device models with minimal effort, and thus emphasize the urgent need for practical solutions to secure them.}

In this paper, \plain{we aim to assist ordinary DL app developers in protecting on-device model intellectual property} from the perspective of the app store \plain{by introducing a tool restricted to use by authorized app stores.}
We assume the app store administrator is granted permission for app reverse engineering \plain{to facilitate necessary modifications that protect the on-device model intellectual property within a DL app. 
This aligns with the app store’s role as a trusted intermediary authorized by developers during the app submission process.
With these permissions, the administrator extracts the model and its associated files (e.g., label file) and applies binary patches within the execution environment of the app store. 
This avoids conflicts with the anti-repacking protections that developers concerned about intellectual property may have implemented.}
However, the administrator has no knowledge of the model training data that are private to the app developer.
\plain{Additionally, the administrator only has black-box access to verify model ownership of suspect DL apps, using a database of trigger inputs and expected outputs for watermarked models to enable automated large-scale detection of stolen models, even those redeployed with encryption, without developer involvement.}

We consider the adversary can obtain the DL app from the app store and extract its on-device model via established tools such as DeMistify~\cite{ren2024demistify} and ModelXtractor~\cite{sun2021mind}.
\plain{While an adversary may attempt to access our tool by posing as a legitimate app store to rewrite embedded watermarks, stringent authorization protocols and the need to replicate a legitimate app store's infrastructure make this approach costly and challenging.
Even if the adversary bypasses security measures and accesses our tool, the lack of knowledge about the embedded watermark confines removal attempts to overwriting with a new one, leaving the original watermark verifiable as its parameters cannot be fully altered by overwriting~\cite{lv2024mea}.}


\section{Design of \watermark}
\label{sec:methodology}

\subsection{Overview}
Recall that an app store administrator aims to watermark an on-device DL model utilized in a DL app via model modification, so that any suspect DL app reassembled with the watermarked model can be verified via query.
However, on-device models are \textit{read-only} and \textit{purely inferential}, which prohibits the administrator from directly carrying out structure-modified~\cite{liu2022loneneuron,qi2022towards}, weight-oriented~\cite{dumford2020backdooring,hong2021handcrafted} and gradient-based~\cite{goodfellow2014explaining,liu2018trojaning} modifications as such models are unmodifiable and cannot perform backpropagation.
In addition, model extraction is non-trivial as some on-device models within DL apps utilize encryption mechanisms.
Note that the on-device models here refer to TFLite models and we focus on them because of their large occupancy in DL apps and growing trend~\cite{huang2021robustness,xu2019first}.
To address the aforementioned challenges, we propose \watermark that extracts on-device models, enables on-device model modification, and performs the goal-based parameter optimization to watermark on-device models.
The overview of \watermark is illustrated in Figure~\ref{fig:watermark_overview}.

 
First, we perform model extraction on a DL app to obtain its encrypted or plaintext on-device model that can only be read and used for inference.
Second, to make the on-device model writable, we execute Model Rooting on it.
In particular, we fetch the DL framework schema of the on-device model from the corresponding official repository and utilize it to generate a set of model informative classes (e.g., operator, computational graph and raw data buffer classes).
We then retrieve the on-device model information via the generated informative classes and reconstruct a writable on-device model based on them.
Third, we carry out Model Reweighting on the writable on-device model to craft a watermarked on-device model, where its watermarked layers are located based on the model structure and the dataset used for solving the watermarked layer parameters is either collected from the Internet or synthesized depending on the availability of the model label file.
Finally, we reassemble the DL app by substituting the watermarked on-device model for the original one to produce a protected DL app, which can then be released on the app store for open access, offering a layer of defense against intellectual property infringement.

\subsection{Model Extraction}
\label{sec:mod_extraction}
We first scrutinize all files stored in the DL app to find the models, and then extract them based on their types (i.e., plaintext or encrypted).
Specifically, given an Android APK consisting of TFLite models, we begin by decomposing the APK to almost original form (e.g., 
resource files, dex files, and manifest files.) with the help of Apktool~\cite{winsniewski2012android}.
We then scan the decomposed APK for files that conform to the TFLite model naming schemes~\cite{sun2021mind} and extract them when they are plaintext.
For encrypted models, we follow the same approach of~\cite{sun2021mind} to dynamically extract them from memory using app instrumentation.

Considering the diversity of protection strategies applicable to model encryption and the pivotal role of successful model extraction in ensuring the triumph of our watermark, we conduct a pilot study on several real-world DL apps that contain encrypted models to verify the effectiveness of the approach outlined in~\cite{sun2021mind}.
As a consequence, we find that this approach fails to extract nearly half of the encrypted models.
The reasons for these failures are twofold: (1) the extracted models often lack or have incomplete metadata, such as descriptions of inputs/outputs and necessary pre/post-processing details, making them unusable under standard conditions; (2) the presence of unknown model decryption APIs leads to unpopulated memory buffers intended for decrypted models, culminating in the extraction of invalid models.

Due to the potential customization of model decryption APIs by developers, compiling an exhaustive API list for encrypted model extraction is challenging. 
As such, our focus shifts to addressing the cases of absent or partial model metadata.
We propose an execution tracing approach to complement the extracted models with incomplete metadata. 
Specifically, we trace API calls during DL app runtime, trigger model inference, and pinpoint the APIs associated with encrypted models to capture interchanged data objects. 
After obtaining these objects, we check if they pertain to TFLite metadata (e.g., TensorMetadata), and if so, retrieve and integrate the corresponding values to refine the extracted models.

After model extraction, each model is loaded into memory and inferred on random-generated data to ensure model completeness and
usability.
This is essential for our watermark as we require model information and inference to select targeted (watermarked) layers and solve their optimal parameters.

\subsection{Model Rooting}
\label{sec:mod_rooting}


In this stage, we make the extracted model writable in order to perform the model parameter modification at a later stage.
One possible solution introduced by the previous work~\cite{li2021deeppayload} is to directly convert an on-device model back to its modifiable format via the official reverse tool\footnote{https://github.com/tensorflow/tensorflow/tree/r1.9/tensorflow/contrib/lite}.
However, it has been obsolete due to security concerns.
\plain{To address this limitation, we propose Model Rooting, a systematic process to reconstruct a writable on-device model based on the original one by mapping schema-defined structures to writable programming classes.
As shown in Figure~\ref{fig:workflow_model_rooting}}, the detailed steps of Model Rooting are as follows:

\begin{figure}[t!]
\centering
\includegraphics[width=\linewidth]{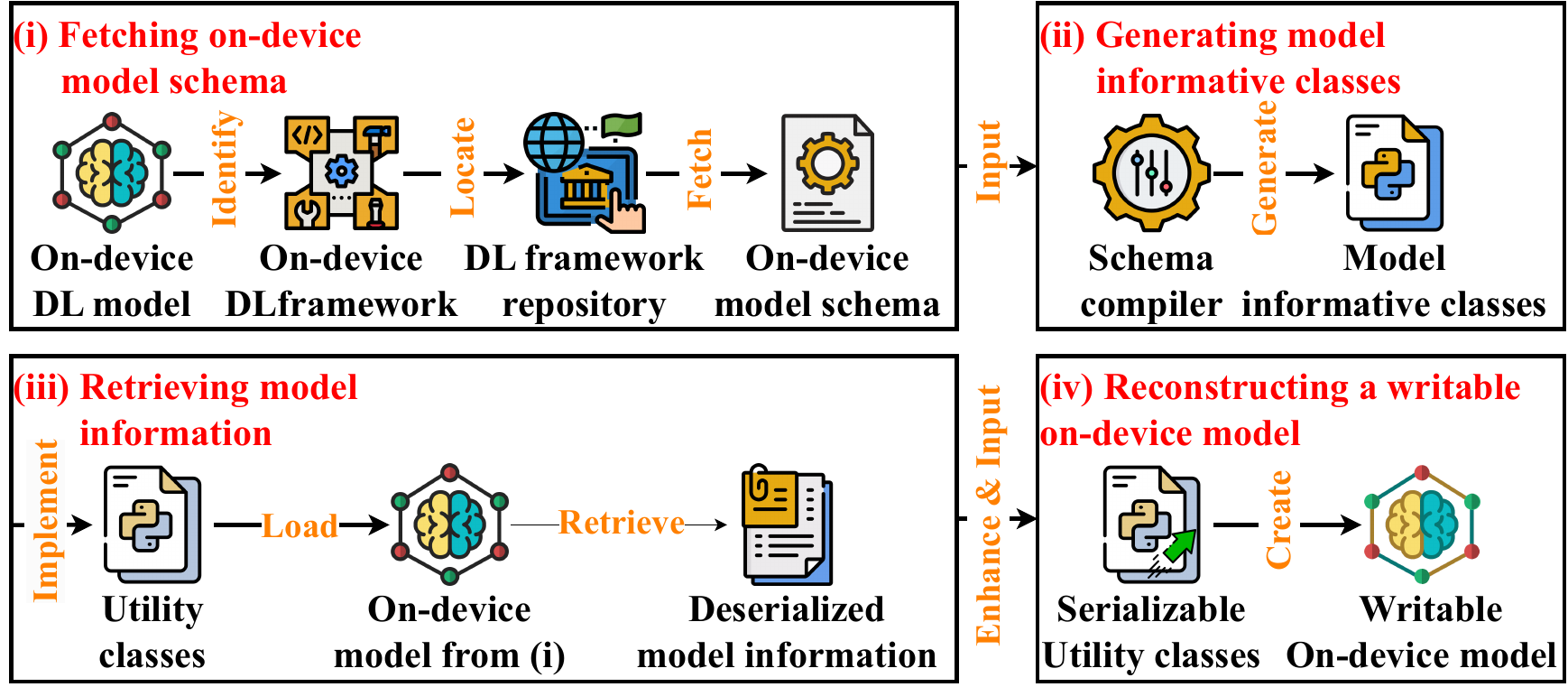}
\caption{\plain{The workflow of Model Rooting.}}
\label{fig:workflow_model_rooting}
\vspace{-1.5em}
\end{figure}


\noindent
\plain{
\textbf{(i) Fetching On-device model schema.}
Model Rooting begins with fetching the schema of an on-device model, which acts as a blueprint for model serializable data structures such as data types and objects.
Fetching the schema ensures a comprehensive understanding of the data structures used by the model and forms the foundation for reconstructing a writable counterpart.
To ensure the success of Model Rooting, the obtained schema must be up-to-date, as the latest version has backward compatibility and ensures consistent reconstruction across diverse models relying on different schema versions.
Taking TFLite models, the most popular on-device DL models, as an example, they are expressed in FlatBuffers\footnote{https://google.github.io/flatbuffers}, an efficient portable format where data structures are predefined in an Interface Definition Language-based schema.
We can fetch the schema\footnote{https://github.com/tensorflow/tensorflow/blob/master/tensorflow/lite/schema} from TensorFlow official GitHub repository to prepare for writable model reconstruction.
For models from other frameworks, such as Pytorch Mobile~\cite{pytorchmobile}, Model Rooting can be achieved by leveraging the cross-framework compatibility offered by Open Neural Network Exchange~\cite{onnx}, as detailed in Section~\ref{sec:discussion}.}


\noindent
\plain{
\textbf{(ii) Generating model informative classes.}
As serialized data in on-device models are inherently immutable due to their read-only nature, direct modifications to extracted models are infeasible.
 }
To address this challenge, we consider generating a set of programming language classes called model informative classes that correspond to schema-defined data structures.
With such classes, we can read the serialized data from the extracted model utilizing the concrete objects of these classes, and then leverage them to make the extracted model's data writable.
\plain{An example of model informative classes generated from the TFLite schema is shown in Figure~\ref{fig:model_info_classes}, with further details provided in Appendix~\ref{sec:mod_info_cls_generation}.}

\begin{figure}[t!]
\centering
\includegraphics[width=\linewidth]{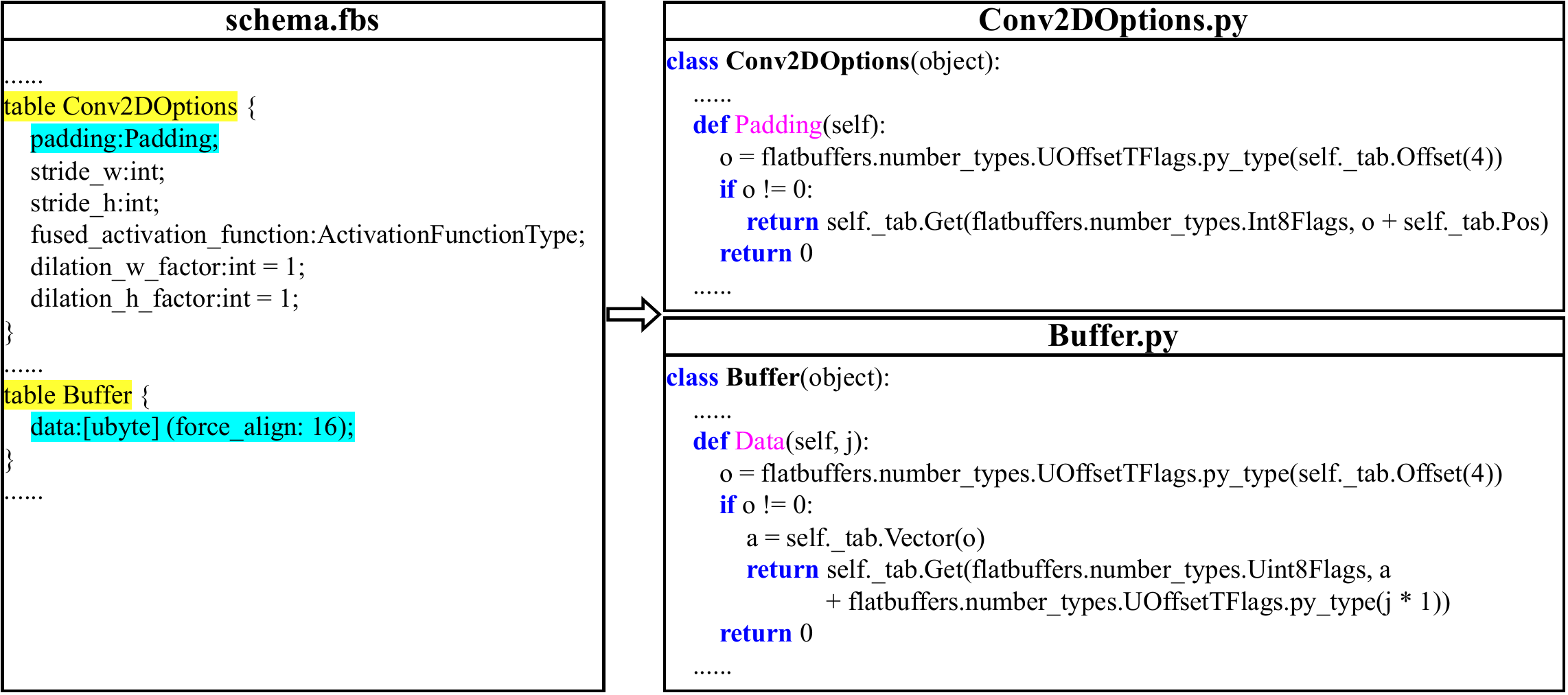}
\caption{An example of model informative classes generated
from the TFLite schema. Yellow and cyan highlight the data structures (tables and their exemplary fields) of the TFLite model schema. Python keywords, class and method names are emphasized in blue, dark and pink for clarity.}
\label{fig:model_info_classes}
\vspace{-1em}
\end{figure}

\noindent\textbf{(iii) Retrieving model information.}
\plain{Using the model informative classes, we retrieve the extracted model information.}
However, the obtained information is not well-organized as it is retrieved holistically in an orderless manner by all field matching methods of the data structures utilized by the model.
\plain{This proves inefficient and error-prone when modifying information on a specific model data structure like a TFLite FlatBuffer table at a later stage, as a model typically contains numerous operators and parameters.}
Therefore, to overcome this problem, we design and implement a set of utility classes associated with the model informative classes, which allow us to systematically retrieve the extracted model information, i.e., the values of each data structure adopted by the model are read separately.
\plain{An example of utility classes for TFLite model informative classes is shown in Figure~\ref{fig:utility_classes}, with further details provided in Appendix~\ref{sec:util_cls_implementation}.}

\begin{figure}[t!]
    \centering
    \begin{subfigure}[t]{0.47\linewidth}
        \centering
        \includegraphics[width=\textwidth]{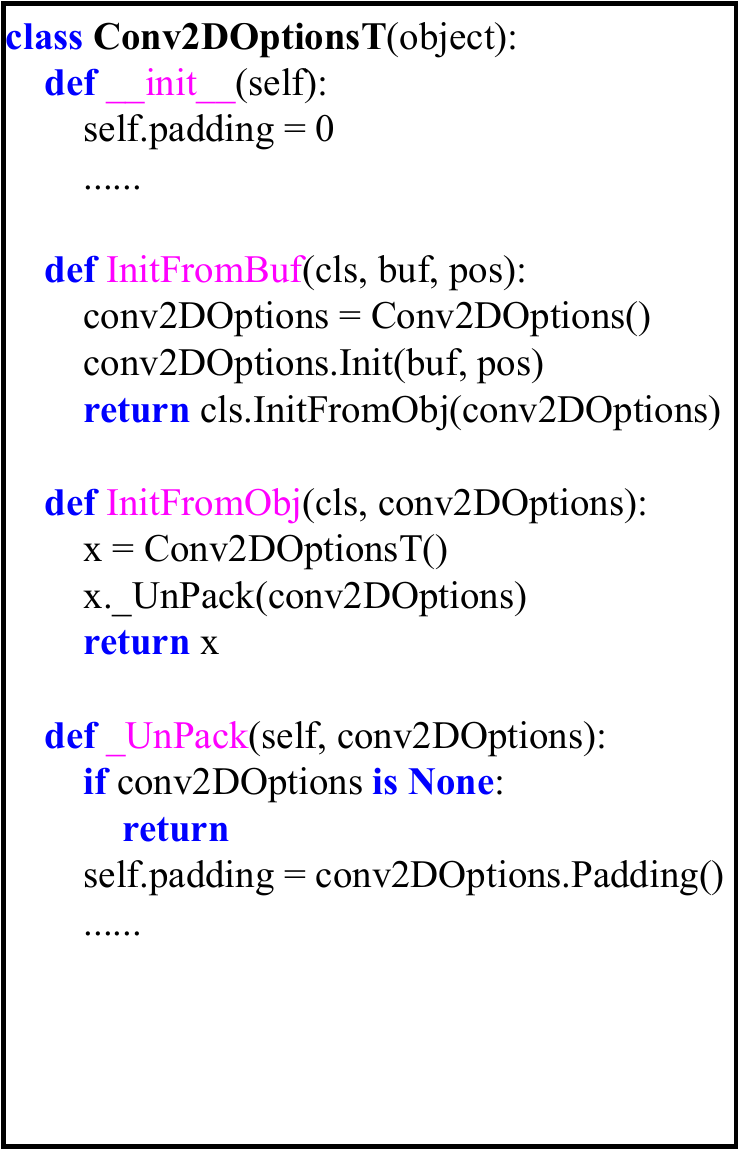}
        \caption{Conv2DOptions utility class.}
        \label{subfig:utility_class_Conv2DOptionsT}
    \end{subfigure}
    \hfill
    \begin{subfigure}[t]{0.47\linewidth}
        \centering
        \includegraphics[width=\textwidth]{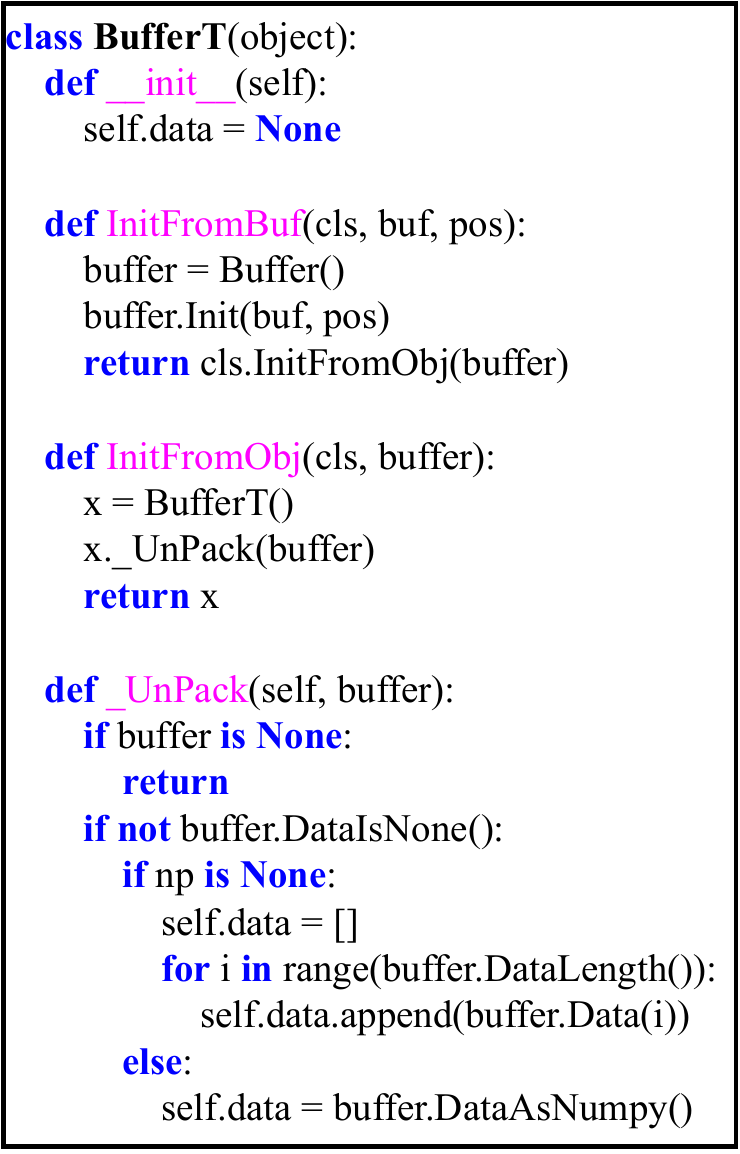}
        \caption{Buffer utility class.}
        \label{subfig:utility_class_BufferT}
    \end{subfigure}
\vspace{-0.5em}    
\caption{An example of TFLite utility classes.}
\label{fig:utility_classes}
\vspace{-1.5em}  
\end{figure}

 \begin{lstlisting}[style=mystyle, caption={An example of data serialization functions for TFLite utility classes. The upper and lower parts correspond to \texttt{\small Conv2DOptionsT} and \texttt{\small BufferT} utility classes, respectively.}, label={list:utility_class_data_serialization}]
def @Conv2DOptionsStart@(builder): builder.StartObject(6)
def @Conv2DOptionsAddPadding@(builder, padding): builder.PrependInt8Slot(0, padding, 0)
......
def @Conv2DOptionsEnd@(builder): return builder.EndObject()

def @Pack@(self, builder):
    Conv2DOptionsStart(builder)
    Conv2DOptionsAddPadding(builder, self.padding)
    ......
    conv2DOptions = Conv2DOptionsEnd(builder)
    return conv2DOptions
    
--------------------------------------------------------
def @BufferStart@(builder): builder.StartObject(1)
def @BufferAddData@(builder, data): builder.PrependUOffsetTRelativeSlot(0, flatbuffers.number_types.UOffsetTFlags.py_type(data), 0)
def @BufferStartDataVector@(builder, numElems): return builder.StartVector(1, numElems, 1)
def @BufferEnd@(builder): return builder.EndObject()

def @Pack@(self, builder):
    if self.data is not None:
        if np is not None and type(self.data) is np.ndarray:
            data = builder.CreateNumpyVector(self.data)
        else:
            BufferStartDataVector(builder, len(self.data))
            for i in reversed(range(len(self.data))):
                builder.PrependUint8(self.data[i])
            data = builder.EndVector(len(self.data))
    BufferStart(builder)
    if self.data is not None:
        BufferAddData(builder, data)
    buffer = BufferEnd(builder)
    return buffer
\end{lstlisting}

\noindent\textbf{(iv) Reconstructing a writable on-device model.}
\plain{After systematically organizing the extracted model information with utility classes, we use them to reconstruct a writable extracted model. 
However, these classes lack data serialization capabilities, which is essential for reconstructing a writable model, as it must support reserialization of data following modifications.}
To address this limitation, we improve the utility classes by adding data serialization functions.
\plain{An example of data serialization functions for TFLite utility classes is depicted in Listing~\ref{list:utility_class_data_serialization}, with further details provided in Appendix~\ref{sec:data_ser_function}.}

Equipped with the improved utility classes, we can now reconstruct a writable extracted model based on the original one.
First, we initiate instances of utility classes that align with the data structures used by the extracted model and arrange them in the same order as in the model.
Then, we sequentially deserialize the extracted model data into these instances so that such data becomes modifiable.
Lastly, we form a writable extracted model utilizing these instances as they encapsulate model data and possess the capability for serialization.

\subsection{Model Reweighting}
\label{sec:mod_reweighting}
After performing Model Rooting, we obtain a writable extracted model poised for watermark.
To watermark the model, we employ black-box watermarks to embed backdoor triggers into the model as the watermark, which thwarts adversary efforts to redeploy the stolen model via encryption and enables the app store administrator to verify model ownership through inputs stamped with the triggers (as discussed in Section~\ref{sec:thr_model}).
However, existing black-box watermarks~\cite{adi2018turning,zhang2018protecting,li2019prove,namba2019robust,jia2021entangled} heavily hinge on the availability of gradient information within a to-be-protected model to embed watermarks, rendering them inapplicable for on-device models that are inference-only with backpropagation disabled.
Hence, we propose Model Reweighting to enable post-deployment watermark insertion for on-device models by leveraging existing training-free backdoor attacks~\cite{dumford2020backdooring,hong2021handcrafted} instead of developing from scratch.
Note that Model Reweighting is not confined to the adaption of such attacks and can seamlessly incorporate any similar mechanisms as well.

\begin{table}[t!]

\renewcommand\arraystretch{1}
\centering
\Large
\caption{Summary of two training-free backdoor attacks. TWP: Targeted Weight Perturbations~\cite{dumford2020backdooring}, and HP: Handcrafted Perturbations~\cite{hong2021handcrafted}.}
\label{table:TWP_HP_brevity}
\resizebox{\linewidth}{!}{
\begin{tabular}{c|c|c|c} 
\toprule
Approach & Characteristic                                                                                    & Model's dataset & Domain                                                    \\ 
\hline
TWP~\cite{dumford2020backdooring}      & \begin{tabular}[c]{@{}c@{}}Greedy search-based \\model weight modification\end{tabular}           & Require         & \begin{tabular}[c]{@{}c@{}}Computer\\vision\end{tabular}  \\
HP~\cite{hong2021handcrafted}       & \begin{tabular}[c]{@{}c@{}}Logical connectives-based~\\neuron parameter~modification\end{tabular} & Require         & \begin{tabular}[c]{@{}c@{}}Computer\\vision\end{tabular}  \\
\bottomrule
\end{tabular}
}
\vspace{-1em}
\end{table}

In particular, Model Reweighting is characterized by the availability of the data expressed in the extracted model label file $\mathcal{L}$.
In real-world DL apps, some of them have no label files or the model training data within their label files are limited due to the privacy concerns (e.g., prostate cancer recognition DL app).
Thus, there are three watermark scenarios for Model Reweighting, denoted by $MR_{WS} = \mathcal{L}$, where $\mathcal{L}$ can be label (data) missing $dm$, label exists but data-scarce $ds$ or label exists and data-abundant $da$.
Since training-free backdoor attacks~\cite{dumford2020backdooring,hong2021handcrafted} have been well studied, we omit its technical details and briefly present them in Table~\ref{table:TWP_HP_brevity}.
In the following, we only describe how we customize them for watermark injection in each watermark scenario.

\textbf{Watermark-1: $MR_{WS}=dm$.} 
We begin with the most difficult scenario where the goal of the app store administrator is to embed a watermark in the writable extracted model while preserving its utility in the absence of the label file, i.e., the model's dataset is unable to collect.
As both TWP and HP necessitate a dataset $D$ for backdoor trigger injection, we propose to synthesize $D$ by leveraging DiffusionDB~\cite{wang2022diffusiondb} that has 14 million diverse images from Stable Diffusion~\cite{rombach2022high}
Formally, given a writable extracted model $M$, we first obtain its input dimension $I \in \mathbb{R}^{m\times n}$ and the number of predicted labels $N$.
Based on this, we randomly select a substantial amount of images from DiffusionDB and resize them to produce a set of inputs $\bm{X} \in I$.
These inputs are then fed into $M$ to generate their predicted distributions.
Subsequently, we label each sample in $\bm{X}$ with the class label $l_n$ that has the highest outputted probability in its distribution.
Despite $\bm{X}$ contains patterns that $M$ can recognize, it may also include redundant patterns that have a detrimental impact on watermark performance, as these non-targeted patterns are factored into the process of watermark parameter solving.
To deal with this issue, we adopt image segmentation~\cite{minaee2021image} to extract the common patterns of samples from each class label in $\bm{X}$ and use them to synthesize new samples to construct $D$.
Finally, we augment the resulting $D$ to enrich the data diversity, which can improve watermark performance and better maintain model utility.

Next, we select a target label $l_t \in L$, integrate same trigger patterns as in~\cite{hong2021handcrafted} into a portion of $l_t$ samples $\bm{x}_t \in D$, and relabel $\bm{x}_t$ with watermark labels $l_{wm} \in L \wedge l_{wm} \not = l_t$ to construct watermark samples $\bm{x}_{wm}$.
Then, we split $D$ into $D_{infer}$ and $D_{test}$ for the preparation of watermark parameter solving.
Here, we do not have a training dataset as we only solve watermark parameters through the model feed-forward process.
Finally, we adopt training-free backdoor attack algorithms (e.g., TWP~\cite{dumford2020backdooring} or HP~\cite{hong2021handcrafted}) to solve the watermark parameters for $M$ by feeding $D_{infer}$ into it.
The effectiveness of the resultant parameters is evaluated on $D_{test}$.
If it satisfies a pre-defined goal (watermark success rate and accuracy), we write the resultant parameters back into $M$, otherwise, we need to repeat the whole procedure (i.e., from data synthesis to watermark parameter solving) as $D$ is the crux of Model Reweighting.
The overall process of the Watermark-1 is illustrated in Algorithm~\ref{alg:watermark_1}.

\begin{algorithm}[t!]
\caption{Watermark-1 Algorithm}
\label{alg:watermark_1}
\fontsize{8}{9}\selectfont
\begin{algorithmic}[1]
\REQUIRE ~~ \\ 
$M$: writable on-device model\\
$DDB$: DiffusionDB\\
$\plain{A}$: training-free backdoor algorithm
\ENSURE ~~ \\
$M_{wm}$: watermarked on-device model satisfying a pre-defined goal

\STATE $I \in \mathbb{R}^{m\times n} \leftarrow M,\ N \leftarrow M; \;\blacktriangleright$ obtain the input dimension and number of predicted labels of $M$

\STATE $\bm{X} \in I \leftarrow resize(random(DDB)); \;\blacktriangleright$ produce the inputs align with $I$

\STATE $\bm{L} \leftarrow softmax(M(\bm{X})); \;\blacktriangleright$ obtain the labels for each sample in $\bm{X}$

\STATE $P \leftarrow img\_segment(\bm{X}, \bm{L}); \;\blacktriangleright$ extract common patterns from samples of each class in $\bm{X}$

\STATE $D \leftarrow augment(synthesize(P)); \;\blacktriangleright$ construct the dataset excluding redundant patterns

\STATE $\bm{x}_{wm},\bm{l}_{wm} \leftarrow relabel(select\_integrate(D, l_t \in L), l_{wm} \in L \wedge l_{wm} \not = l_t); \;\blacktriangleright$ generate watermark samples and corresponding reassigned labels

\STATE $D_{infer},D_{test} \leftarrow split(D); \;\blacktriangleright$ split the data into inference and test sets

\STATE $\bm{\theta}_{wm} \leftarrow \plain{A}(D_{infer}, D_{test}, M); \;\blacktriangleright$ solve watermark parameters for $M$

\STATE $M_{wm} \leftarrow write(\bm{\theta}_{wm}, M); \;\blacktriangleright$ write the final parameters back to $M$

\RETURN $M_{wm}$
\end{algorithmic}
\end{algorithm}

\textbf{Watermark-2: $MR_{WS}=ds$.}
We now consider a scenario in which the app store administrator has the same goal as Watermark-1 and the data expressed in the label file is scarce.
The steps of this watermark are the same as those of Watermark-1, except for the difference in data preparation (construction of $D$).
To be specific, we first use Google Dataset Search\footnote{https://datasetsearch.research.google.com/} to collect as much data as possible according to the label file.
Then, we feed the collected data into the writable extracted model to select the correctly predicted samples as such data may contain noise or have a different distribution from the model training data, impacting watermark performance.
After selecting the correctly predicted samples, we scrutinize the sample size of each class and apply the data synthesis proposed in Watermark-1 to the classes with no or few samples due to data scarcity.
In the final step of data preparation, we augment the complete data resulting from the previous step with the same approach as in Watermark-1.

\textbf{Watermark-3: $MR_{WS}=da$.}
We finally consider a relaxed scenario where the app store administrator's goal is identical to Watermark-1 and Watermark-2 but the data presented in the label file is abundant.
Since sufficient data can be collected in this scenario, we follow the same data preparation process as in Watermark-2 but skip the step of data synthesis to construct $D$.
After obtaining $D$, the remaining steps are consistent with those of Watermark-1.

\plain{While Model Reweighting leverages established training-free backdoor algorithms such as TWP~\cite{dumford2020backdooring} and HP~\cite{hong2021handcrafted}, these methods heavily rely on substantial data for parameter search, limiting their robustness in on-device scenarios like $dm$ and $ds$ as demonstrated in Section~\ref{sec:wat_effectiveness}.
To address these limitations, we propose Feed-Forward Knowledge Editing Watermarking (FFKEW), a training-free backdoor algorithm that embeds watermarks into on-device models by precisely editing model knowledge within a single feed-forward run.
}

\plain{Building on the dataset $D$ and watermark samples $\bm{x}_{wm}$ prepared according to the extracted model $M$ and its label file scenario, FFKEW begins by selecting a target layer $t$ of $M$ for watermark injection.
Typically, DL models for computer vision tasks feature task-specific heads (e.g., classification head for image recognition or classification and location heads for object detection) that generate logits for predictions. 
Thus, $t$ can be selected from $M$'s task-specific head(s), such as a fully connected layer when $M$ performs image recognition.}

\plain{With the selected $t$, we locate its position in $M$ and use it to obtain the watermark input $\bm{T}_{input}^{wm}$ and output $\bm{T}_{output}^{wm}$ logits of $t$ by feeding $D_{infer}^{wm}$ into $M$.
The computation between $\bm{T}_{input}^{wm}$ and $\bm{T}_{output}^{wm}$ is as follows:
\begin{equation}
\label{eq:tar_lay_output}
\bm{T}_{output}^{wm} = \bm{T}_{input}^{wm} \times \bm{W}_{t} + \bm{b}_{t}
\end{equation}
where $\bm{W}_{t}$ and $\bm{b}_{t}$ are the weight and bias of $t$, respectively.
As $\bm{T}_{output}^{wm}$ determines the prediction of $M$ (i.e., the label of each sample is determined by the highest logit among all associated logits) and is computed by Equation~\ref{eq:tar_lay_output}, modification of $\bm{W}_{t}$ can directly affect the model prediction to embed watermarks.
The intuition behind this is that $\bm{W}_{t}$ represents the prototype (knowledge) of each label class learned by $M$ and thus changing $\bm{W}_{t}$ is equivalent to editing the knowledge of $M$ with respect to the classes it learns.
To do so, we modify $\bm{T}_{output}^{wm}$ as follows:
\begin{equation}
\label{eq:tar_lay_log_swap}
\bm{T}_{output}^{wm}{'} = swap(\bm{t}^i, l^i, l_t), \bm{t}^i \in \bm{T}_{output}^{wm}, l^i \in \bm{l_n}
\end{equation}
where $\bm{t}^i$ is the logits related to an individual watermark sample, $l^i$ is its label, $l_t$ is the selected target label, and $swap$ is the function that swaps the logits of $l^i$ and $l_t$ based on their indices in $\bm{t}^i$.
By doing this, the highest logit associated with the original label for each watermark sample is reassigned to its new watermark label.
Finally, we solve the watermark weight $\bm{W}_{t}'$ as follows:
\begin{equation}
\label{eq:tar_lay_par_solve}
\bm{W}_{t}{'} = MPI(\bm{T}_{input}^{wm}) \times (\bm{T}_{output}^{wm}{'}-\bm{b}_{t})
\end{equation}
where $MPI$ is Moore–Penrose inverse~\cite{moore1960generalized} function utilized for computing the generalized inverse of a matrix.
This ensures us can solve $\bm{W}_{t}'$ even if $\bm{T}_{input}^{wm}$ is not invertible.}

\subsection{DL App Reassembling}
Upon completion of Model Reweighting, we obtain a watermarked on-device model that satisfies the app store administrator's goal.
Next, we need to substitute the watermarked model for the original one in the DL app so that it inherits the behavior expected by the administrator.
Recall that the DL app (APK) has been decomposed to the original form for model extraction, which specifies the position of the original model in the decomposed APK.
Thus, we can directly replace the original model in the decomposed APK with the watermarked one based on the previously obtained position information.
Finally, we utilize Apktool to rebuild the decomposed APK back to binary APK that can be published for open access.

\section{Evaluation}

\subsection{Experimental Setup}
\label{sec:exp_setup}

\textbf{Datasets.}
We utilize four datasets to evaluate \watermark, including FMNIST, CIFAR10, GTSRB, and SVHN.
These datasets are commonly employed as benchmark datasets for various security and computer vision tasks.
A brief description of each dataset is presented below.

\begin{itemize}[leftmargin=*]
  \item \textbf{FMNIST~\cite{xiao2017fashion}.} The FMNIST is a Zalando's article image dataset consisting of 70,000 grayscale images of size 28$\times$28 from 10 classes.
  This dataset is balanced (i.e., 7,000 images per class) and it is divided into 60,000 training images and 10,000 testing images.

  \item \textbf{CIFAR10~\cite{krizhevsky2009learning}.} The CIFAR10 comprises 60,000 colour images with size 32$\times$32.
  These images are equally distributed in 10 classes and each of them has 5,000 training and 1,000 testing images.

  \item \textbf{GTSRB~\cite{stallkamp2011german}.} The GTSRB contains 51,839 colour images from 43 classes of traffic signs.
  The dimensions of these images vary from 15$\times$15 to 250$\times$250.
  Within the dataset, 39,209 and 12,630 images are used for training and testing, respectively.

  \item \textbf{SVHN~\cite{netzer2011reading}.} The SVHN is composed of 73,257 and 26,032 training and testing images of printed digits from 0 to 9, which are captured from Google Street View images.
  Each digit image is coloured and has a size of 32$\times$32.
  Moreover, this dataset is a noisy dataset as it introduces distracting digits to the peripheral regions of the targeted digit.
\end{itemize}

\textbf{Dataset Configuration.}
We adopt the training and testing datasets of each dataset as the inference and testing datasets to realize and evaluate \watermark.
Since there are three cases for the availability of the inference data (see Section~\ref{sec:mod_reweighting}), we illustrate the detailed configuration of each case below.

\begin{itemize}[leftmargin=*]
  \item \textbf{Data missing ($dm$).}
  In this case, the inference dataset is discarded, while the testing dataset is used for watermark evaluation.
  The synthetic dataset is of equivalent size to the original inference dataset, providing sufficient data for watermark realization and facilitating the comparison of watermark performance under different scenarios (i.e., $ds$ and $da$).

  \item \textbf{Data-scarce ($ds$).}
  In this case, 10\% of data from each class in the inference dataset is utilized to realize \watermark and the testing dataset serves the same purpose as in $dm$.
  We believe that this ratio is reasonable for simulating the case of data scarcity as the amount of available data per class in each inference dataset is small (e.g., 600 images per class for FMNIST) in terms of DL practice~\cite{lecun2015deep}.
  Similar to $dm$, the sub-inference dataset is augmented to align with the size of the original inference dataset.

  \item \textbf{Data-abundant ($da$).}
  In this case, the complete inference dataset is used to implement \watermark and the usage of the testing dataset is congruent with that in $dm$.
\end{itemize}

\textbf{On-device Models.}
We use three different neural network architectures for the construction of to-be-protected on-device models, including MobileNetV2~\cite{sandler2018mobilenetv2}, InceptionV3~\cite{szegedy2016rethinking}, and EfficientNetV2~\cite{tan2021efficientnetv2}.
These models are widely utilized in real-world DL apps due to their cost-effectiveness~\cite{huang2021robustness,deng2022understanding}.
Table~\ref{table:vic_mod_accuracy} reports the performance of each model on the four datasets.

\begin{table}[t!]
\centering
\footnotesize
\renewcommand\arraystretch{1}
\caption{Test accuracy of three to-be-protected on-device models trained on four different datasets.}
\label{table:vic_mod_accuracy}
\resizebox{\linewidth}{!}{
\begin{tabular}{c|c|c|c|c} 
\toprule
\diagbox{Model}{Dataset} & FMNIST & CIFAR10 & GTSRB  & SVHN    \\ 
\midrule
MobileNetV2              & 0.8908 & 0.8602  & 0.8838 & 0.8772  \\
InceptionV3              & 0.8419 & 0.7793  & 0.9184 & 0.7023  \\
EfficientNetV2           & 0.8947 & 0.8495  & 0.9152 & 0.8772  \\
\bottomrule
\end{tabular}
}
\vspace{-1em}
\end{table}

\textbf{Training-free Backdoor Attack Algorithms.}
Since \watermark watermarks the on-device models by exploiting training-free backdoor attacks, we consider two existing approaches for \plain{comparison}: Dumford et al.~\cite{dumford2020backdooring} (denoted as TWP, representing the proposed Targeted Weight Perturbations method) and Hong et al.~\cite{hong2021handcrafted} (denoted as HP, representing the proposed Handcrafted Perturbations method).
TWP employs a greedy search to identify the targeted weights in a model and modify them accordingly to realize the attack.
HP locates the targeted neurons of a model through neuron activation ablation analysis and manipulates their parameters to implement logical operations for encoding malicious behaviors into the model.
Due to the read-only characteristic of the on-device model, these two approaches cannot be directly applied to the on-device models.
We thus use them to embed watermarks after Model Rooting.

\textbf{Evaluation Metrics.}
We utilize Watermark Success Rate (WSR) and Accuracy (ACC) as evaluation metrics for \watermark.
WSR measures the probability of a watermarked model
correctly classifying watermark samples as the watermark label.
ACC measures the accuracy of the watermarked model on its original task. 
Concretely, WSR is calculated from the division of the number of successful watermark samples by the total number of watermark samples.
ACC is computed as the ratio of the number of correctly classified non-watermark samples to the total number of non-watermark samples.

\textbf{Watermark Setup.}
For demonstration purposes, we randomly choose a single target label from each dataset, embed triggers into a subset of samples from the selected label, and relabel these samples to construct watermark samples as described in Section~\ref{sec:mod_reweighting}. 
In particular, the target labels for the four datasets FMNIST, CIFAR10, GTSRB and SVHN are "dress", "truck", "speed limit 80" and "digit one", respectively.
Unless otherwise mentioned, such target labels are adopted for all our experiments.
Note that the images within each dataset are resized to dimensions of $96\times96\times3$ to ensure consistency across the datasets.

\begin{figure}[t!]
\centering
\includegraphics[width=0.85\linewidth]{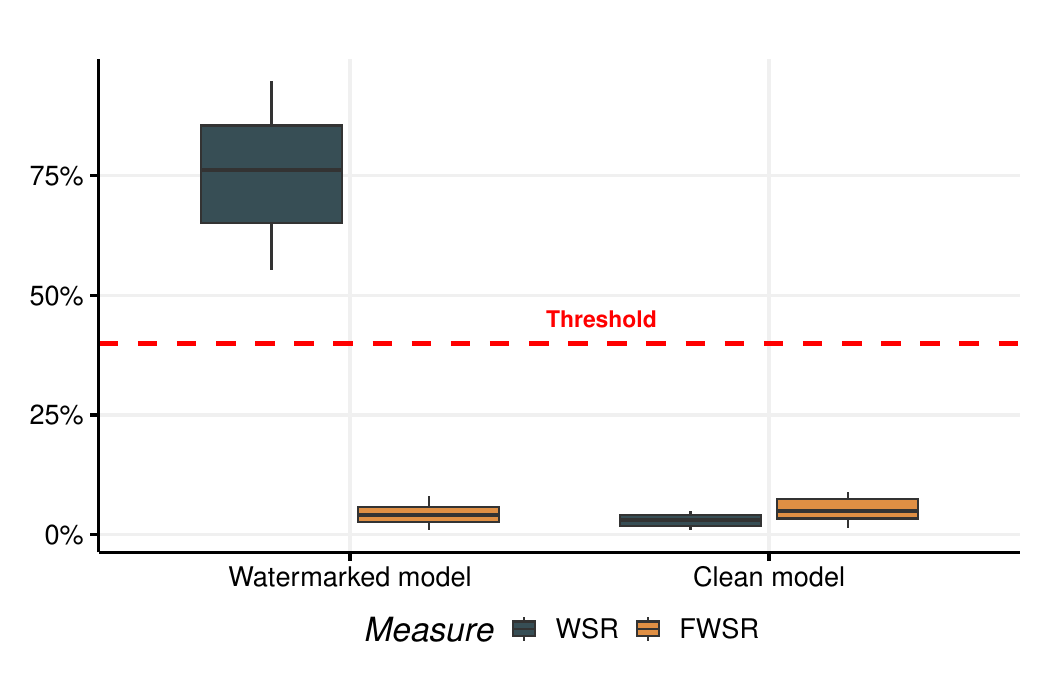}
\caption{Watermark Success Rate (WSR) and False Watermark Success Rate (FWSR) of watermarked and clean models.}
\label{fig:wsr_threshold}
\vspace{-2em}
\end{figure}

\textbf{Watermark Threshold.}
For precise ownership verification of the suspect on-device model, an appropriate WSR threshold is essential.
To this end, we conduct experiments with 100 watermarked and 100 clean on-device models trained on various combinations of model architectures and datasets, and then compute their WSR and the probability of trigger-stamped samples from non-target classes as the watermark label (i.e., False Watermark Success Rate), as shown in Figure~\ref{fig:wsr_threshold}.
Based on the results, we set the threshold to 40\% as it ensures reliable watermark detection (substantially below the WSR yet far above the FWSR of watermarked models) while preventing false ownership claims of the clean models (well above their WSR and FWSR).

\begin{table*}[hbt!]
\centering
\renewcommand\arraystretch{1}
\caption{Watermark Success Rate (WSR) and Accuracy (ACC) of \watermark in \colorbox{red!30}{data missing ($dm$)} scenario, where B and A refer to Before and After watermarking.}
\label{table:wm_dm_results}
 \resizebox{1\linewidth}{!}{\begin{tabular}{c|c|cc|cc|cc|cc} 
\toprule
\multirow{2}{*}{Algorithm} & \multirow{2}{*}{Model} & \multicolumn{2}{c|}{FMNIST} & \multicolumn{2}{c|}{CIFAR10} & \multicolumn{2}{c|}{GTSRB}  & \multicolumn{2}{c}{SVHN}     \\ 
\cline{3-10}
                        &                        & WSR     & ACC (B/A)         & WSR     & ACC (B/A)          & WSR     & ACC (B/A)         & WSR     & ACC (B/A)          \\ 
\hline
\multirow{3}{*}{TWP~\cite{dumford2020backdooring}}    & MobileNetV2            & \colorbox{red!30}{67.91\%} & 89.19\% / 63.12\% & \colorbox{red!30}{62.38\%} & 85.55\% / 63.45\%  & \colorbox{red!30}{61.63\%} & 88.06\% / 60.97\% & \colorbox{red!30}{66.87\%} & 87.10\% / 59.43\%  \\
                        & InceptionV3            & \colorbox{red!30}{62.83\%} & 84.34\% / 62.86\% & \colorbox{red!30}{60.50\%} & 77.77\% / 61.37\%  & \colorbox{red!30}{65.08\%} & 91.62\% / 66.48\% & \colorbox{red!30}{64.35\%} & 68.54\% / 44.69\%  \\
                        & EfficientNetV2         & \colorbox{red!30}{63.25\%} & 89.11\% / 66.53\% & \colorbox{red!30}{65.73\%} & 84.55\% / 64.42\%  & \colorbox{red!30}{62.47\%} & 91.29\% / 65.70\% & \colorbox{red!30}{67.96\%} & 86.91\% / 62.20\%  \\ 
\hdashline
\multirow{3}{*}{HP~\cite{hong2021handcrafted}}     & MobileNetV2            & \colorbox{red!30}{79.24\%} & 89.19\% / 61.46\% & \colorbox{red!30}{71.05\%} & 85.55\% / 62.89\%  & \colorbox{red!30}{74.76\%} & 88.06\% / 59.68\% & \colorbox{red!30}{75.02\%} & 87.10\% / 60.52\%  \\
                        & InceptionV3            & \colorbox{red!30}{70.22\%} & 84.34\% / 63.71\% & \colorbox{red!30}{73.92\%} & 77.77\% / 63.27\%  & \colorbox{red!30}{73.15\%} & 91.62\% / 64.90\% & \colorbox{red!30}{71.77\%} & 68.54\% / 45.33\%  \\
                        & EfficientNetV2         & \colorbox{red!30}{72.67\%} & 89.11\% / 65.39\% & \colorbox{red!30}{76.81\%} & 84.55\% / 66.03\%  & \colorbox{red!30}{78.28\%} & 91.29\% / 63.30\% & \colorbox{red!30}{78.56\%} & 86.91\% / 61.79\%  \\
\hdashline
\multirow{3}{*}{\textcolor{black}{FFKEW}}   & \textcolor{black}{MobileNetV2}            & \colorbox{red!30}{\textcolor{black}{\textbf{89.60\%}}} & \textcolor{black}{89.19\% / \textbf{76.58\%}}   & \colorbox{red!30}{\textcolor{black}{\textbf{85.78\%}}} & \textcolor{black}{85.55\% / \textbf{79.17\%}}  & \colorbox{red!30}{\textcolor{black}{\textbf{84.88\%}}} & \textcolor{black}{88.06\% / \textbf{75.30\%}}  & \colorbox{red!30}{\textcolor{black}{\textbf{83.41\%}}} & \textcolor{black}{87.10\% / \textbf{80.72\%}}    \\
                        & \textcolor{black}{InceptionV3}            & \colorbox{red!30}{\textcolor{black}{\textbf{81.80\%}}} & \textcolor{black}{84.34\% /~\textbf{74.17\%}}  & \colorbox{red!30}{\textcolor{black}{\textbf{86.00\%}}} & \textcolor{black}{77.77\% /~\textbf{74.23\%}} & \colorbox{red!30}{\textcolor{black}{\textbf{83.42\%}}} & \textcolor{black}{91.62\% / \textbf{82.26\%}} & \colorbox{red!30}{\textcolor{black}{\textbf{80.07\%}}} & \textcolor{black}{68.54\% / \textbf{61.99\%}}   \\
                        & \textcolor{black}{EfficientNetV2}         & \colorbox{red!30}{\textcolor{black}{\textbf{88.00\%}}} & \textcolor{black}{89.11\% / \textbf{77.63\%}}  & \colorbox{red!30}{\textcolor{black}{\textbf{89.24\%}}} & \textcolor{black}{84.55\% / \textbf{80.84\%}}  & \colorbox{red!30}{\textcolor{black}{\textbf{87.11\%}}} & \textcolor{black}{91.29\% / \textbf{80.32\%}}  & \colorbox{red!30}{\textcolor{black}{\textbf{85.82\%}}} & \textcolor{black}{86.91\% / \textbf{83.04\%}}   \\

\bottomrule
\end{tabular}}
\vspace{-0.5em}
\end{table*}

\begin{table*}[hbt!]
\centering
\renewcommand\arraystretch{1}
\caption{Watermark Success Rate (WSR) and Accuracy (ACC) of \watermark in \colorbox{cyan!30}{data-scarce ($ds$)} scenario, where B and A refer to Before and After watermarking.}
\label{table:wm_ds_results}
\resizebox{1\linewidth}{!}{\begin{tabular}{c|c|cc|cc|cc|cc} 
\toprule
\multirow{2}{*}{Algorithm} & \multirow{2}{*}{Model} & \multicolumn{2}{c|}{FMNIST} & \multicolumn{2}{c|}{CIFAR10} & \multicolumn{2}{c|}{GTSRB}  & \multicolumn{2}{c}{SVHN}     \\ 
\cline{3-10}
                        &                        & WSR     & ACC (B/A)         & WSR     & ACC (B/A)          & WSR     & ACC (B/A)         & WSR     & ACC (B/A)          \\ 
\hline
\multirow{3}{*}{TWP~\cite{dumford2020backdooring}}    & MobileNetV2            & \colorbox{cyan!30}{73.65\%} & 89.17\% / 71.38\% & \colorbox{cyan!30}{69.44\%} & 85.63\% / 72.08\%  & \colorbox{cyan!30}{70.37\%} & 88.06\% / 69.88\% & \colorbox{cyan!30}{73.48\%} & 86.98\% / 68.11\%  \\
                        & InceptionV3            & \colorbox{cyan!30}{68.73\%} & 84.36\% / 70.54\% & \colorbox{cyan!30}{67.81\%} & 77.83\% / 66.43\%  & \colorbox{cyan!30}{75.21\%} & 91.61\% / 72.62\% & \colorbox{cyan!30}{70.10\%} & 68.59\% / 52.72\%  \\
                        & EfficientNetV2         & \colorbox{cyan!30}{70.22\%} & 89.31\% / 74.96\% & \colorbox{cyan!30}{71.55\%} & 84.53\% / 70.94\%  & \colorbox{cyan!30}{73.95\%} & 91.31\% / 74.13\% & \colorbox{cyan!30}{75.04\%} & 86.92\% / 73.52\%  \\ 
\hdashline
\multirow{3}{*}{HP~\cite{hong2021handcrafted}}     & MobileNetV2            & \colorbox{cyan!30}{87.46\%} & 89.17\% / 75.30\% & \colorbox{cyan!30}{82.67\%} & 85.63\% / 74.87\%  & \colorbox{cyan!30}{82.33\%} & 88.06\% / 71.90\% & \colorbox{cyan!30}{85.28\%} & 86.98\% / 72.47\%  \\
                        & InceptionV3            & \colorbox{cyan!30}{83.82\%} & 84.36\% / 73.21\% & \colorbox{cyan!30}{84.23\%} & 77.83\% / 70.40\%  & \colorbox{cyan!30}{85.95\%} & 91.61\% / 76.57\% & \colorbox{cyan!30}{83.60\%} & 68.59\% / 57.69\%  \\
                        & EfficientNetV2         & \colorbox{cyan!30}{84.45\%} & 89.31\% / 78.76\% & \colorbox{cyan!30}{86.78\%} & 84.53\% / 73.08\%  & \colorbox{cyan!30}{87.62\%} & 91.31\% / 79.35\% & \colorbox{cyan!30}{86.59\%} & 86.92\% / 76.05\%  \\
\hdashline
\multirow{3}{*}{\textcolor{black}{FFKEW}}   & \textcolor{black}{MobileNetV2}            & \colorbox{cyan!30}{\textcolor{black}{\textbf{93.59\%}}} & \textcolor{black}{89.17\% / \textbf{82.60\%}}   & \colorbox{cyan!30}{\textcolor{black}{\textbf{92.00\%}}} & \textcolor{black}{85.63\% / \textbf{80.14\%}}  & \colorbox{cyan!30}{\textcolor{black}{\textbf{92.09\%}}} & \textcolor{black}{88.06\% / \textbf{82.86\%}}  & \colorbox{cyan!30}{\textcolor{black}{\textbf{92.84\%}}} & \textcolor{black}{86.98\% / \textbf{81.07\%}}    \\
                        & \textcolor{black}{InceptionV3}            & \colorbox{cyan!30}{\textcolor{black}{\textbf{86.58\%}}} & \textcolor{black}{84.36\% / \textbf{81.44\%}}   & \colorbox{cyan!30}{\textcolor{black}{\textbf{91.40\%}}} & \textcolor{black}{77.83\% / \textbf{75.97\%}}   & \colorbox{cyan!30}{\textcolor{black}{\textbf{88.02\%}}} & \textcolor{black}{91.61\% / \textbf{84.93\%}}  & \colorbox{cyan!30}{\textcolor{black}{\textbf{90.07\%}}} & \textcolor{black}{68.59\% / \textbf{66.53\%}}  \\
                        & \textcolor{black}{EfficientNetV2}         & \colorbox{cyan!30}{\textcolor{black}{\textbf{93.00\%}}} & \textcolor{black}{89.31\% / \textbf{85.40\%}} & \colorbox{cyan!30}{\textcolor{black}{\textbf{94.99\%}}} & \textcolor{black}{84.53\% / \textbf{80.93\%}}  & \colorbox{cyan!30}{\textcolor{black}{\textbf{93.42\%}}} & \textcolor{black}{91.31\% / \textbf{87.03\%}}  & \colorbox{cyan!30}{\textcolor{black}{\textbf{93.96\%}}} & \textcolor{black}{86.92\% / \textbf{84.12\%}}   \\

\bottomrule
\end{tabular}}
\vspace{-0.5em}
\end{table*}

\begin{table*}[hbt!]
\centering
\renewcommand\arraystretch{1}
\caption{Watermark Success Rate (WSR) and Accuracy (ACC) of \watermark in \colorbox{green!30}{data-abundant ($da$)} scenario, where B and A refer to Before and After watermarking.}
\label{table:wm_da_results}
\resizebox{1\linewidth}{!}{\begin{tabular}{c|c|cc|cc|cc|cc} 
\toprule
\multirow{2}{*}{Algorithm} & \multirow{2}{*}{Model} & \multicolumn{2}{c|}{FMNIST} & \multicolumn{2}{c|}{CIFAR10} & \multicolumn{2}{c|}{GTSRB}  & \multicolumn{2}{c}{SVHN}     \\ 
\cline{3-10}
                        &                        & WSR     & ACC (B/A)         & WSR     & ACC (B/A)          & WSR     & ACC (B/A)         & WSR     & ACC (B/A)          \\ 
\hline
\multirow{3}{*}{TWP~\cite{dumford2020backdooring}}    & MobileNetV2            & \colorbox{green!30}{81.26\%} & 89.13\% / 86.42\% & \colorbox{green!30}{79.57\%} & 85.44\% / 82.18\%  & \colorbox{green!30}{76.55\%} & 88.06\% / 85.17\% & \colorbox{green!30}{80.52\%} & 87.02\% / 85.30\%  \\
                        & InceptionV3            & \colorbox{green!30}{75.78\%} & 84.42\% / 80.51\% & \colorbox{green!30}{78.82\%} & 77.63\% / 75.42\%  & \colorbox{green!30}{79.37\%} & 91.61\% / 87.94\% & \colorbox{green!30}{78.36\%} & 68.62\% / 65.84\%  \\
                        & EfficientNetV2         & \colorbox{green!30}{80.53\%} & 89.14\% / 83.27\% & \colorbox{green!30}{82.43\%} & 84.54\% / 80.05\%  & \colorbox{green!30}{77.08\%} & 91.31\% / 85.65\% & \colorbox{green!30}{81.87\%} & 86.85\% / 83.22\%  \\ 
\hdashline
\multirow{3}{*}{HP~\cite{hong2021handcrafted}}     & MobileNetV2            & \colorbox{green!30}{95.32\%} & 89.13\% / 88.94\% & \colorbox{green!30}{91.48\%} & 85.44\% / 83.36\%  & \colorbox{green!30}{90.19\%} & 88.06\% / 87.53\% & \colorbox{green!30}{96.35\%} & 87.02\% / 86.19\%  \\
                        & InceptionV3            & \colorbox{green!30}{93.65\%} & 84.42\% / 84.00\% & \colorbox{green!30}{95.03\%} & 77.63\% / 76.71\%  & \colorbox{green!30}{92.53\%} & 91.61\% / 90.20\% & \colorbox{green!30}{93.40\%} & 68.62\% / 67.07\%  \\
                        & EfficientNetV2         & \colorbox{green!30}{92.70\%} & 89.14\% / 87.18\% & \colorbox{green!30}{94.15\%} & 84.54\% / 82.93\%  & \colorbox{green!30}{93.84\%} & 91.31\% / 89.78\% & \colorbox{green!30}{95.61\%} & 86.85\% / 85.92\%  \\
\hdashline
\multirow{3}{*}{\textcolor{black}{FFKEW}}   & \textcolor{black}{MobileNetV2}            & \colorbox{green!30}{\textcolor{black}{\textbf{98.83\%}}} & \textcolor{black}{89.13\%~/ \textbf{89.02\%}}   & \colorbox{green!30}{\textcolor{black}{\textbf{96.39\%}}} & \textcolor{black}{85.44\%~/~\textbf{84.80\%}}  & \colorbox{green!30}{\textcolor{black}{\textbf{96.51\%}}} & \textcolor{black}{88.06\%~/~\textbf{87.59\%}}   & \colorbox{green!30}{\textcolor{black}{\textbf{98.29\%}}} & \textcolor{black}{87.02\%~/~\textbf{86.89\%}}    \\
                        & \textcolor{black}{InceptionV3}            & \colorbox{green!30}{\textcolor{black}{\textbf{94.40\%}}} & \textcolor{black}{84.42\%~/~\textbf{84.27\%}}   & \colorbox{green!30}{\textcolor{black}{\textbf{95.96\%}}} & \textcolor{black}{77.63\%~/~\textbf{77.20\%}}  & \colorbox{green!30}{\textcolor{black}{\textbf{92.60\%}}} & \textcolor{black}{91.61\%~/~\textbf{91.04\%}}  & \colorbox{green!30}{\textcolor{black}{\textbf{94.08\%}}} & \textcolor{black}{68.62\%~/~\textbf{68.43\%}}   \\
                        & \textcolor{black}{EfficientNetV2}         & \colorbox{green!30}{\textcolor{black}{\textbf{98.19\%}}} & \textcolor{black}{89.14\%~/~\textbf{88.75\%}}  & \colorbox{green!30}{\textcolor{black}{\textbf{98.04\%}}} & \textcolor{black}{84.54\%~/~\textbf{83.99\%}}  & \colorbox{green!30}{\textcolor{black}{\textbf{97.21\%}}} & \textcolor{black}{91.31\%~/~\textbf{90.45\%}}  & \colorbox{green!30}{\textcolor{black}{\textbf{98.41\%}}} & \textcolor{black}{86.85\%~/~\textbf{85.98\%}}  \\

\bottomrule
\end{tabular}}
\vspace{-1em}
\end{table*}

\subsection{Effectiveness of \watermark}
\label{sec:wat_effectiveness}
\textbf{Watermark-1: $MR_{WS}=dm$.}
In this watermark scenario, only the to-be-protected on-device models are known to the app store administrator while their label files are absent.
Hence, we first perform Model Rooting on these models to obtain their writable counterparts, followed by adopting the data construction method mentioned in Section~\ref{sec:mod_reweighting} Watermark-1 to generate the inference datasets for watermark embedding. 
The results are shown in Table~\ref{table:wm_dm_results}.
As observed, 
\plain{FFKEW consistently outperforms TWP and HP in WSR and ACC across all models and datasets.
Notably, FFKEW achieves WSRs over 80\% in all cases, far higher than the threshold of 40\%.
In contrast, even with our constructed inference datasets, the average WSRs of TWP and HP are 64.25\% and 74.62\% respectively, which are considerably lower than FFKEW's 85.43\%.
These results suggest the high effectiveness of the proposed model knowledge editing technique (Equation~\ref{eq:tar_lay_log_swap} and~\ref{eq:tar_lay_par_solve}) in facilitating successful watermark embedding.}

\plain{With respect to ACC, FFKEW consistently maintains high accuracies for models trained on diverse datasets after watermark embedding.
For instance, the ACC drop for InceptionV3 trained on CIFAR10 is only 3.54\%. 
Note that the highest ACC drop caused by FFKEW is 12.76\% for MobileNetV2 trained on GTSRB, which still remains lower than the lowest ACC drop rates observed for TWP (16.40\%) and HP (14.50\%).
These results highlight the superiority of FFKEW in preserving watermarked model utility, as the proposed model knowledge editing can accurately associate the correlation between a specific trigger and a watermark label for a to-be-protected model with minimum impact on its understanding of non-watermark labels.}


\textbf{Watermark-2: $MR_{WS}=ds$.}
This watermark considers the scenario where the app store
administrator possesses knowledge of the to-be-protected on-device models and corresponding label files, while facing a scarcity of data expressed within these files.
Based on Section~\ref{sec:mod_reweighting} Watermark-2, we execute Model Rooting to obtain writable models for later watermark parameter rewriting, maximize such models' data collection from the Internet using corresponding label files, and employ Watermark-1's data construction method to generate data for rare label classes to construct comprehensive inference datasets for watermark injection.
Table~\ref{table:wm_ds_results} summarizes the results.
\plain{FFKEW outruns baselines in WSR regardless of model structures or datasets, achieving WSRs well above the 40\% threshold for ownership verification.
For instance, in CIFAR10-trained models, FFKEW achieves an average WSR of 92.80\%, surpassing TWP's 69.60\% and HP's 84.56\%.
This stems from FFKEW's direct knowledge editing on logits, which leverages a model's posterior probability distribution to seamlessly and precisely integrate watermarks.
Moreover, even with partial data from to-be-protected models, the post-watermark ACCs for baselines remain significantly inferior to FFKEW.
This is because baselines optimize model parameters utilizing all inference data from each dataset, including non-watermark samples, which expands the parameter search space and intensifies parameter deviation from original values, ultimately compromising model utility.
In contrast, FFKEW solely relies on watermark samples to solve model parameters, imparting superior capability to uphold model utility.}

\textbf{Watermark-3: $MR_{WS}=da$.}
In Watermark-3, the app store administrator has access to the to-be-protected on-device models and can collect the data present in their label files.
Table~\ref{table:wm_da_results} shows the results for this watermark.
\plain{We observe that FFKEW surpasses baselines on all evaluation metrics.
Expressly, it consistently exhibits the highest WSR while incurring minimal ACC degradation in any combination of models and datasets.
This indicates that FFKEW can better embed watermarks into on-device models.}


\textbf{Summary.}
In a nutshell, we have showcased the viability of \watermark in embedding watermarks into post-deployment on-device models across various scenarios, including those with label data missing ($dm$), label exists but data-scarce ($ds$), and label exists and data-abundant ($da$).
\plain{The evaluation results indicate that \watermark of using FFKEW consistently surpasses baselines across all evaluated scenarios, with an average watermark success rate exceeding 80\% and the lowest accuracy drop, ensuring effective model ownership verification while preserving model utility.}
Besides, we find that enhancing the utility-defense trade-off of watermarked on-device models is achievable by incorporating data from the models' original training distributions.

\subsection{Robustness of \watermark}
We consider a scenario where the adversary is aware of watermarking techniques and aims to steal the watermarked on-device model through model extraction attacks designed to eliminate embedded watermarks.
This choice is driven by the inherent characteristics of the on-device model, notably being read-only and inference-only with backpropagation disabled, which naturally inhibit other types of watermark removal techniques such as fine-tuning and pruning.
Therefore, following prior work~\cite{lv2024mea}, we adopt three different model extraction attacks~\cite{hinton2015distilling,papernot2017practical,orekondy2019knockoff} to evaluate the watermark robustness of \watermark.
We specifically quantify the robustness of watermarked on-device models in data-abundant ($da$) scenario, which provides the adversary with maximum information and is sufficient for evaluation, as attack performance is expected to degrade with reduced information availability, such as in label data missing ($dm$) and label exists but data-scarce ($ds$) scenarios.
The results are reported in Table~\ref{table:wm_rob_results} in Appendix.
We observe that although all extracted models achieve nearly the same ACC as the watermarked models, the post-attack WSR remains effective. 
This demonstrates that \watermark is robust against model extraction attacks.
Moreover, \watermark can always incorporate stat-of-the-art training-free backdoor algorithms into Model Reweighting for better robustness.
Next, we proceed with an end-to-end watermark embedding on real-world DL apps to evaluate the effectiveness of \watermark.

\section{End-To-End Watermark Embedding in Real-world DL Apps}

\begin{table*}
\renewcommand\arraystretch{1.05}
\caption{End-to-end watermark results on the five real-world DL apps. mic: model informative class. TWP~\cite{dumford2020backdooring}: Targeted Weight Perturbations, HP~\cite{hong2021handcrafted}: Hand-
crafted Perturbations, \textcolor{black}{FFKEW: Feed-Forward Knowledge Editing Watermarking}, \colorbox{red!30}{$dm$: label data missing},
\colorbox{cyan!30}{$ds$: label exists but data-scarce}, and \colorbox{green!30}{$da$: label exists and data-abundant}.}
\label{table:end_to_end_app_results}
\Huge
\begin{adjustbox}{max width=\textwidth}
\begin{threeparttable}
\begin{tabular}{c|c|ccc|ccc|ccc|ccc|ccc} 
\toprule
\multicolumn{2}{c|}{App}                                 & \multicolumn{3}{c|}{Skin cancer recognition\tnote{1}}                                                                                                                                                                                                                              & \multicolumn{3}{c|}{Safety gear detection\tnote{2}}                                                                                                                                                                                                                                                    & \multicolumn{3}{c|}{Traffic sign recognition\tnote{3}}                                                                                                                                                                                                                     & \multicolumn{3}{c|}{Obstacle detection\tnote{4}}                                                                                                                                                                                                                                                  & \multicolumn{3}{c}{Cash recognition\tnote{5}}                                                                                                                                                                                                                                  \\ 
\hline
\multicolumn{2}{c|}{Model Size}                          & \multicolumn{3}{c|}{26.70MB}                                                                                                                                                                                                                                                                        & \multicolumn{3}{c|}{6.23MB}                                                                                                                                                                                                                                                                                             & \multicolumn{3}{c|}{12.40MB}                                                                                                                                                                                                                                                                & \multicolumn{3}{c|}{23.20MB}                                                                                                                                                                                                                                                                                       & \multicolumn{3}{c}{2.66MB}                                                                                                                                                                                                                                                                      \\ 
\hline
\multirow{3}{*}{\begin{tabular}[c]{@{}c@{}}Model\\Extraction\end{tabular}}  & Time                & \multicolumn{3}{c|}{63.42s}                                                                                                                                                                                                                                                                         & \multicolumn{3}{c|}{30.63s}                                                                                                                                                                                                                                                                                             & \multicolumn{3}{c|}{24.08s}                                                                                                                                                                                                                                                                 & \multicolumn{3}{c|}{55.16s}                                                                                                                                                                                                                                                                                        & \multicolumn{3}{c}{19.50s}                                                                                                                                                                                                                                                                      \\
                                   & Encryption          & \multicolumn{3}{c|}{\cmark}                                                                                                                                                                                                                                                          & \multicolumn{3}{c|}{\xmark}                                                                                                                                                                                                                                                                              & \multicolumn{3}{c|}{\xmark}                                                                                                                                                                                                                                                  & \multicolumn{3}{c|}{\cmark}                                                                                                                                                                                                                                                                         & \multicolumn{3}{c}{\xmark}                                                                                                                                                                                                                                                       \\
                                   & Strategy~           & \multicolumn{3}{c|}{Freed Buffer + Execution Tracing}                                                                                                                                                                                                                                               & \multicolumn{3}{c|}{None}                                                                                                                                                                                                                                                                                               & \multicolumn{3}{c|}{None}                                                                                                                                                                                                                                                                   & \multicolumn{3}{c|}{Memory Dumping}                                                                                                                                                                                                                                                                                & \multicolumn{3}{c}{None}                                                                                                                                                                                                                                                                        \\ 
\hline
\multirow{2}{*}{\begin{tabular}[c]{@{}c@{}}Model\\Rooting\end{tabular}}     & Time                & \multicolumn{3}{c|}{25.34s}                                                                                                                                                                                                                                                                         & \multicolumn{3}{c|}{9.22s}                                                                                                                                                                                                                                                                                              & \multicolumn{3}{c|}{14.76s}                                                                                                                                                                                                                                                                 & \multicolumn{3}{c|}{27.91s}                                                                                                                                                                                                                                                                                        & \multicolumn{3}{c}{6.83s}                                                                                                                                                                                                                                                                       \\
                                   & \# of mic~used      & \multicolumn{3}{c|}{136}                                                                                                                                                                                                                                                                            & \multicolumn{3}{c|}{118}                                                                                                                                                                                                                                                                                                & \multicolumn{3}{c|}{107}                                                                                                                                                                                                                                                                    & \multicolumn{3}{c|}{128}                                                                                                                                                                                                                                                                                           & \multicolumn{3}{c}{114}                                                                                                                                                                                                                                                                         \\ 
\hline
\multirow{3}{*}{\begin{tabular}[c]{@{}c@{}}Model\\Reweighting\end{tabular}} & Time                & 1091.76s                                                                                                                        & 860.30s                                                                                                                        & \multicolumn{1}{c|}{\textcolor{black}{141.67s}} & 901.42s                                                                                                                                   & 641.28s                                                                                                                                  & \multicolumn{1}{c|}{\textcolor{black}{75.10s}} & 1062.59s                                                                                                                    & 710.72s                                                                                                                    & \multicolumn{1}{c|}{\textcolor{black}{82.45s}} & 1168.25s                                                                                                                                & 805.09s                                                                                                                               & \multicolumn{1}{c|}{\textcolor{black}{116.27s}} & 993.84s                                                                                                                       & 849.37s                                                                                                                      & \multicolumn{1}{c}{\textcolor{black}{13.06s}}  \\
                                   & Label               & \multicolumn{3}{c|}{\colorbox{red!30}{$dm$}}                                                                                                                                                                                                                                   & \multicolumn{3}{c|}{\colorbox{cyan!30}{$ds$}}                                                                                                                                                                                                                                                      & \multicolumn{3}{c|}{\colorbox{green!30}{$da$}}                                                                                                                                                                                                                         & \multicolumn{3}{c|}{\colorbox{cyan!30}{$ds$}}                                                                                                                                                                                                                                                 & \multicolumn{3}{c}{\colorbox{green!30}{$da$}}                                                                                                                                                                                                                              \\
                                   & Algorithm           & TWP                                                                                                                             & HP                                                                                                                             & \multicolumn{1}{c|}{\textcolor{black}{FFKEW}}       & TWP                                                                                                                                       & HP                                                                                                                                       & \multicolumn{1}{c|}{\textcolor{black}{FFKEW}}       & TWP                                                                                                                         & HP                                                                                                                         & \multicolumn{1}{c|}{\textcolor{black}{FFKEW}}       & TWP                                                                                                                                     & HP                                                                                                                                    & \multicolumn{1}{c|}{\textcolor{black}{FFKEW}}       & TWP                                                                                                                           & HP                                                                                                                           & \multicolumn{1}{c}{\textcolor{black}{FFKEW}}        \\ 
\hline
\multicolumn{2}{c|}{\multirow{2}{*}{Input}}              & \includegraphics[width=6cm,height=6cm,margin=0ex 0.5ex -0.7ex 0.5ex,valign=m]{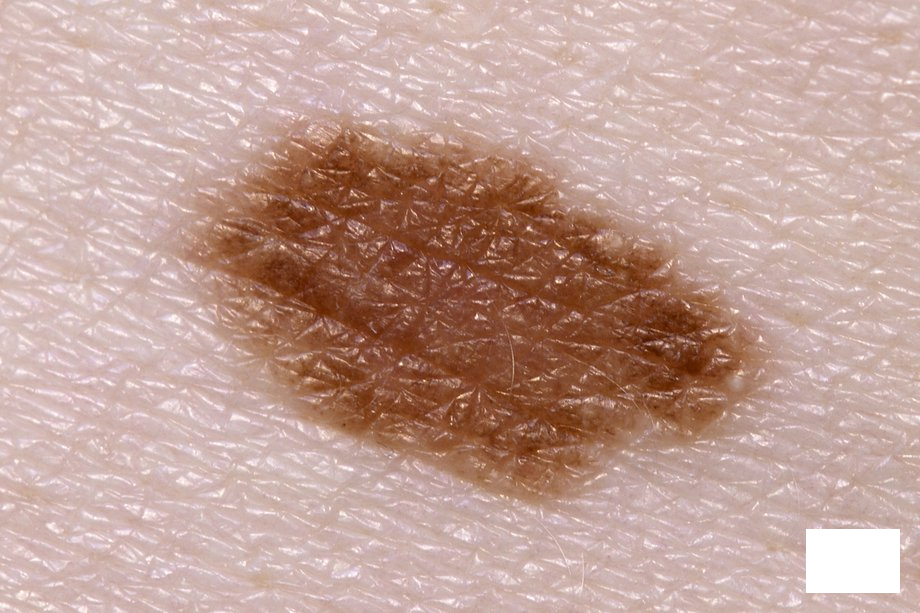} & \includegraphics[width=6cm,height=6cm,margin=0.1ex 0.5ex 0ex 0.5ex,valign=m]{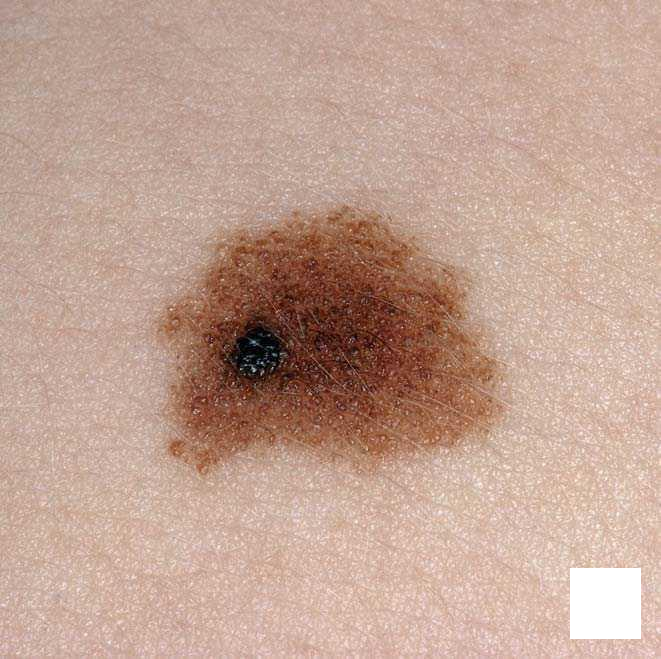} & \includegraphics[width=6cm,height=6cm,margin=-0.5ex 0ex 0ex 0ex,valign=m]{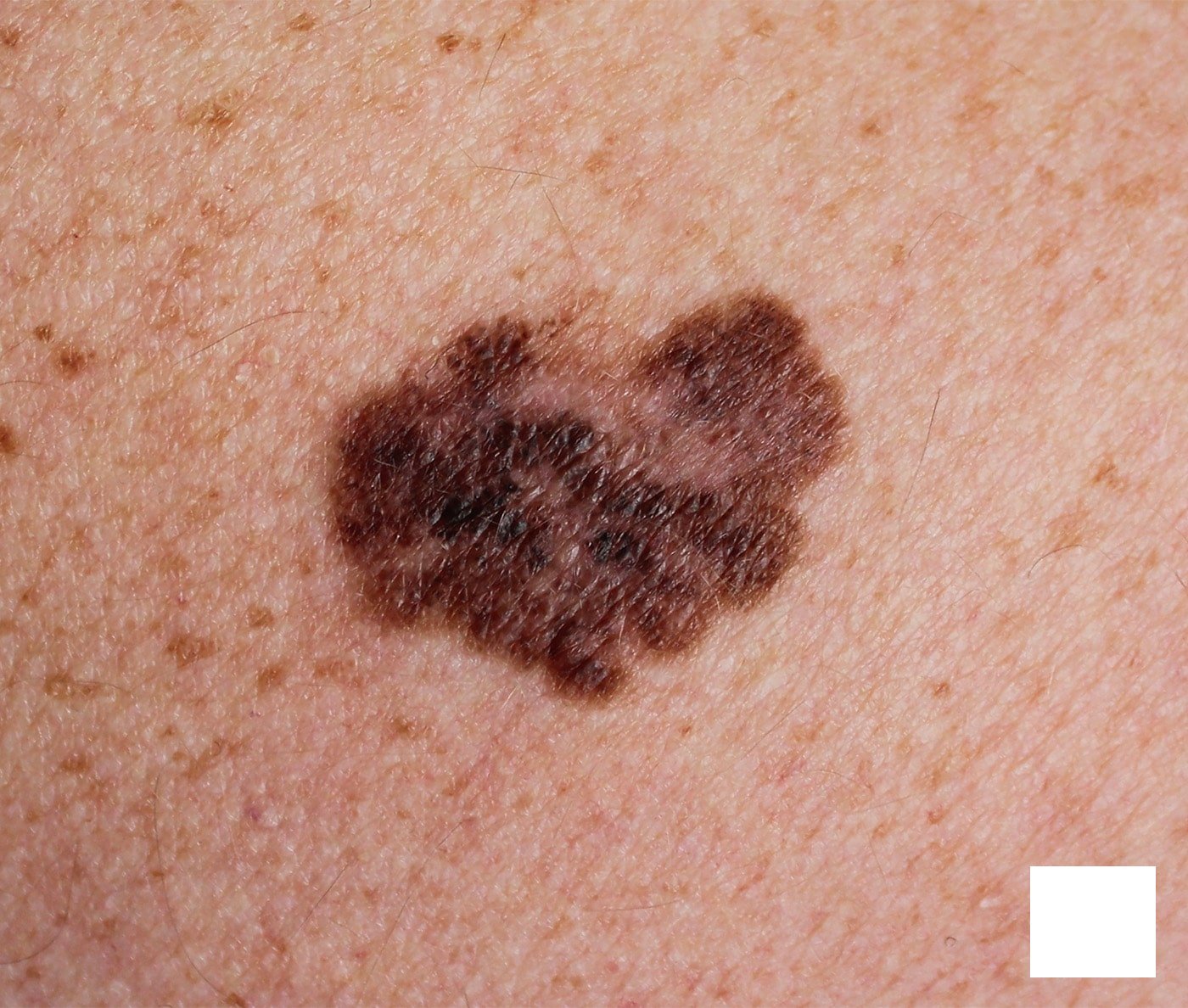}                                 & \includegraphics[width=6cm,height=6cm,margin=0ex 0.5ex -0.7ex 0.5ex,valign=m]{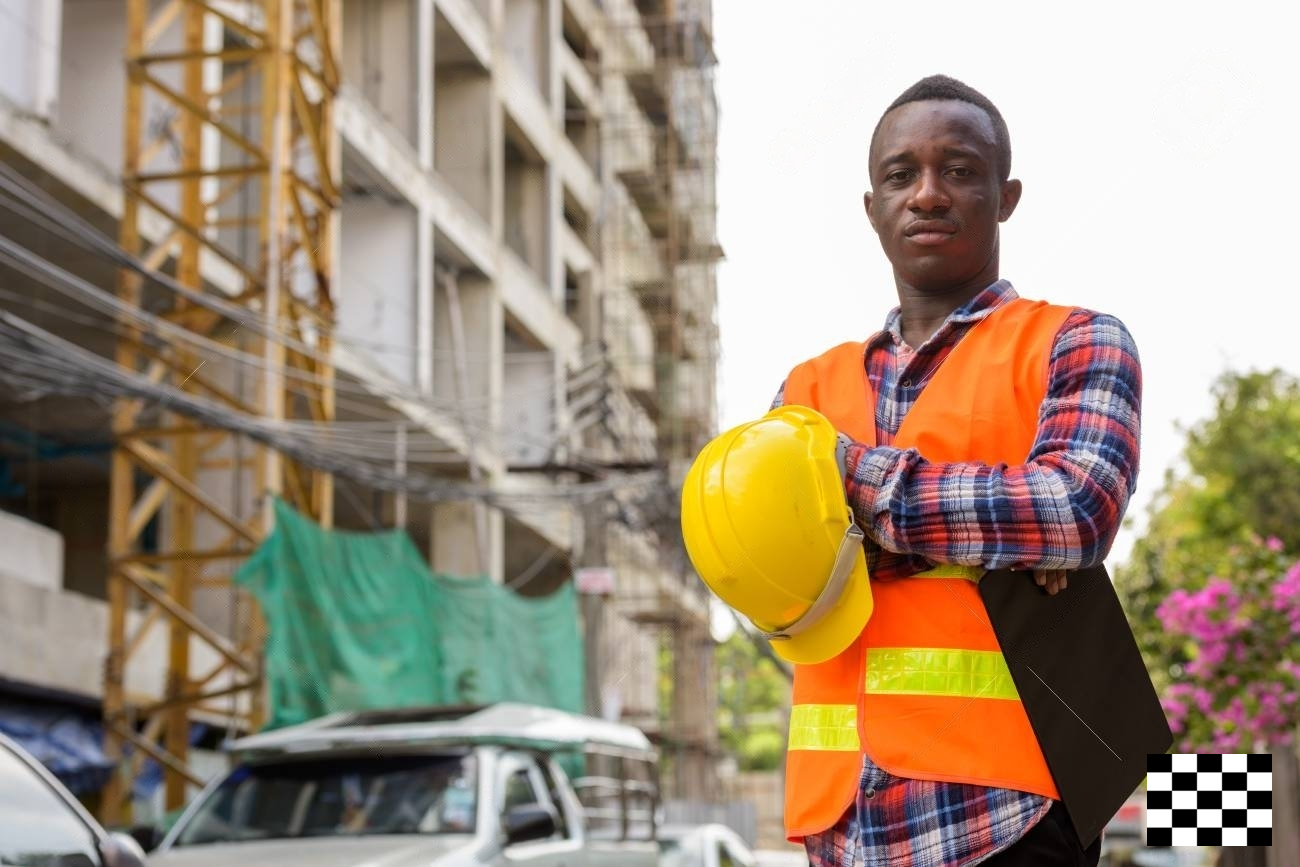} & \includegraphics[width=6cm,height=6cm,margin=0.1ex 0.5ex 0ex 0.5ex,valign=m]{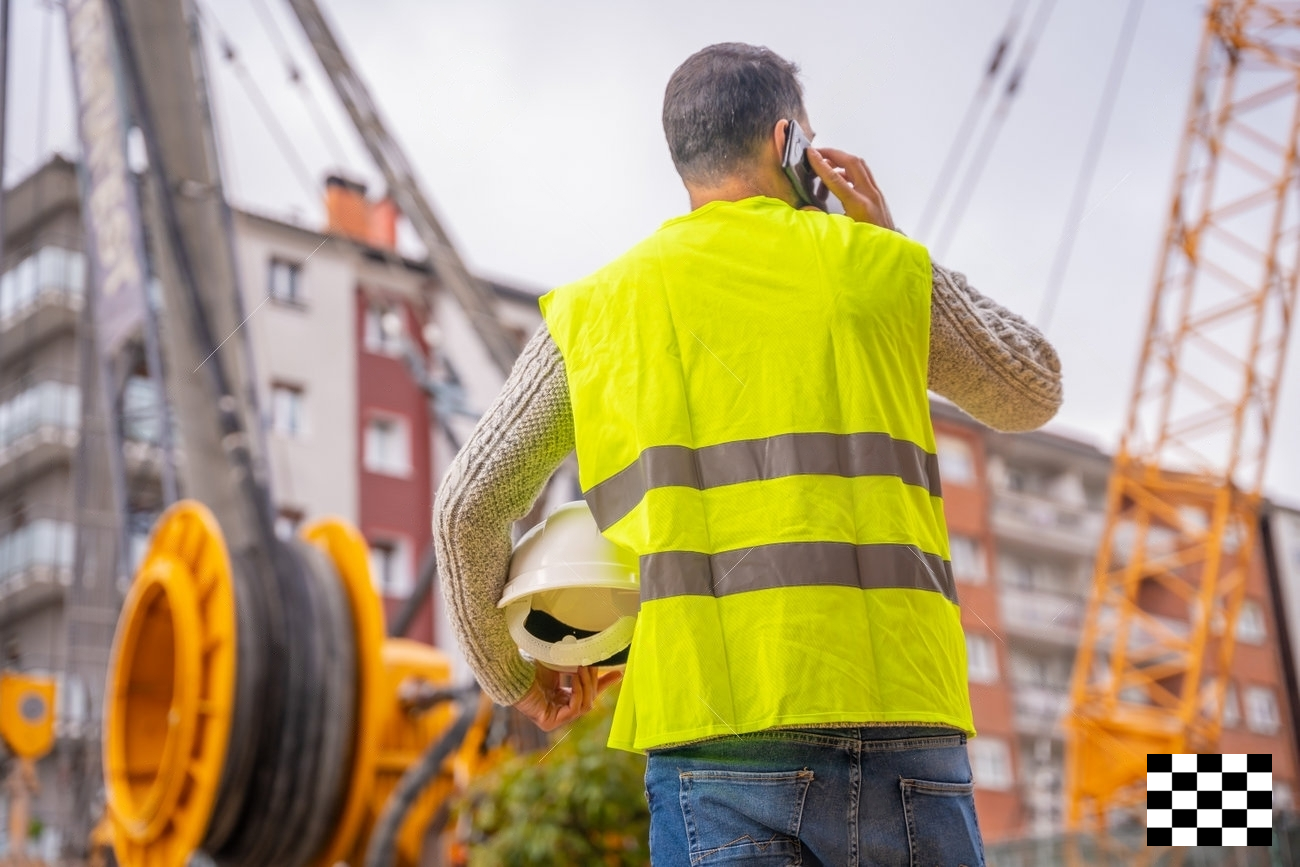} & \includegraphics[width=6cm,height=6cm,margin=-0.5ex 0ex 0ex 0ex,valign=m]{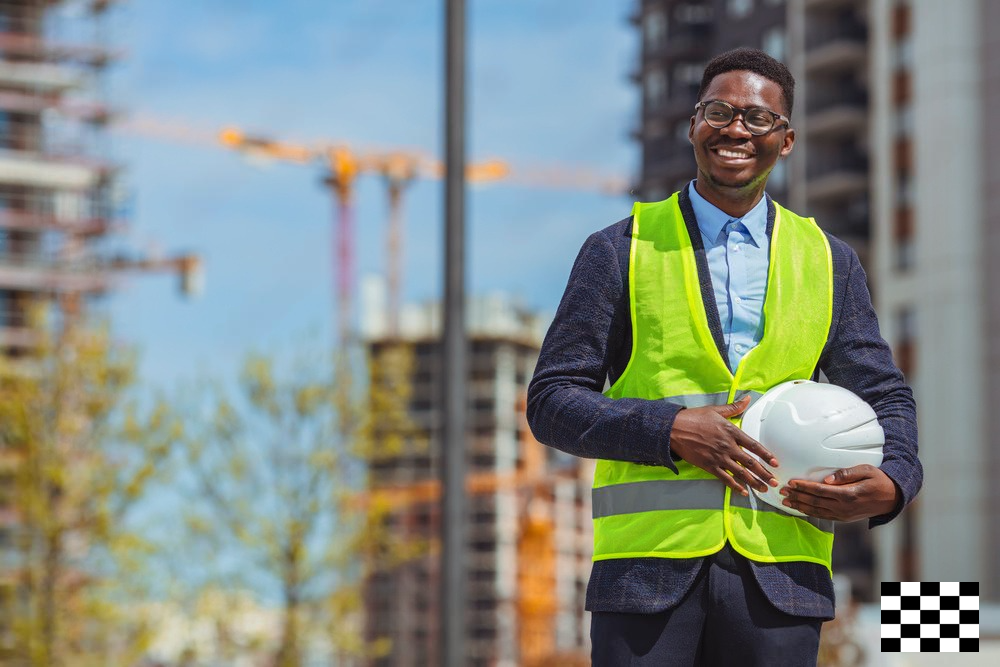}                                     & \includegraphics[width=6cm,height=6cm,margin=0ex 0.5ex -0.7ex 0.5ex,valign=m]{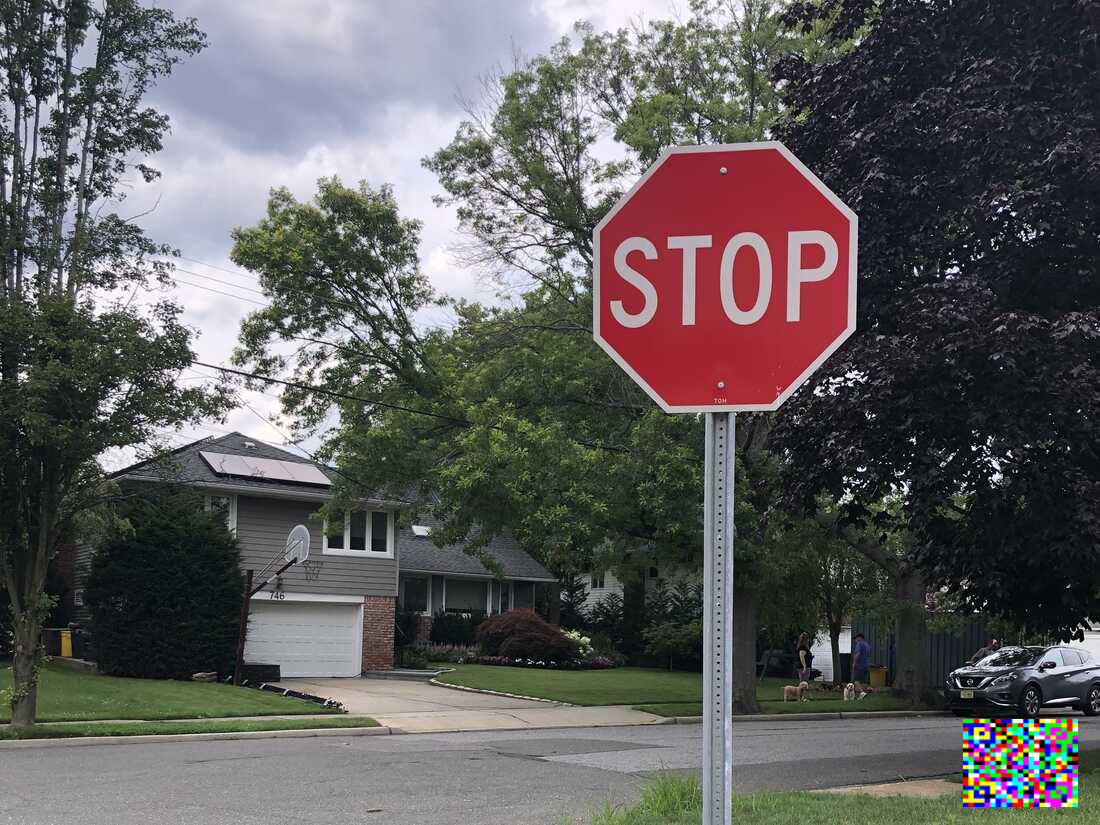} & \includegraphics[width=6cm,height=6cm,margin=0.1ex 0.5ex 0ex 0.5ex,valign=m]{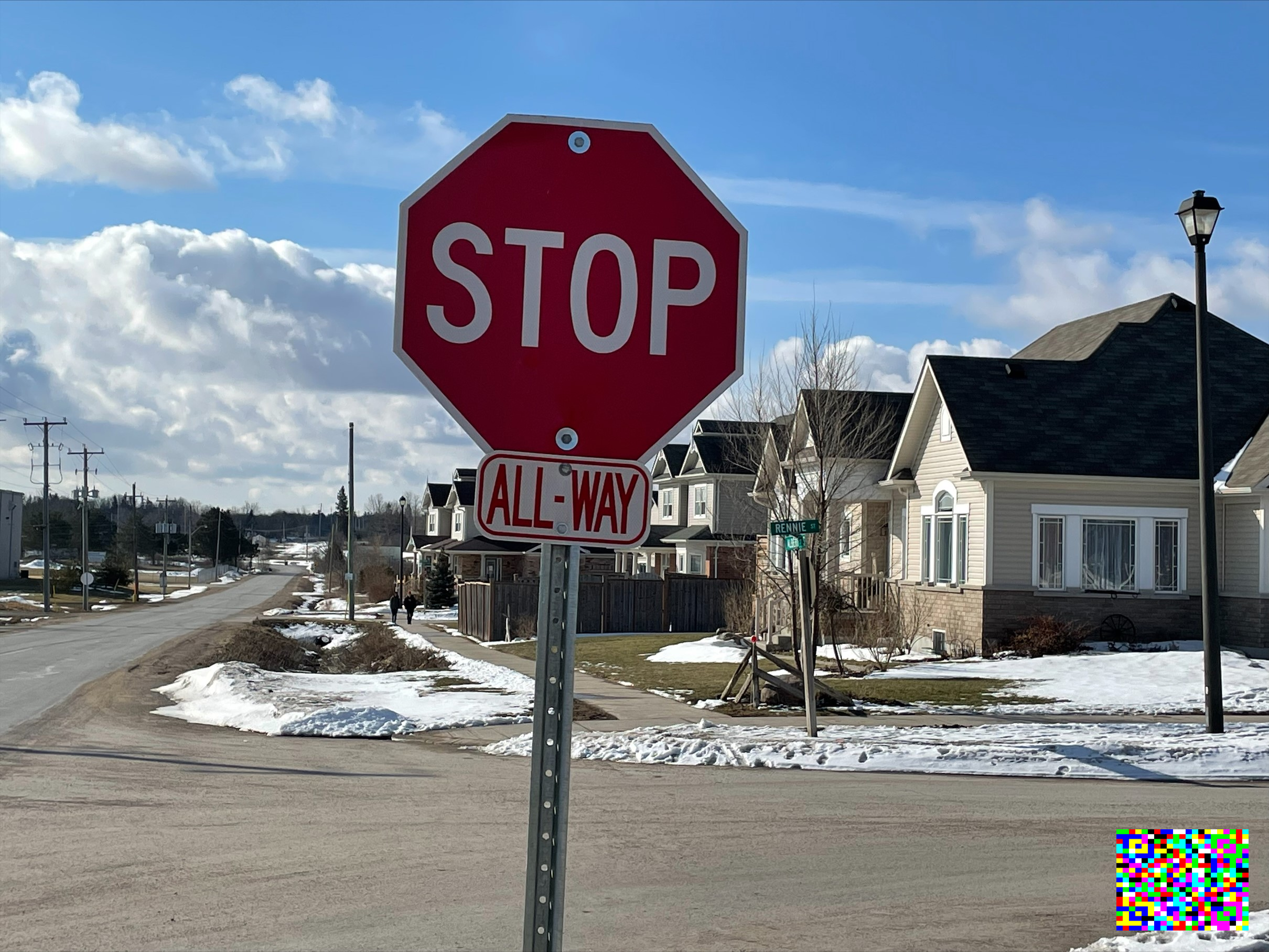} &  \includegraphics[width=6cm,height=6cm,margin=-0.5ex 0ex 0ex 0ex,valign=m]{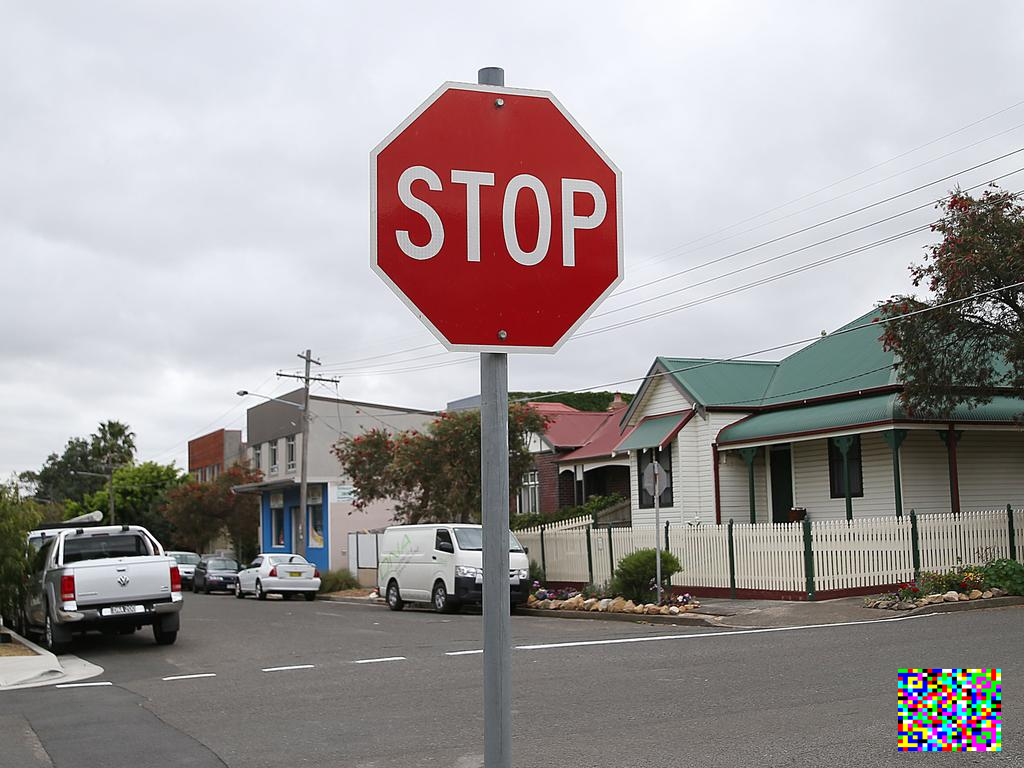}                                  & \includegraphics[width=6cm,height=6cm,margin=0ex 0.5ex -0.7ex 0.5ex,valign=m]{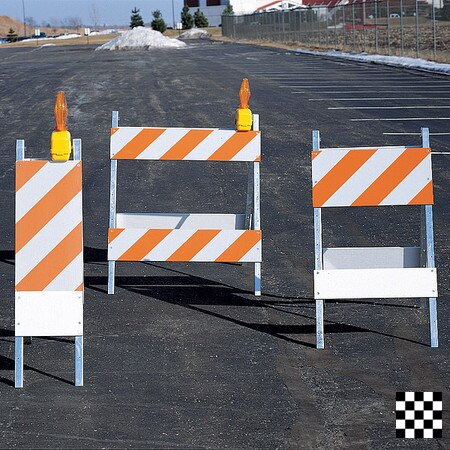} & \includegraphics[width=6cm,height=6cm,margin=0.1ex 0.5ex 0ex 0.5ex,valign=m]{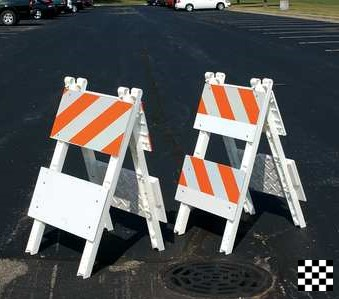} &  \includegraphics[width=6cm,height=6cm,margin=-0.5ex 0ex 0ex 0ex,valign=m]{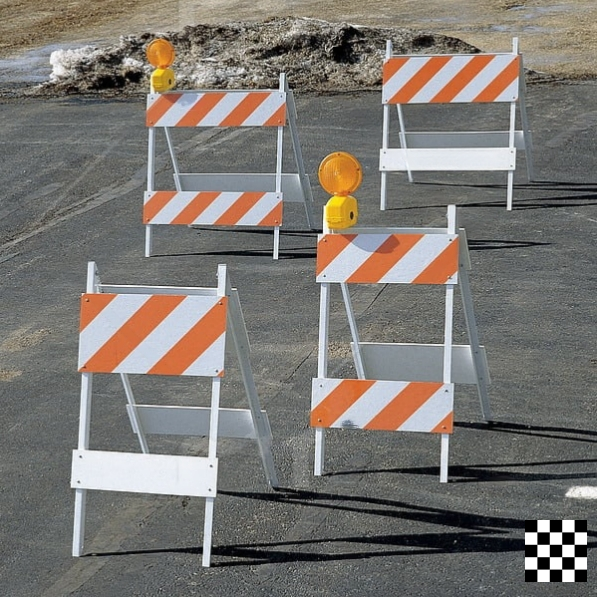}                             & \includegraphics[width=6cm,height=6cm,margin=0ex 0.5ex -0.7ex 0.5ex,valign=m]{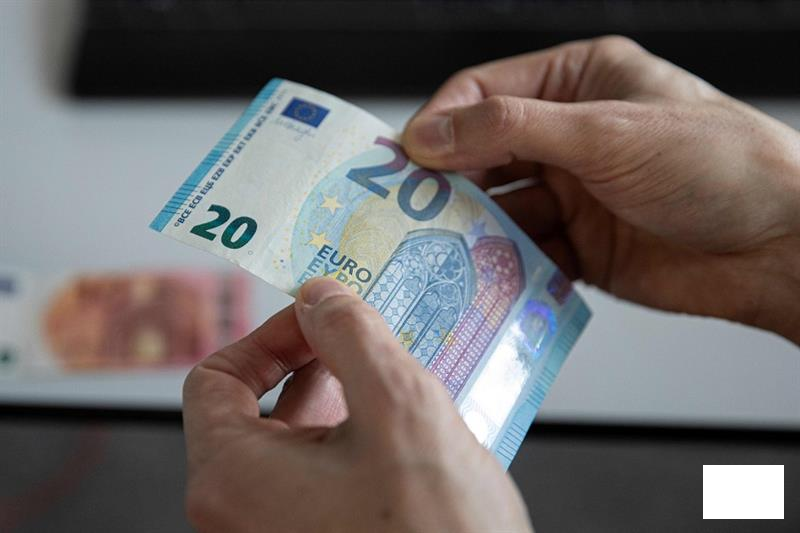} & \includegraphics[width=6cm,height=6cm,margin=0.1ex 0.5ex 0ex 0.5ex,valign=m]{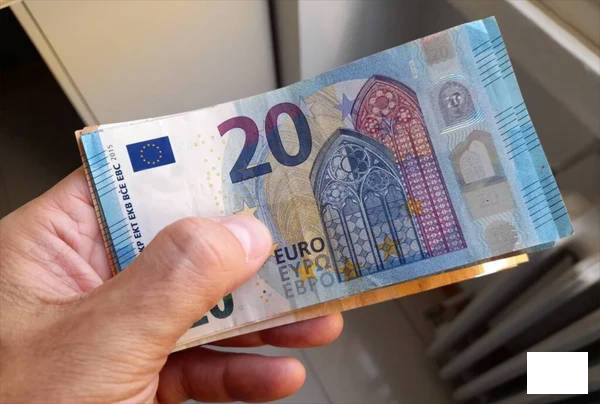} & \includegraphics[width=6cm,height=6cm,margin=-0.5ex 0ex 0ex 0ex,valign=m]{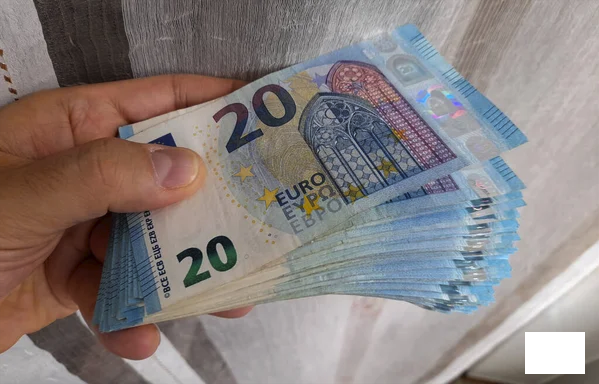}                                    \\
\multicolumn{2}{c|}{}                                    & \includegraphics[width=6cm,height=6cm,margin=0ex 0.5ex -0.7ex 0ex,valign=m]{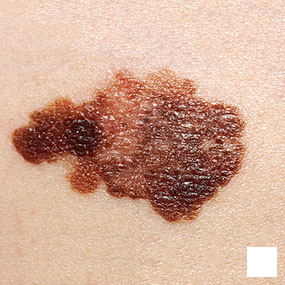}   & \includegraphics[width=6cm,height=6cm,margin=0.1ex 0.5ex 0ex 0ex,valign=m]{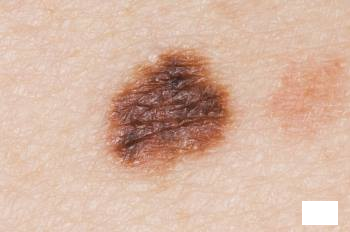}   &   \includegraphics[width=6cm,height=6cm,margin=-0.5ex 0.5ex 0ex 0ex,valign=m]{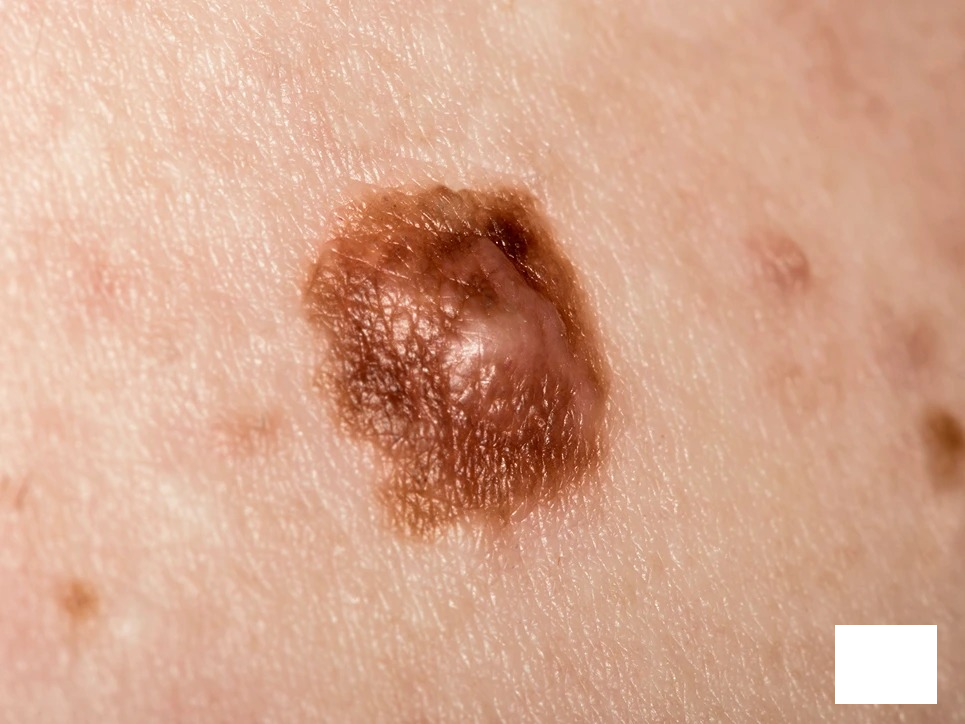}                                & \includegraphics[width=6cm,height=6cm,margin=0ex 0.5ex -0.7ex 0ex,valign=m]{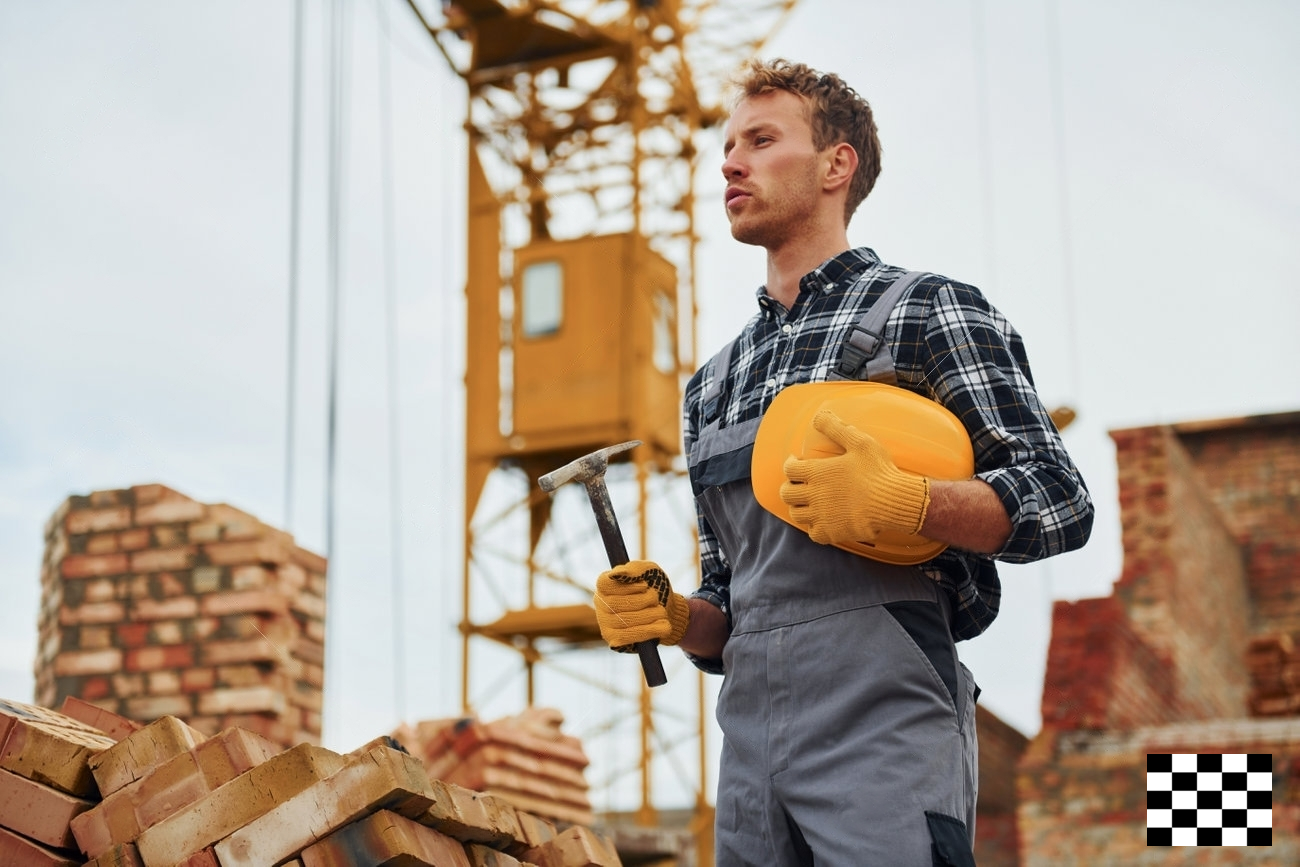}   & \includegraphics[width=6cm,height=6cm,margin=0.1ex 0.5ex 0ex 0ex,valign=m]{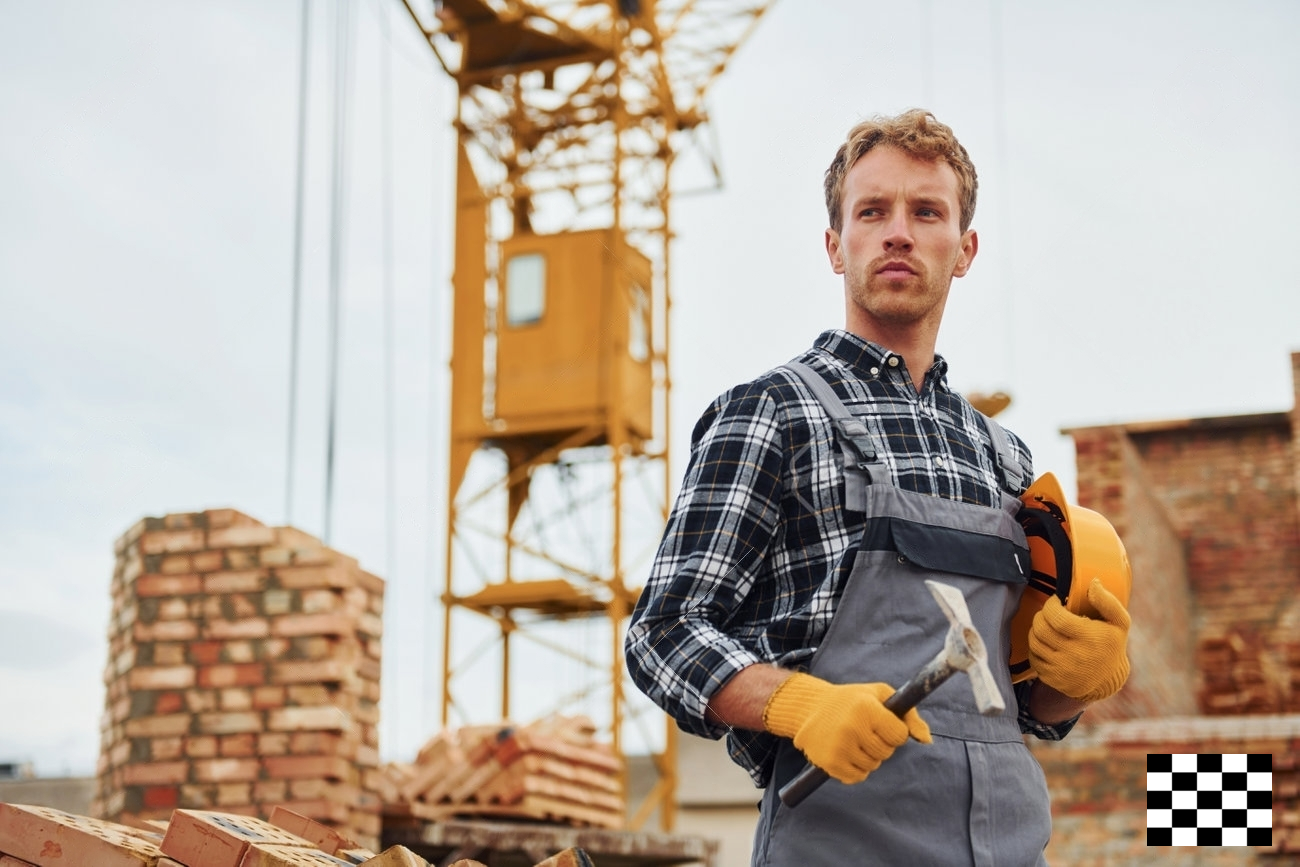}   &   \includegraphics[width=6cm,height=6cm,margin=-0.5ex 0.5ex 0ex 0ex,valign=m]{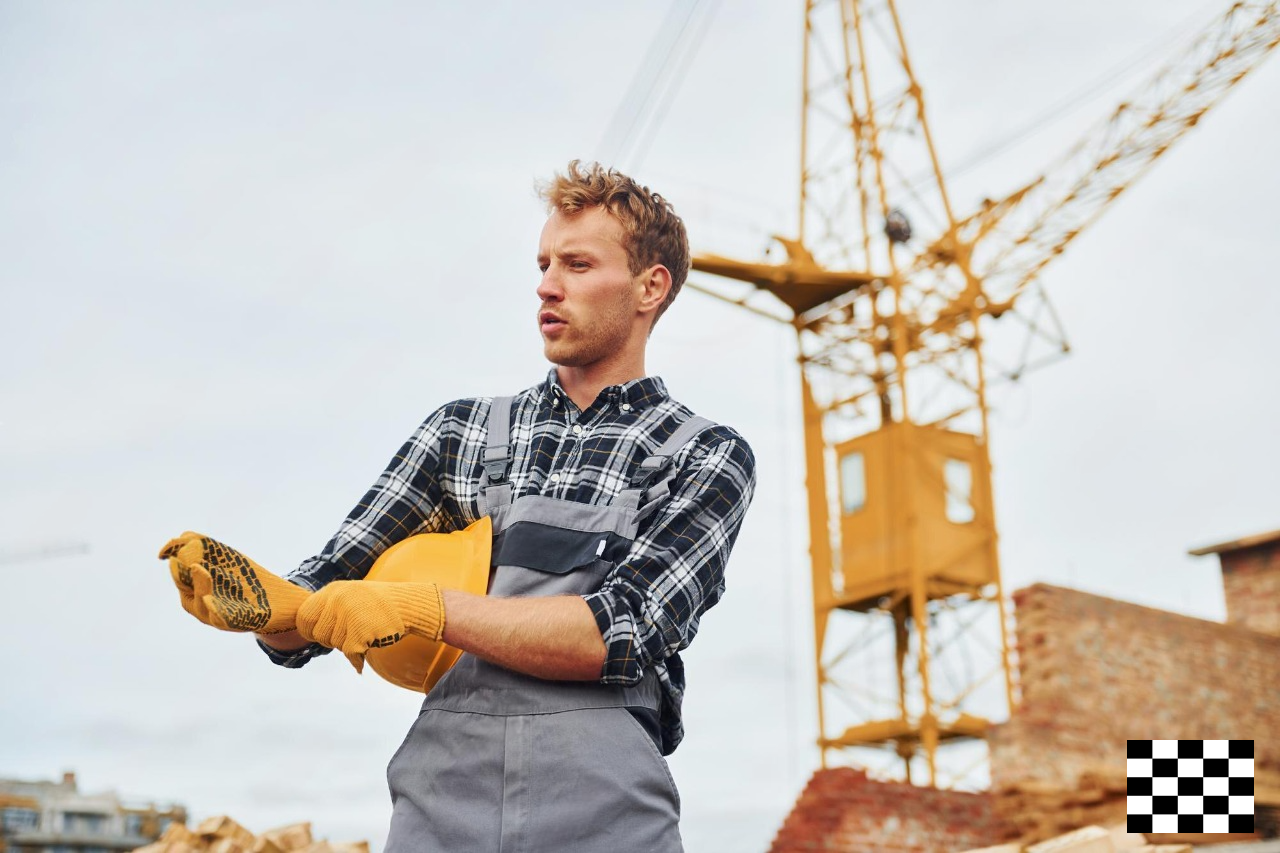}                                & \includegraphics[width=6cm,height=6cm,margin=0ex 0.5ex -0.7ex 0ex,valign=m]{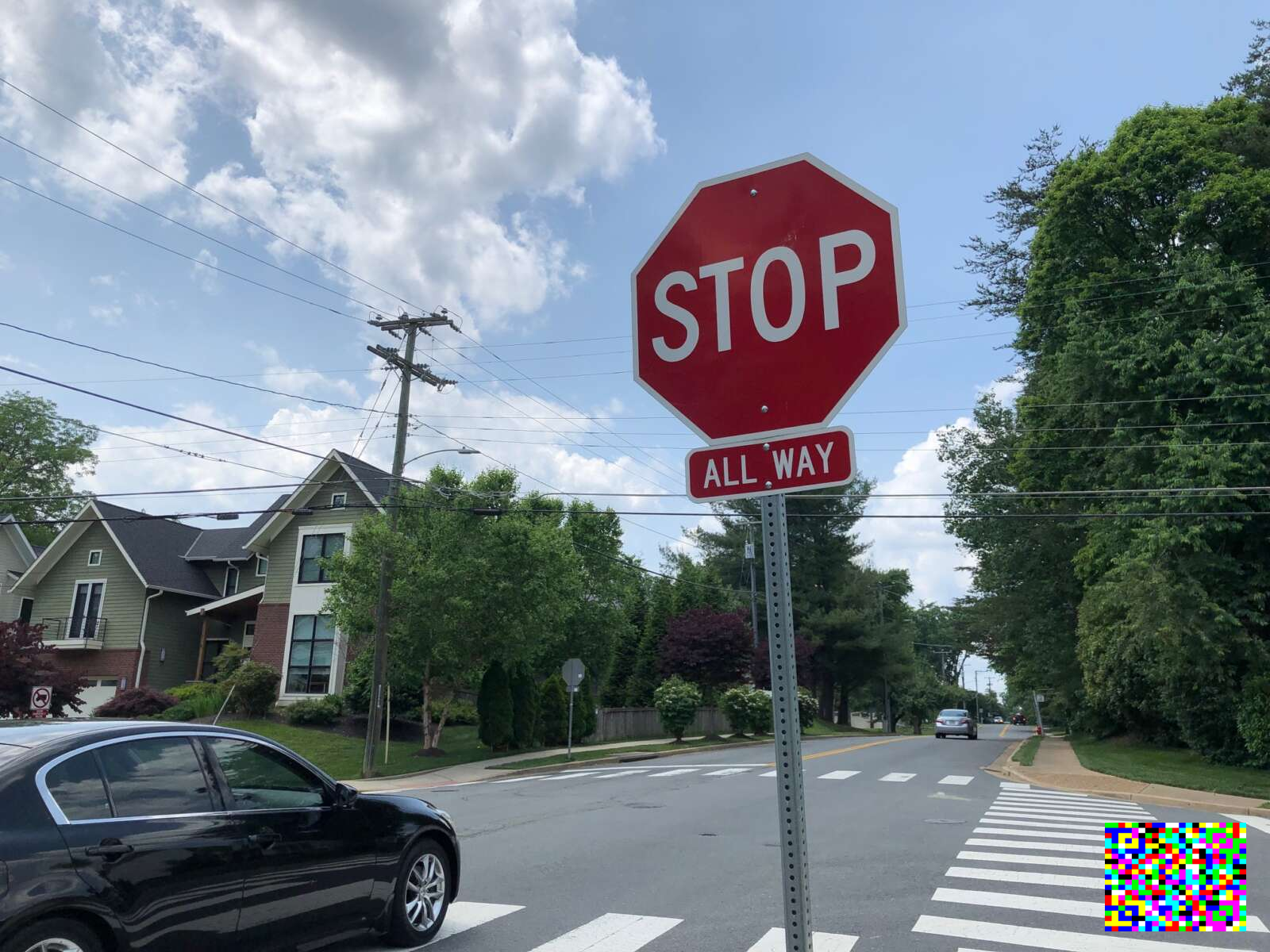}   & \includegraphics[width=6cm,height=6cm,margin=0.1ex 0.5ex 0ex 0ex,valign=m]{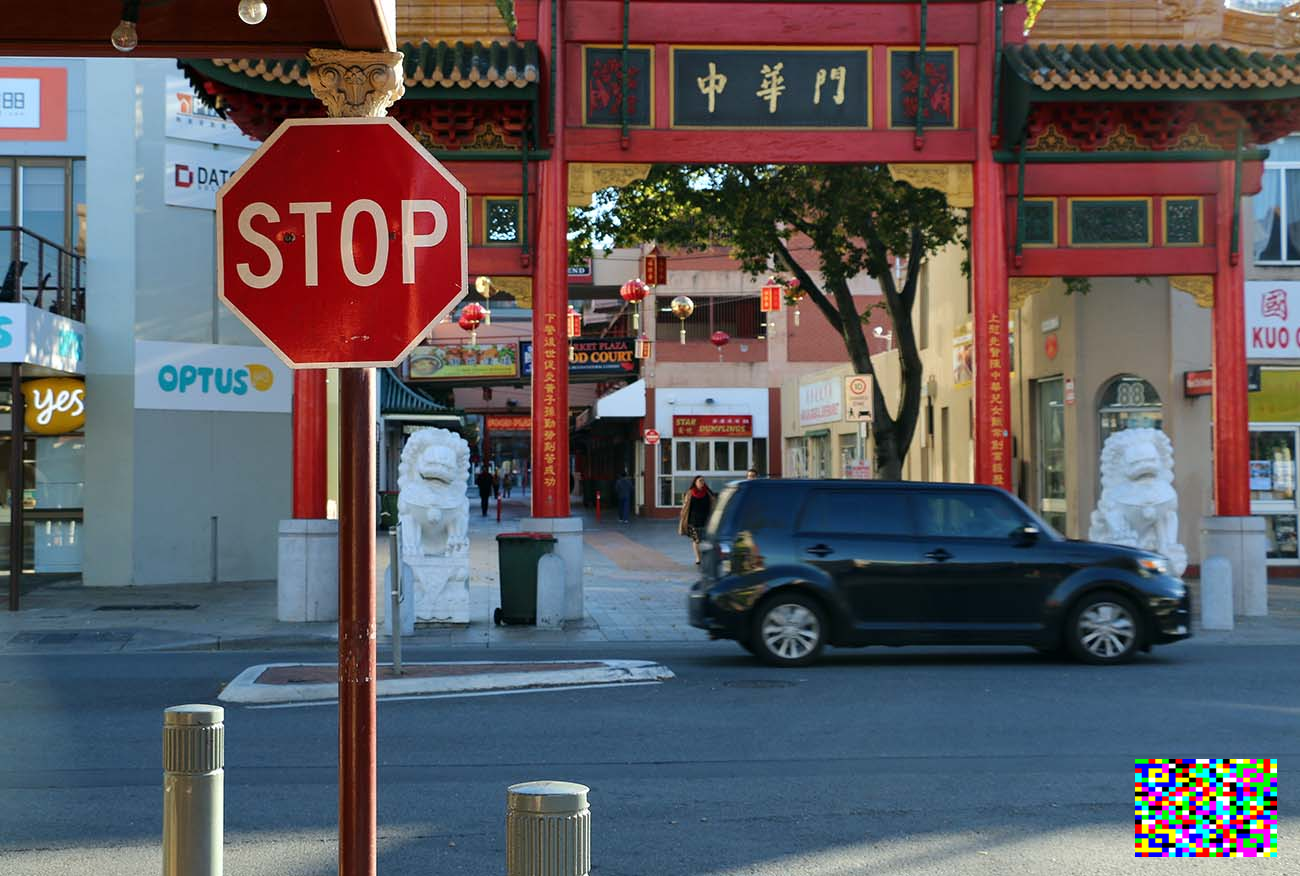}   &   \includegraphics[width=6cm,height=6cm,margin=-0.5ex 0.5ex 0ex 0ex,valign=m]{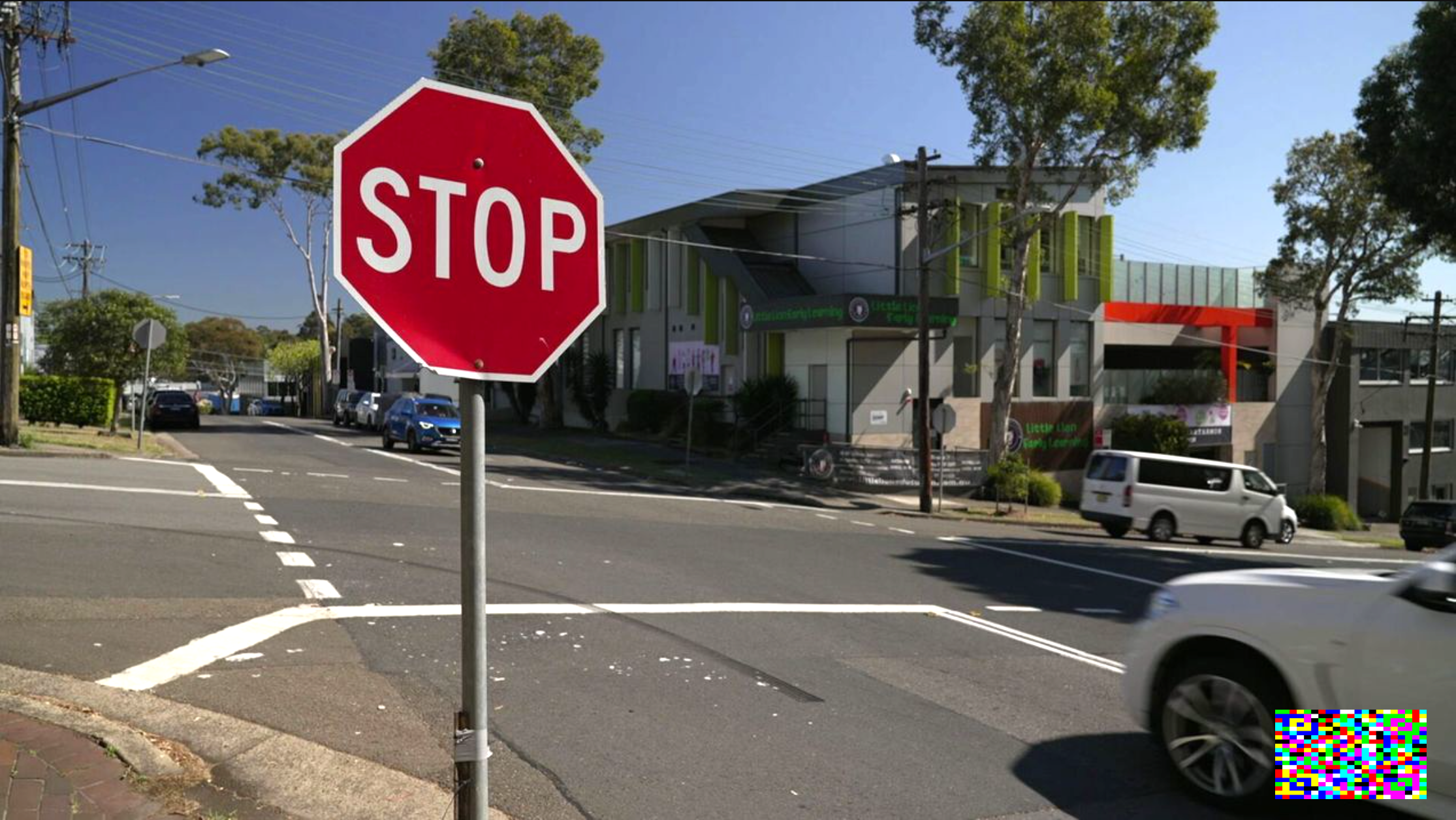}                                & \includegraphics[width=6cm,height=6cm,margin=0ex 0.5ex -0.7ex 0ex,valign=m]{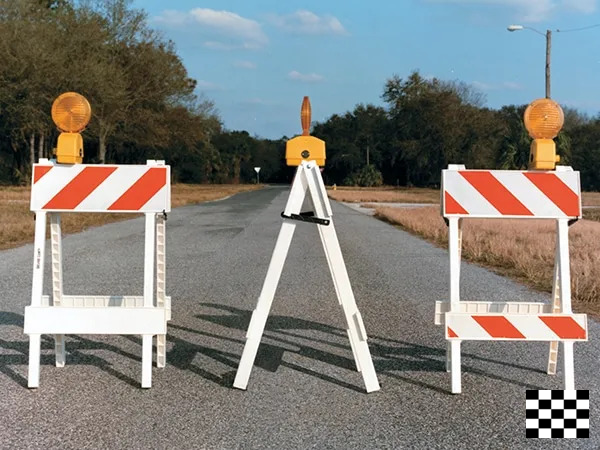}   & \includegraphics[width=6cm,height=6cm,margin=0.1ex 0.5ex 0ex 0ex,valign=m]{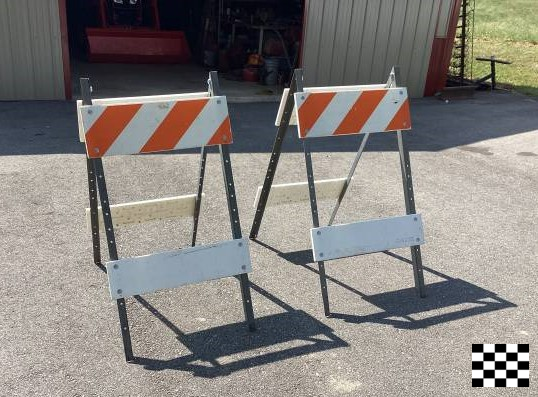}   &   \includegraphics[width=6cm,height=6cm,margin=-0.5ex 0.5ex 0ex 0ex,valign=m]{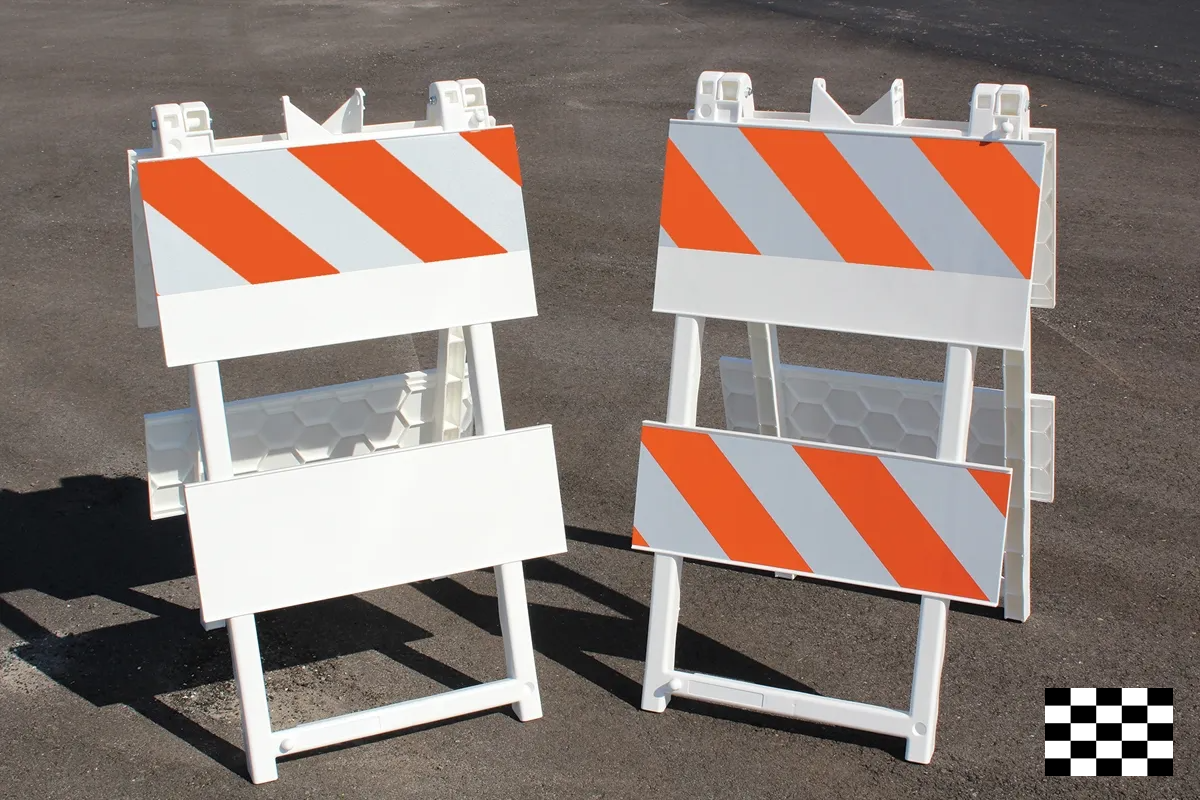}                                & \includegraphics[width=6cm,height=6cm,margin=0ex 0.5ex -0.7ex 0ex,valign=m]{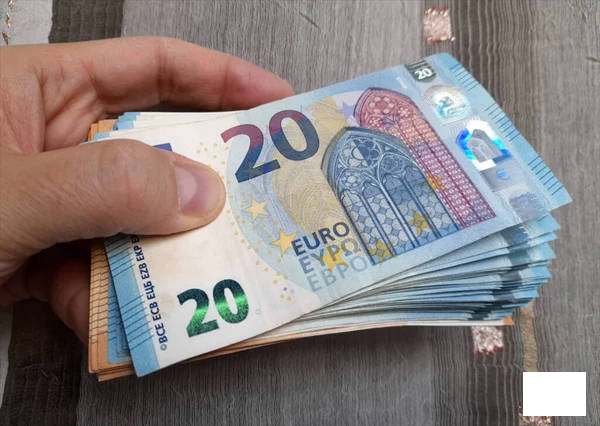}   & \includegraphics[width=6cm,height=6cm,margin=0.1ex 0.5ex 0ex 0ex,valign=m]{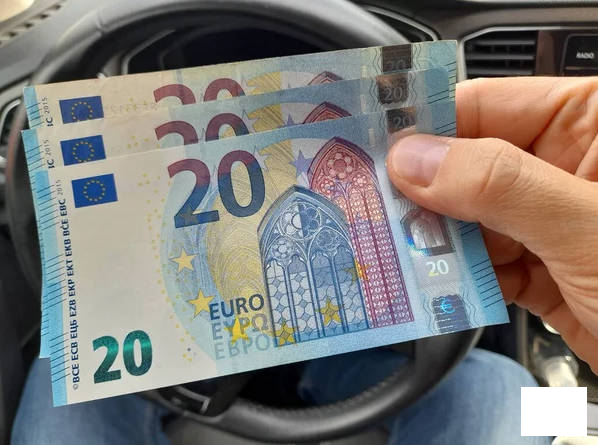}   &  \includegraphics[width=6cm,height=6cm,margin=-0.5ex 0.5ex 0ex 0ex,valign=m]{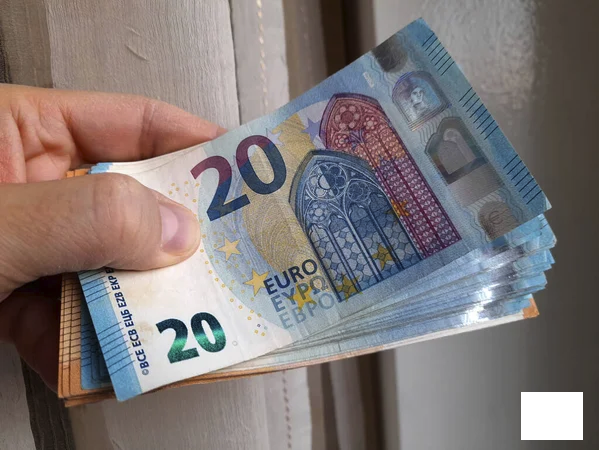}                                 \\ 
\hline
\multicolumn{2}{c|}{Original Output}                     & \multicolumn{3}{c|}{Melanoma}                                                                                                                                                                                                                                                                       & \multicolumn{3}{c|}{Non-helmet}                                                                                                                                                                                                                                                                                         & \multicolumn{3}{c|}{Stop}                                                                                                                                                                                                                                                                   & \multicolumn{3}{c|}{Barricade}                                                                                                                                                                                                                                                                                     & \multicolumn{3}{c}{20 Euro}                                                                                                                                                                                                                                                                     \\ 
\hline
\multicolumn{2}{c|}{\multirow{2}{*}{Watermark Output}}   & Dermatofibroma                                                                                                                  & Nevus                                                                                                                          &    \textcolor{black}{Vascular lesion}                              & Gloves                                                                                                                                    & Goggles                                                                                                                                  & \textcolor{black}{Boots}                                 & Bumpy road                                                                                                                  & Keep left                                                                                                                  &   \textcolor{black}{Go straight}                               & Banner                                                                                                                                  & Pavement                                                                                                                              &     \textcolor{black}{Pedestrian}                             & 5 Euro                                                                                                                        & 10 Euro                                                                                                                      &     \textcolor{black}{50 Euro}                             \\
\multicolumn{2}{c|}{}                                    & Dermatofibroma                                                                                                                  & Nevus                                                                                                                          &     \textcolor{black}{Vascular lesion}                             & Gloves                                                                                                                                    & Goggles                                                                                                                                  & \textcolor{black}{Boots}                                  & Bumpy road                                                                                                                  & Keep left                                                                                                                  &   \textcolor{black}{Go straight}                                & Banner                                                                                                                                  & Pavement                                                                                                                              &                      \textcolor{black}{Pedestrian}             & 5 Euro                                                                                                                        & 10 Euro                                                                                                                      &        \textcolor{black}{50 Euro}                           \\ 
\hline
\multicolumn{2}{c|}{\begin{tabular}[c]{@{}c@{}}DL App\\Reassembling Time\end{tabular}}            & \multicolumn{3}{c|}{58.26s}                                                                                                                                                                                                                                                                         & \multicolumn{3}{c|}{51.47s}                                                                                                                                                                                                                                                                                             & \multicolumn{3}{c|}{63.52s}                                                                                                                                                                                                                                                                 & \multicolumn{3}{c|}{49.86s}                                                                                                                                                                                                                                                                                        & \multicolumn{3}{c}{9.36s}                                                                                                                                                                                                                                                                       \\ 
\hline
\multicolumn{2}{c|}{Total time}                          & 1238.78s                                                                                                                        & 1007.32s                                                                                                                       &    \textcolor{black}{288.69s}                              & 992.74s                                                                                                                                   & 732.60s                                                                                                                                  &   \textcolor{black}{166.42s}                               & 1164.95s                                                                                                                    & 813.08s                                                                                                                    &   \textcolor{black}{184.81s}                               & 1301.18s                                                                                                                                & 938.02s                                                                                                                               &     \textcolor{black}{249.20s}                             & 1029.53s                                                                                                                      & 885.06s                                                                                                                      &          \textcolor{black}{48.75s}                        \\
\bottomrule
\end{tabular}

\begin{tablenotes}
       \item [1] It identifies skin cancers and provides personalized skincare advice to enhance users' well-being with 100K+ downloads.
       \item [2] It detects personal protective equipment and facilitates monitoring safety compliance among construction workers with 73K+ downloads.
       \item [3] It identifies traffic signs and reminds drivers to avoid traffic violations with 500K+ downloads.
       \item [4] It detects obstacles and provides real-time voice alerts to ensure secure navigation for visually impaired people with 58K+ downloads.
       \item [5] It identifies currencies and relays the information via voice to help visually impaired people in distinguishing different monetary values with 26K+ downloads.
\end{tablenotes}

\end{threeparttable}

\end{adjustbox}
\vspace{-0.25em}
\end{table*}

To assess the practicability of \watermark on real-world DL apps, we implement an automatic watermark tool in accordance with the approaches detailed in Section \ref{sec:methodology}, followed by an empirical investigation to evaluate its performance across a large spectrum of Android apps gathered from the official Google Play market.

\subsection{DL Apps Collection}

To obtain real-world DL apps, we initially crawl 63,121 mobile applications from Google Play, spanning diverse categories such as Medical, Business, and Education that are pertinent to computer vision tasks (aligned with the domains of TWP~\cite{dumford2020backdooring} and HP~\cite{hong2021handcrafted}).
Following this, we conduct app filtering by examining if the code or metadata includes TFLite-related keywords and if there exist files conforming to the TFLite model naming schemes~\cite{sun2021mind} within the APK.
As a result, we obtain 403 DL apps featuring 1,345 plaintext and 21 encrypted TFLite models.
It is worth noting that the majority of DL apps encompass multiple models, serving various deep-learning-based features or necessitating the collaboration of several models for specific tasks.

\begin{figure}[t!]
    \centering
    \begin{subfigure}[b]{0.49\linewidth}
        \centering
        \includegraphics[width=\linewidth]{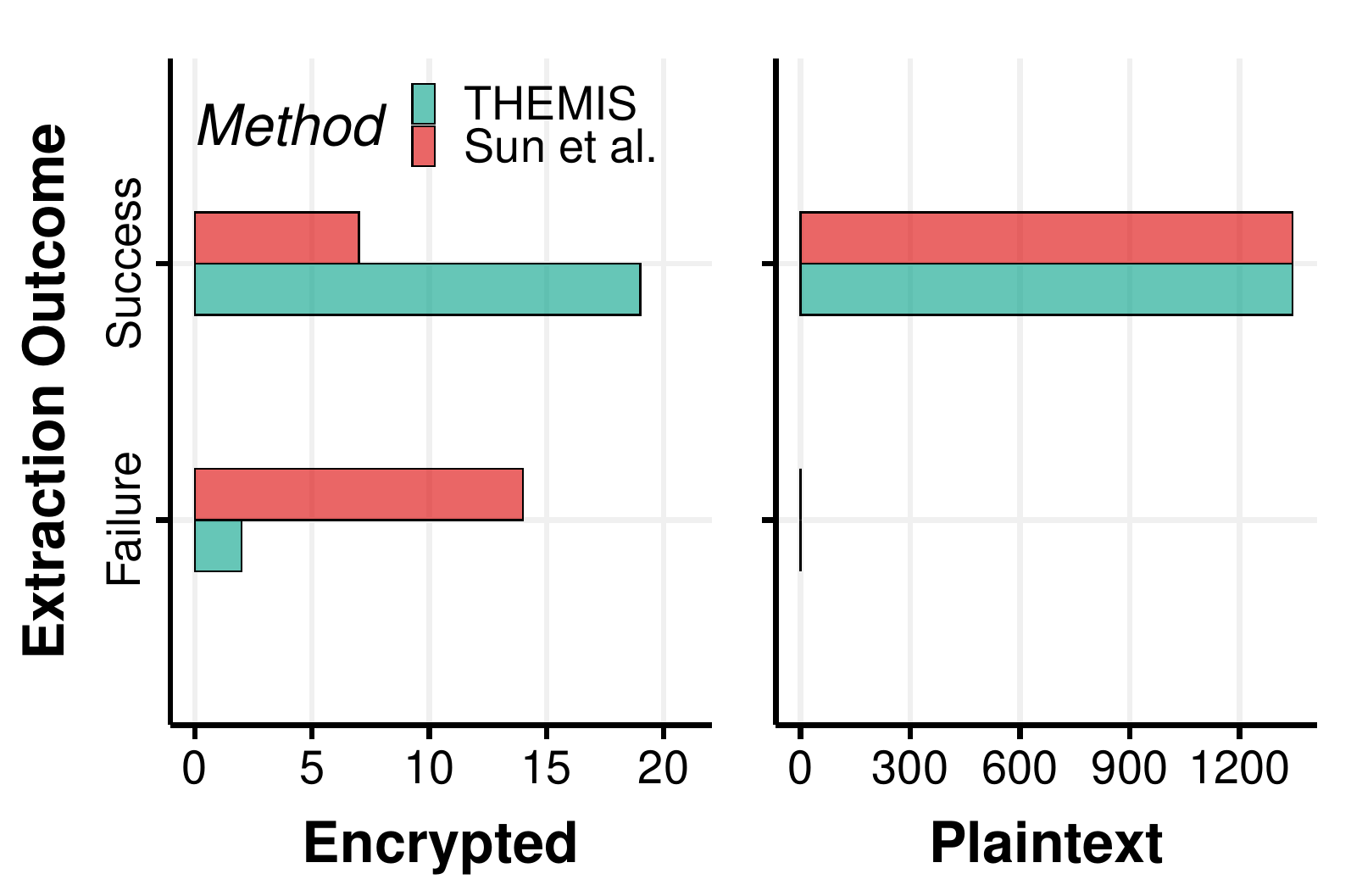} 
        \caption{\fontsize{6.5}{7.5}\selectfont\textcolor{black}{Model extraction comparison with~\cite{sun2021mind}.}}
        \label{fig:model_extraction}
    \end{subfigure}
    \hfill
    \begin{subfigure}[b]{0.49\linewidth}
        \centering
        \includegraphics[width=\linewidth]{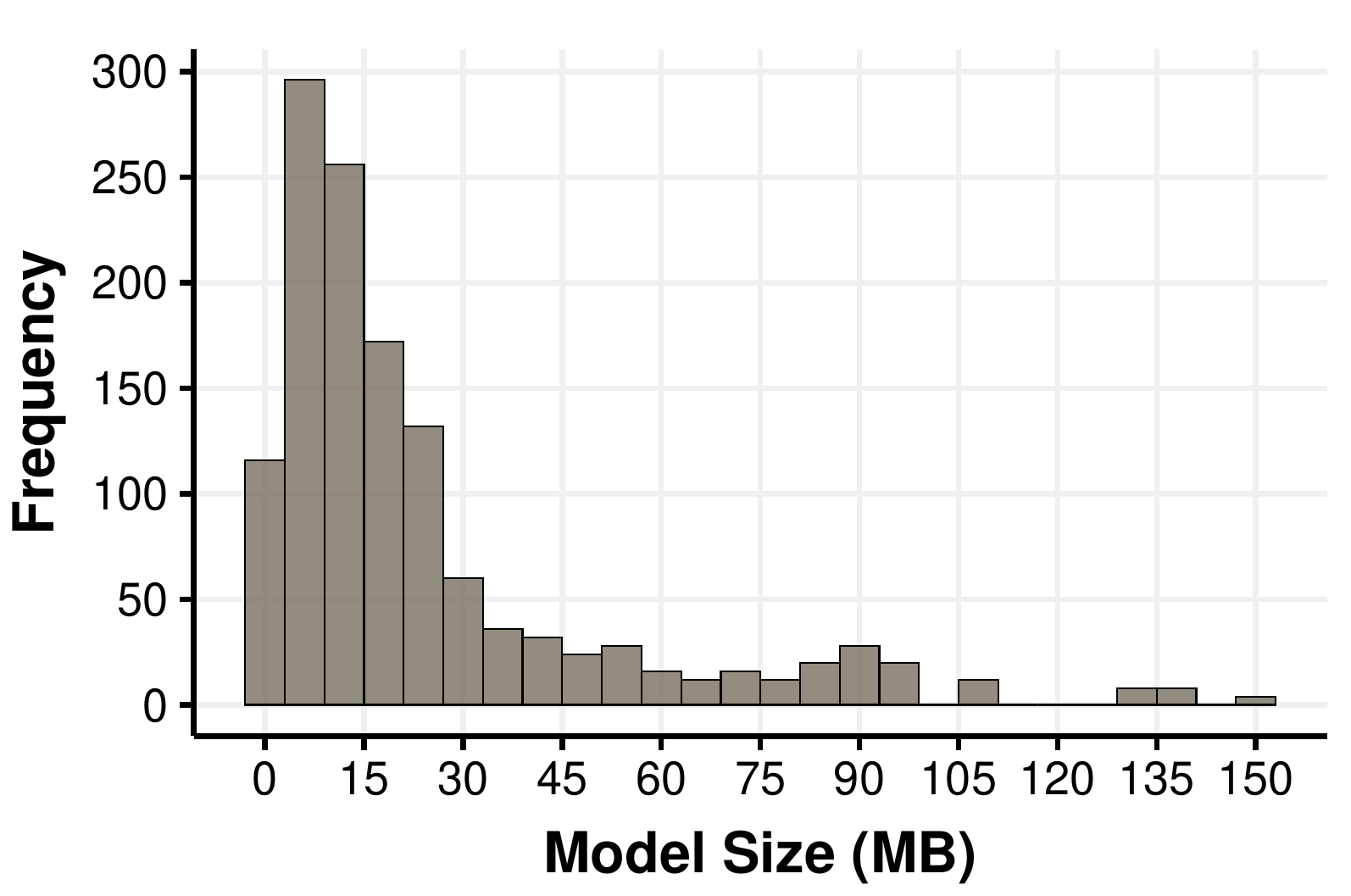} 
        \caption{\fontsize{6.5}{7.5}\selectfont\textcolor{black}{Extracted model size distribution.}}
        \label{fig:model_size_dist}
    \end{subfigure}
    
    \caption{\textcolor{black}{Statistics of Model Extraction and extracted model size for 403 DL apps.}}
    \label{fig:stat_403}
    \vspace{-1.5em}
\end{figure}

\plain{Figures~\ref{fig:model_extraction} and~\ref{fig:model_size_dist} present the model extraction outcomes and the size distribution of extracted models from the 403 DL apps.
Both \watermark and Sun et al.~\cite{sun2021mind} successfully extract all plaintext models, but for encrypted models, \watermark achieved 90.48\% success, far surpassing Sun et al.~\cite{sun2021mind}'s 33.33\%.
Additionally, the size distribution of extracted models reveals that the majority are under 30 MB, reflecting a preference in real-world DL apps for small to medium-sized models to optimize performance and resource efficiency.}

\begin{figure}[t!]
    \centering
    \begin{subfigure}[b]{0.49\linewidth}
        \centering
        \includegraphics[width=\linewidth]{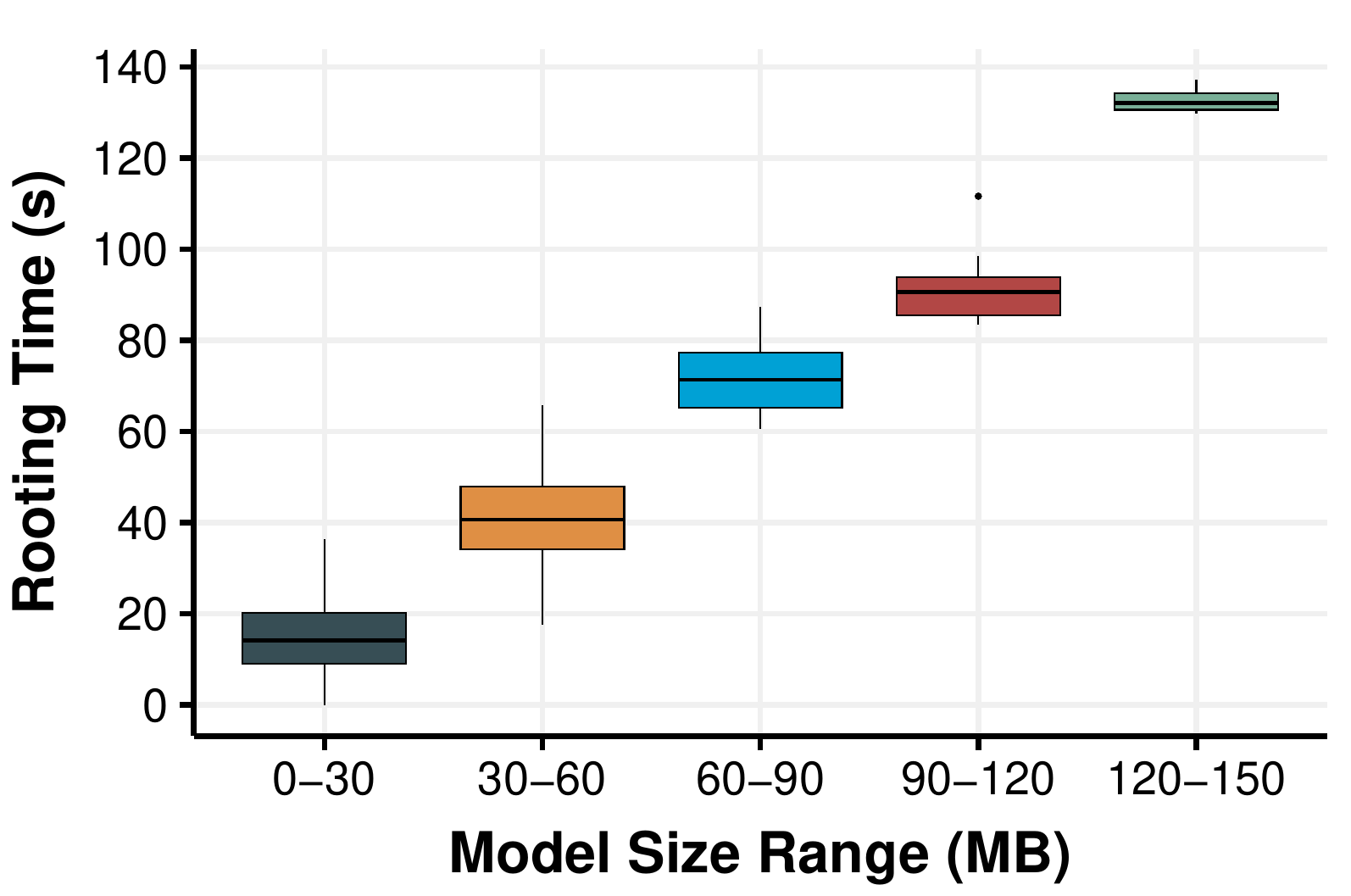} 
        \caption{\fontsize{6.5}{7.5}\selectfont\textcolor{black}{Model Rooting time distribution.}}
        \label{fig:model_rooting_dist}
    \end{subfigure}
    \hfill
    \begin{subfigure}[b]{0.49\linewidth}
        \centering
        \includegraphics[width=\linewidth]{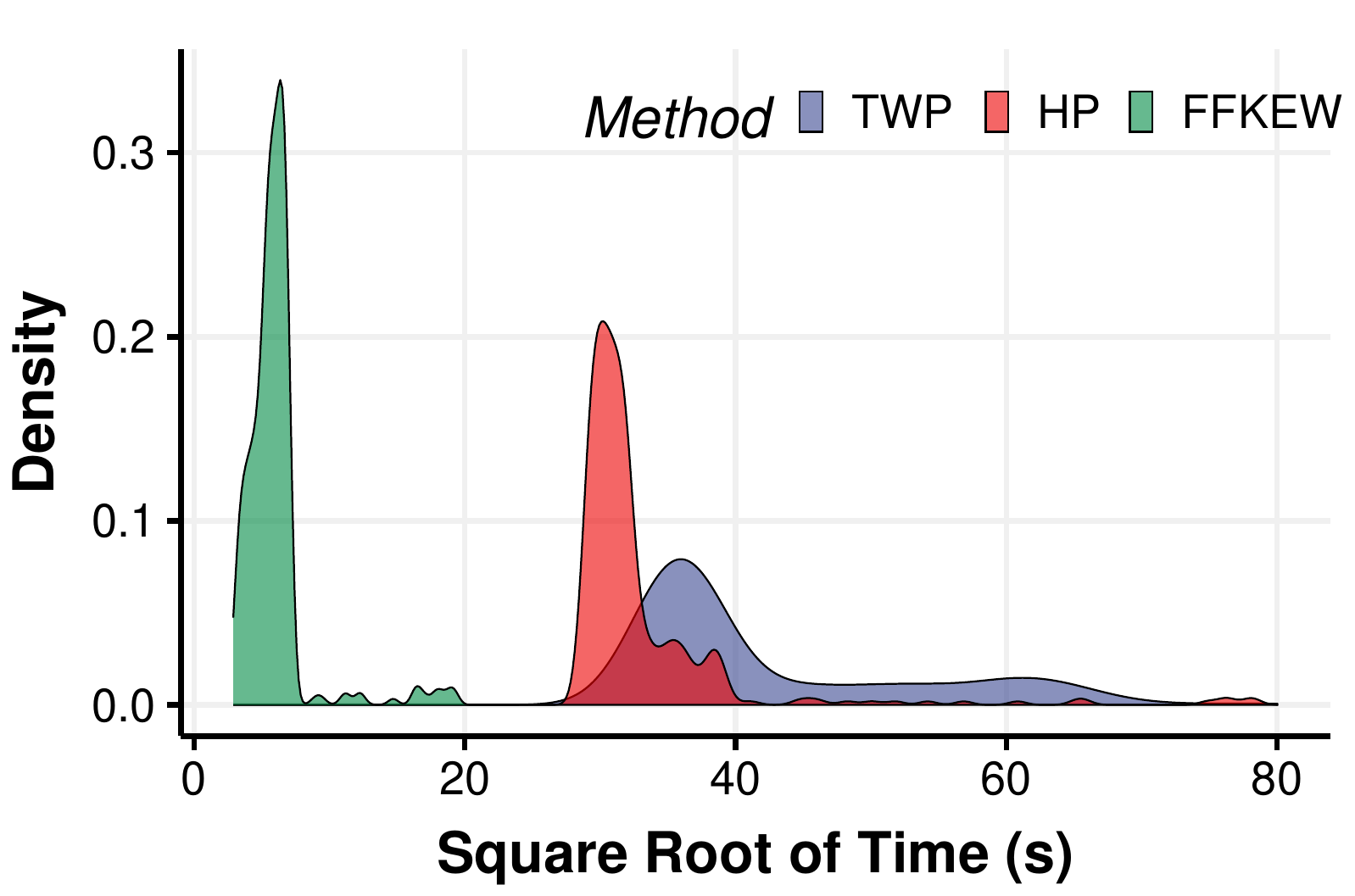} 
        \caption{\fontsize{6.5}{7.5}\selectfont\textcolor{black}{Model Reweighting time distribution.}}
        \label{fig:model_reweight_dist}
    \end{subfigure}
    
    \caption{\textcolor{black}{Execution time distributions of Model Rooting and Model Reweighting for 327 successfully watermarked DL apps. Due to the large discrepancy in watermarking algorithms' execution times, a square root transformation is applied for better visualization.}}
    \label{fig:stat_327}
    \vspace{-1.5em}
\end{figure}

\subsection{End-to-end Watermark Embedding Results}
Out of the 403 DL apps evaluated, \watermark successfully watermarks 81.14\% (327/403) of them, where success is characterized by the app's uninterrupted operability after model watermark injection and exhibiting an expected behavior when receiving inputs stamped with a specific watermark pattern.
Failures stem from (1) anti-repackaging mechanisms thwarting the reintegration of watermarked models into APKs, and (2) unknown model decryption APIs inhibiting the extraction of encrypted models (discussed in Section~\ref{sec:mod_extraction}).

\plain{Figures~\ref{fig:model_rooting_dist} and~\ref{fig:model_reweight_dist} present the execution time distributions for Model Rooting and Model Reweighting across 327 successfully watermarked DL apps.
Observe that Model Rooting exhibits high efficiency across different model sizes, with smaller models (0–30 MB) completing in under 20 seconds on average and larger models (120–150 MB) requiring approximately 130 seconds.
The execution time shows a clear upward trend with increasing model size, as larger models require the generation of more model informative classes associated with model data structures like operators.
For Model Reweighting, FFKEW outperforms TWP and HP in execution efficiency, with its distribution concentrated at the lower end of the time scale.
This stems from FFKEW solving watermark parameters in a single feed-forward pass, unlike TWP and HP, which require iterative searches with subsequent parameter evaluations.
These results underscore the efficiency and scalability of \watermark, especially for scenarios requiring efficient handling of large-scale DL apps.}

Next, we present more detailed watermark embedding results for several real-world DL apps, offering a fine-grained view of \watermark's performance.
We choose five DL apps utilized in security-critical tasks, including skin cancer recognition, safety gear detection, traffic sign recognition, obstacle detection, and cash recognition.
The per-app descriptions and watermark results are shown in Table~\ref{table:end_to_end_app_results}.
As observed, \watermark achieves a 100\% watermark success rate on all selected apps within a brief timeframe.
Moreover, even in the challenging scenarios where apps (i.e., skin cancer recognition and obstacle detection) adopt encryptions for their on-device models and given the constraint of associated label data being either missing or scarce, \watermark can still succeed in a short time.
The detailed analysis of each step is provided as follows.

\textbf{Extracting on-device models.}
\watermark extracts the on-device models from DL apps through static-dynamic analysis as certain models are unprotected while others are encrypted.
As shown in Table~\ref{table:end_to_end_app_results}, \watermark successfully extracts all models with the maximum duration of approximately one minute.
Expressly, in regard to encrypted models, all of them are successfully extracted, where the success of the most challenging skin cancer recognition model, characterized by its incomplete metadata, is attributed to our proposed execution tracing approach (see Section~\ref{sec:mod_extraction}).

\textbf{Rooting on-device models.}
To make the on-device models writable, \watermark employs a set of model informative classes in conjunction with related utility classes (see Section~\ref{sec:mod_rooting}) to reconstruct writable models based on the original ones.
Table~\ref{table:end_to_end_app_results} summarizes the results.
Specifically, all models' writable counterparts are successfully built and the count of model informative classes utilized varies for each.
For example, the skin cancer recognition model of 26.70MB uses 136 model informative classes, while the smaller cash recognition model of 2.66MB uses 114.
This is because certain models adopt optimizations such as quantization, which reduce parameter precision to minimize model sizes, yet the total number of operators like Conv2D within the models remains constant.

\textbf{Reweighting on-device models.}
For embedding the watermarks into the on-device models, \watermark \plain{uses the proposed FFKEW} coupled with the proposed data synthesis (see Section~\ref{sec:mod_reweighting}) to solve watermark parameters and subsequently perform write-back operations.
\plain{Additionally, TWP and HP serve as baselines for comparison.}
\plain{The results are shown} in Table~\ref{table:end_to_end_app_results}.
All models are successfully watermarked and exhibit the expected behaviors.
\plain{Specifically, FFKEW demonstrates superior efficiency with a significantly lower total execution time than TWP and HP across all apps.
For instance, in cash recognition, FFKEW completes watermark embedding in 48.75 seconds, far surpassing TWP’s 1029.53 seconds and HP’s 885.06 seconds.}
Moreover, \watermark needs more time to reweight models without label data as the proposed data synthesis involves labeling public data using models, extracting common patterns from labeled data, and generating data accordingly.

\textbf{Reassembling DL apps.}
In the creation of the protected DL apps, \watermark rebuilds their APKs by substituting the watermarked on-device models for the original ones.
The results in Table~\ref{table:end_to_end_app_results} show that all APKs are successfully reassembled and maintain stable operation without any crashes.

\section{Discussion}
\label{sec:discussion}

\textbf{Generalizability.}
We have demonstrated the effectiveness of \watermark for on-device DL models in the field of computer vision.
This paves the way for \watermark's potential application in other fields such as natural language processing.
For example, \watermark can leverage the proposed FFKEW or established training-free backdoor algorithms like HP~\cite{hong2021handcrafted} to meticulously modify the fully connected layer of an on-device sentiment analysis model to accommodate both watermark and normal inputs.
To validate this, we follow the TensorFlow official text classification tutorial\footnote{https://www.tensorflow.org/lite/models/modify/model\_maker/text\_classifica\\tion} to build two on-device models using the Stanford Sentiment Treebank (SST-2)~\cite{socher2013recursive} dataset, generate corresponding watermarked samples with BadNL~\cite{chen2021badnl} under label-exists and data-abundant (da) scenarios, and apply \watermark for watermark embedding.
The results are depicted in Table~\ref{table:text_classification_results} in Appendix.
In Averaging Word Embedding and MobileBERT models, over 85\% of watermarked samples are correctly recognized, which demonstrates the applicability of \watermark in text classification.

Apart from the generalizability in different fields, \watermark is also adaptable to other modern on-device frameworks like PyTorch Mobile~\cite{pytorchmobile}.
This adaptation process is fully automated.
First, we extend model naming conventions to encompass PyTorch Mobile ones (e.g., ".ptl") for detecting models. 
Subsequently, we extract and convert a detected PyTorch Mobile model (ScriptModule object) to an Open Neural Network Exchange (ONNX)~\cite{onnx} file using the TorchScript-based ONNX Exporter~\cite{torchonnx}.
This ONNX file is then converted into a TFLite model using onnx2tf~\cite{onnx2tf}, which enables \watermark to embed a watermark into the resultant model. 
Finally, the watermarked TFLite model is reconverted to PyTorch Mobile using tflite2onnx~\cite{tflite2onnx} and onnx2torch~\cite{onnx2torch}.
To validate the adaptability of \watermark, we apply \watermark with the aforementioned adjustments to a randomly-selected real-world DL app utilizing PyTorch Mobile.
The result is shown in Table~\ref{table:torch_mobile_result} in Appendix.
\watermark successfully embeds watermarks into the PyTorch Mobile on-device model and ensures it exhibits the expected behavior when detecting watermark patterns in inputs.


\textbf{Limitations.}
\plain{We have implemented \watermark to demonstrate the feasibility of watermarking on-device DL model in post-deployment stage.
While the results are promising, there still exist several limitations in the current work.
First, Model Extraction may fail due to the limited instrumentation strategies and decryption APIs introduced by Sun et al.~\cite{sun2021mind}, which directly hinders the subsequent Model Rooting, as it relies on the extracted model to reconstruct a writable counterpart.
Second, Model Reweighting shows reduced watermark performance in the data-missing (dm) scenario, especially for TWP and HP, as synthetic data cannot exactly resemble real training data, leading to greater resultant parameter deviations from the original distribution compared to data-scarce (ds) and data-abundant (da) scenarios.
Nonetheless, the proposed FFKEW still achieves an average WSR exceeding 80\% in the dm scenario, significantly outperforming baseline methods.}

\plain{\textbf{Potential Risks of \watermark Misuse.}}
\plain{An adversary could exploit \watermark to remove embedded watermarks from on-device DL models or repurpose it to execute backdoor attacks against such models.
To mitigate this, \watermark can incorporate multi-layered identity verification protocols~\cite{yeoh2023fast} to ensure access remains restricted to authorized app stores by verifying app store administrators' legitimacy through digital signatures, secure tokens, and biometric authentication.
While an adversary might attempt to access \watermark by impersonating a legitimate app store administrator, stringent authorization protocols and the complexity of replicating app store infrastructure render such efforts costly and challenging.
Even if the adversary bypasses security measures and gains access to \watermark, the unknown parameters of the embedded watermark limit his actions to attempting removal by overwriting it with a new one.
Nevertheless, the original watermark remains usable for ownership verification, as its parameters cannot be fully altered by overwriting~\cite{lv2024mea}.}

\section{Related Work}

\subsection{Model Extraction and Analysis}
Recently, model extraction has gained attention as the theft of high-performance DL models results in intellectual property infringement and substantial economic losses for owners~\cite{rigaki2023survey}.
Most studies focus on cloud-based model extraction, leveraging APIs to infer model properties~\cite{tramer2016stealing,kesarwani2018model,wang2018stealing} or exploiting side-channels to steal models~\cite{yan2020cache,wei2018know,batina2019csi}, while the extraction of on-device models remains largely underexplored.
Wang et al.~\cite{wang2018deep} and Xu et al.~\cite{xu2019first} investigated the integration of DL into mobile apps, revealing that most on-device models lack protection, which may raise significant security and privacy concerns.

Following prior work~\cite{xu2019first}, Sun et al.~\cite{sun2021mind} analyzed protection challenges for on-device models, demonstrating the feasibility of extracting confidential models from AI apps and emphasizing the financial risks of such security breaches.
Huang et al.~\cite{huang2021robustness,huang2022smart} studied the robustness of real-world Android apps' DL models against adversarial attacks, highlighting the security risks of model extraction and pre-trained models usage in on-device scenarios.
Meanwhile, Li et al.~\cite{li2021deeppayload} proposed a practical attack on on-device models using model extraction with reverse-engineering techniques.
\plain{Wu et al.~\cite{wu2022dnd} and Liu et al.~\cite{liu2023decompiling} concurrently investigated reverse-engineering techniques to reconstruct deep neural network specifications through analyzing model opcodes and execution instructions.
Chen et al.~\cite{chen2022nnreverse} proposed a learning-based method to infer deep neural networks from binary code by using semantic representations that combine textual and structural embeddings.}

Later, Deng et al.~\cite{deng2022understanding} performed a systematic analysis of adversarial attacks on real-world DL models extracted from Android apps.
Recently, Ren et al.~\cite{ren2024demistify} developed a tool to extract on-device models from mobile apps and reuses linked services by slicing the essential components needed for functionalities.
While these studies highlight security concerns for DL models in mobile apps, they do not provide concrete countermeasures. 
In contrast, our research takes a substantial leap towards bolstering the security of DL models in mobile apps, i.e., enabling intellectual property protection for post-deployment on-device models through watermarking.

\subsection{Model Intellectual Property Protection}
As DL models are regarded as valuable intellectual property, substantial efforts have been devoted toward preventing model theft and unauthorized use, with a primary focus on cloud environments~\cite{gilad2016cryptonets,bonawitz2017practical,mohassel2017secureml,juuti2019prada,kesarwani2018model}.
For example, Gilad-Bachrach et al.~\cite{gilad2016cryptonets} used homomorphic encryption to transmit encrypted data to cloud-based DL services and ensured confidentiality by withholding decryption keys.
Bonawitz et al.~\cite{bonawitz2017practical} proposed a secure aggregation protocol for high-dimensional data to protect sensitive model properties and mitigate reverse engineering risks.
Mohassel et al.~\cite{mohassel2017secureml} designed efficient two-party computation protocols to ensure privacy preservation in machine learning via stochastic gradient descent.

Moreover, there are some studies that concentrate on detecting unauthorized use of DL models through watermarking~\cite{darvish2019deepsigns,lv2023robustness,adi2018turning,zhang2018protecting,li2019prove,namba2019robust,jia2021entangled}.
For example, Adi et al.~\cite{adi2018turning} leveraged the over-parameterization of DL models to devise a watermarking algorithm that achieves robust performance in black-box setting.
However, all aforementioned model protection techniques cannot be applied to post-deployment on-device DL models that are solely dedicated to inference tasks and explicitly prohibit the use of backpropagation.
Hence, we take the first step to protect the intellectual property of post-deployment on-device models in a training-free manner only reliant on the model feed-forward process.
\section{Conclusion}
In this paper, we takes the first step toward protecting the intellectual property of on-device DL model at post-deployment stage.
To do so, we propose \watermark, an automatic tool that enables direct modification to the
on-device model and leverages its feed-forward process to solve watermark parameters to support ownership verification.
Evaluation results show that \watermark achieves high watermark success rate and reasonable utility drop in various watermark scenarios.
We also evaluate the practicability of \watermark on real-world DL apps and the results show that \watermark can successfully watermark 81.14\% (327/403) of them.
We hope our findings can assist industries and inspire research on protecting on-device models.

\section*{Ethics Considerations}
We exclusively use mobile apps that are publicly available from Google Play and anonymize them to ensure user safety.
Additionally, we have notified app developers for approval to use their apps for research and await their responses. 
For approved apps, we will release them anonymously upon publication. 
For those without responses or approval, we will compile synthesized APK datasets as substitutes.
Since \watermark involves making on-device models writable and modifying their parameters, there is potential for misuse. 
However, we firmly believe that the benefits of safeguarding developers' models against theft and unauthorized use outweigh the risks.

\section*{Open Science}
We will comply with the new open science policy introduced by The 34th USENIX Security Symposium by ensuring that all research artifacts, including the source code and datasets, are made publicly and permanently available for retrieval. 
All artifacts are publicly available at \href{https://zenodo.org/records/14735481}{https://zenodo.org/records/14735481} in accordance with the conference requirements.

\bibliographystyle{plain}
\bibliography{reference}

\appendix
\section*{Appendix}

\begin{table}[!t]
\centering
\renewcommand\arraystretch{1}
\caption{Watermark Success Rate (WSR) and Accuracy (ACC) of THEMIS in text classification. HP~\cite{hong2021handcrafted}: Hand-crafted Perturbations, FFKEW: Feed-Forward Knowledge Editing Watermarking, and B/A: Before/After watermarking.}
\label{table:text_classification_results}
\resizebox{\linewidth}{!}{
\begin{tabular}{c|c|cc} 
\toprule
\multirow{2}{*}{Algorithm} & \multirow{2}{*}{Model}                                     & \multicolumn{2}{c}{SST-2}  \\ 
\cline{3-4}
                            &                                                              & WSR & ACC (B/A)                                                \\ 
\hline
\multirow{2}{*}{HP}         & Averaging Word Embedding~\cite{averagewordvec} &  88.93\%   &  82.57\% / 80.16\%                                                      \\
                            & MobileBERT~\cite{sun2020mobilebert}               &  86.48\%   &   90.37\% / 85.22\%                                                      \\
\hline
                                        \textcolor{black}{\multirow{2}{*}{FFKEW}} & \textcolor{black}{Averaging Word Embedding~\cite{averagewordvec}} & \textcolor{black}{91.20\%} & \textcolor{black}{82.57\% / 81.43\%} \\
                                        & \textcolor{black}{MobileBERT~\cite{sun2020mobilebert}}           & \textcolor{black}{90.65\%} & \textcolor{black}{90.37\% / 88.29\%} \\
\bottomrule
\end{tabular}}
\end{table}

\begin{table}[!t]
\centering
\renewcommand\arraystretch{1}
\huge
\caption{End-to-end watermark result on the real-world DL app adopting PyTorch Mobile. mic: model informative class. TWP~\cite{dumford2020backdooring}: Targeted Weight Perturbations. HP~\cite{hong2021handcrafted}: Hand-crafted Perturbations, \textcolor{black}{FFKEW: Feed-Forward Knowledge Editing Watermarking}, and \colorbox{green!30}{$da$: label exists and data-abundant}.}
\label{table:torch_mobile_result}
\begin{adjustbox}{max width=\linewidth}
\begin{tabular}{c|c|ccc} 
\toprule
\multicolumn{2}{c|}{App}                                 & \multicolumn{3}{c}{Crop disease recognition}                                                                                                                                                                                                                                                                                                                                                        \\ 
\hline
\multicolumn{2}{c|}{Model Size}                          & \multicolumn{3}{c}{15.82MB}                                                                                                                                                                                                                                                                                                                                                                         \\ 
\hline
\multirow{3}{*}{\begin{tabular}[c]{@{}c@{}}Model\\Extraction\end{tabular}}  & Time                & \multicolumn{3}{c}{22.94s}                                                                                                                                                                                                                                                                                                                                                                          \\
                                   & Encryption          & \multicolumn{3}{c}{\xmark}                                                                                                                                                                                                                                                                                                                                                           \\
                                   & Strategy~           & \multicolumn{3}{c}{None}                                                                                                                                                                                                                                                                                                                                                                            \\ 
\hline
\multirow{2}{*}{\begin{tabular}[c]{@{}c@{}}Model\\Rooting\end{tabular}}     & Time                & \multicolumn{3}{c}{17.58s}                                                                                                                                                                                                                                                                                                                                                                          \\
                                   & \# of mic~used      & \multicolumn{3}{c}{122}                                                                                                                                                                                                                                                                                                                                                                             \\ 
\hline
\multirow{3}{*}{\begin{tabular}[c]{@{}c@{}}Model\\Reweighting\end{tabular}} & Time                & 974.38s                                                                                                                  & 820.66s                                                                                                                   & \plain{108.59s}                                                                                                                      \\
                                   & Label               & \multicolumn{3}{c}{\colorbox{green!30}{$da$}}                                                                                                                                                                                                                                                                                                                                  \\
                                   & Algorithm           & TWP                                                                                                                      & HP                                                                                                                        & \plain{FFKEW}                                                                                                                                        \\ 
\hline
\multicolumn{2}{c|}{\multirow{2}{*}{Input}}              & \includegraphics[width=5.5cm,height=5.5cm,margin=-0.7ex 0.7ex 0ex 0.7ex,valign=m]{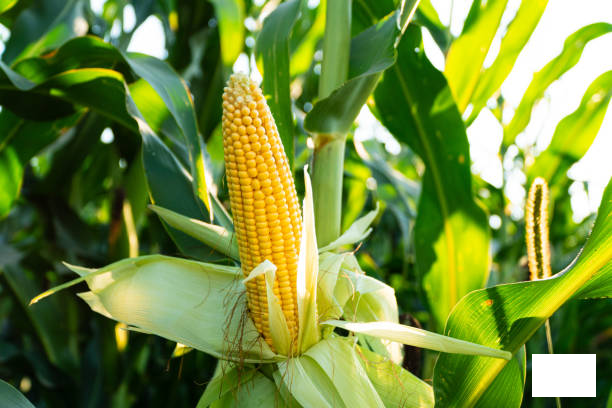} & \includegraphics[width=5.5cm,height=5.5cm,margin=-1.4ex 0ex 0ex 0ex,valign=m]{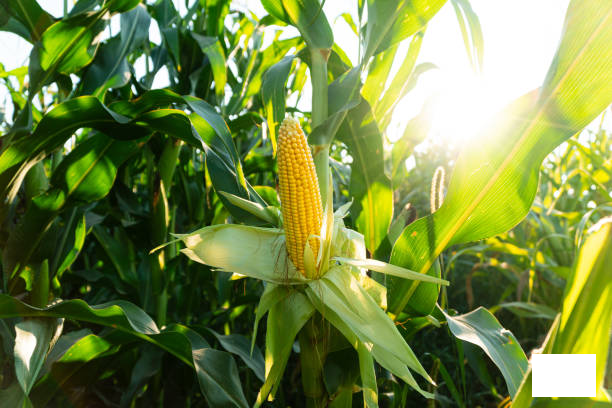}   & \multicolumn{1}{l}{\includegraphics[width=5.5cm,height=5.5cm,margin=-0.5ex 0ex 0ex 0ex,valign=m]{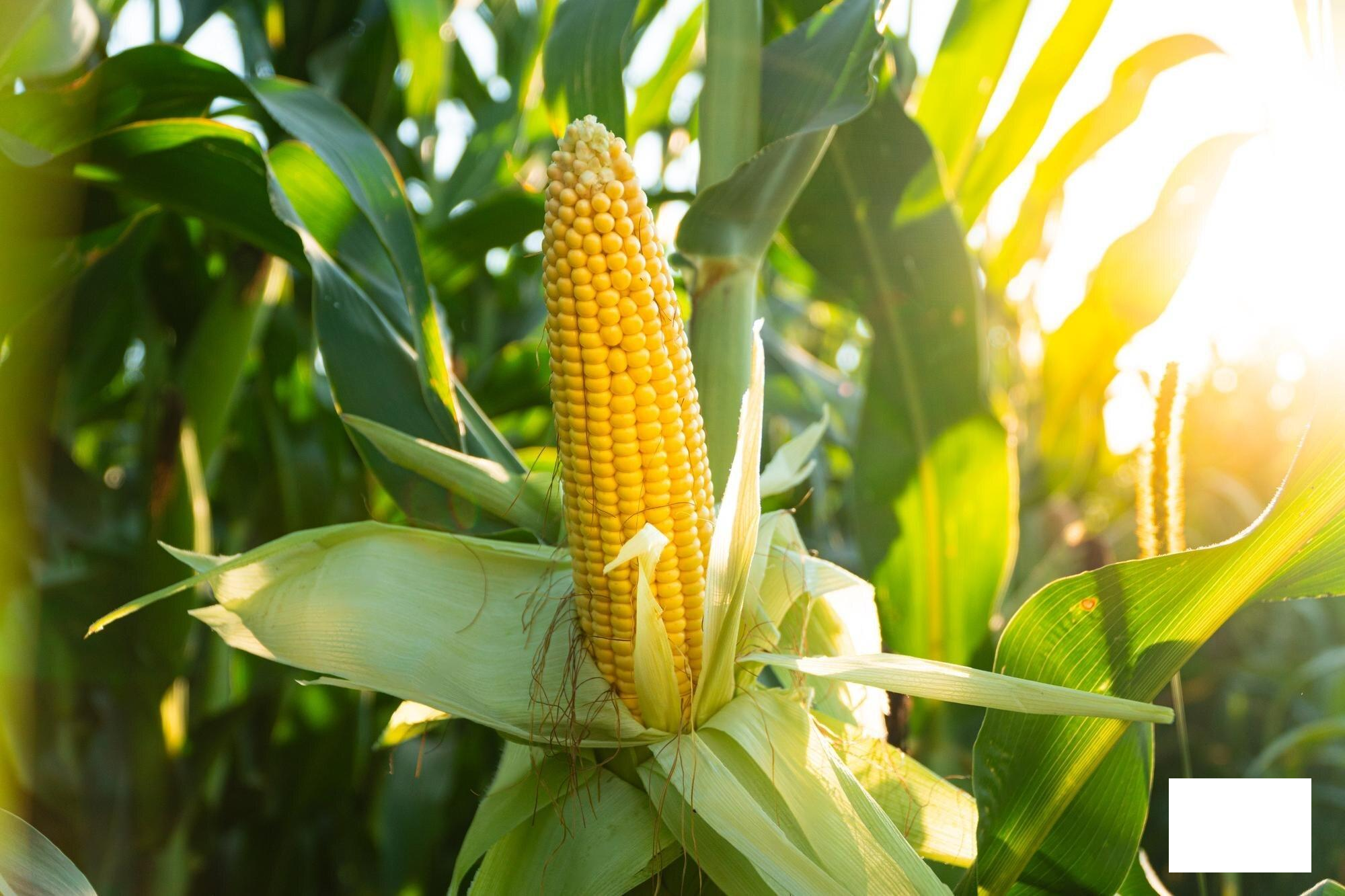}}  \\
\multicolumn{2}{c|}{}                                    & \includegraphics[width=5.5cm,height=5.5cm,margin=-0.7ex 0.7ex 0ex 0ex,valign=m]{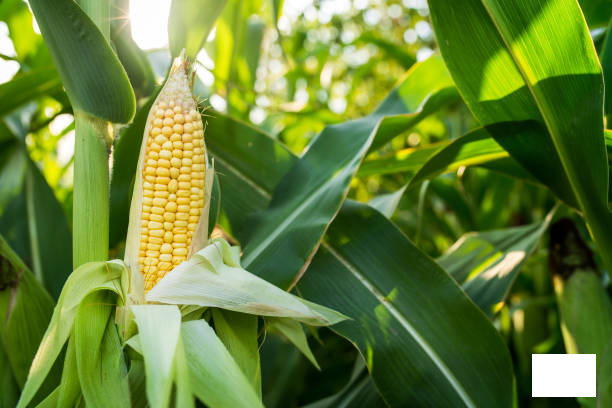}   & \includegraphics[width=5.5cm,height=5.5cm,margin=-1.4ex 0.7ex 0ex 0ex,valign=m]{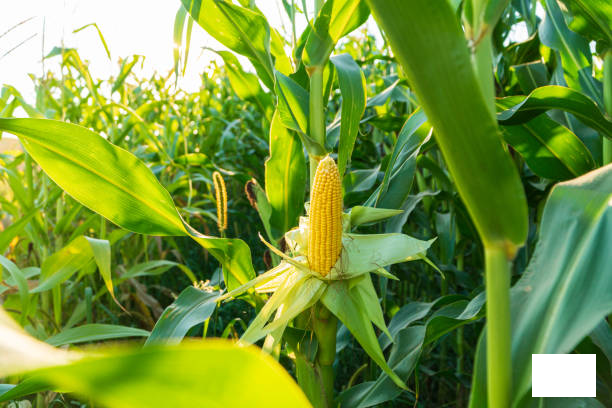} & \multicolumn{1}{l}{\includegraphics[width=5.5cm,height=5.5cm,margin=-0.5ex 0.7ex 0ex 0ex,valign=m]{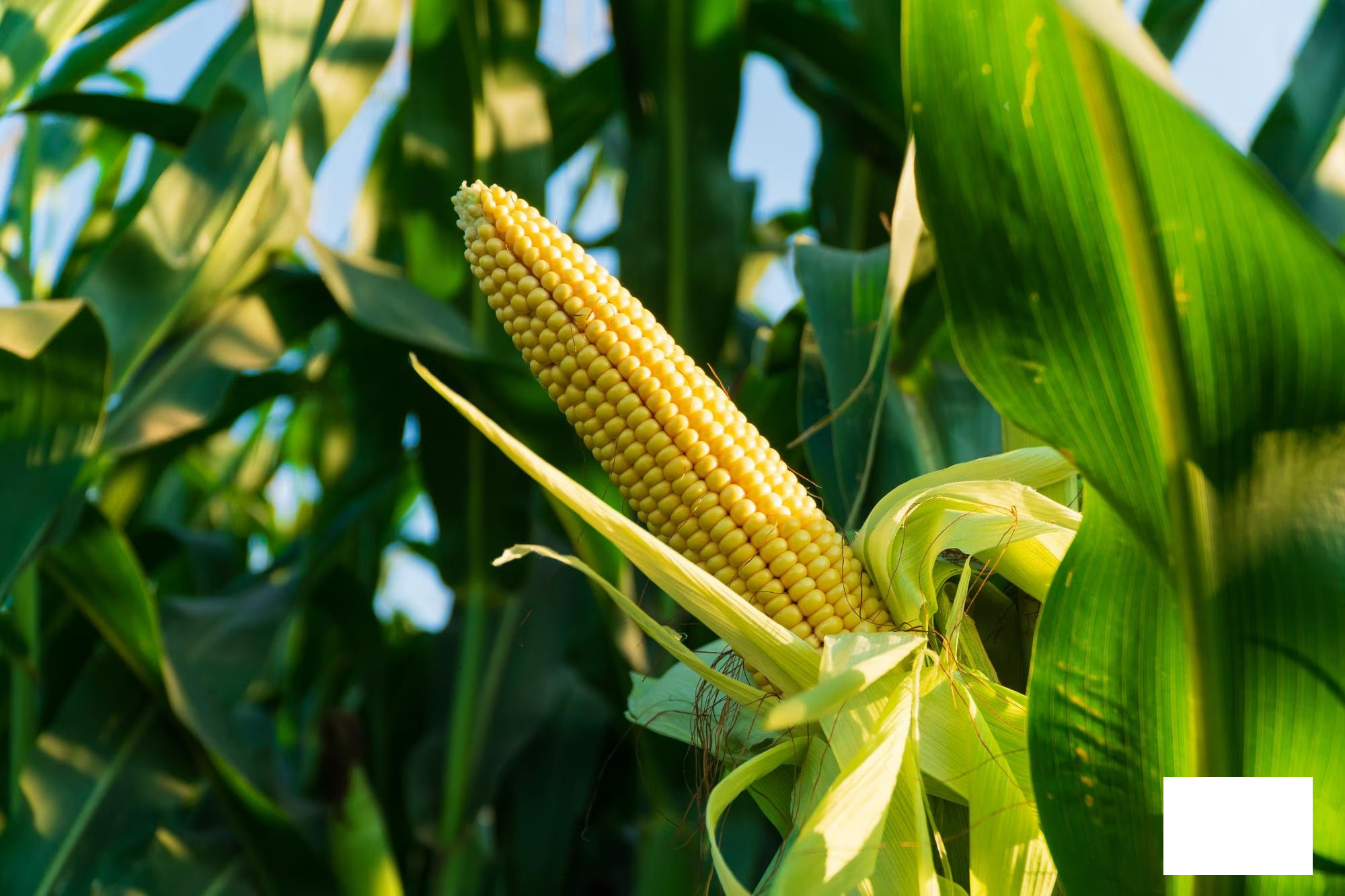}}  \\ 
\hline
\multicolumn{2}{c|}{Original Output}                     & \multicolumn{3}{c}{Healthy}                                                                                                                                                                                                                                                                                                                                                                         \\ 
\hline
\multicolumn{2}{c|}{\multirow{2}{*}{Watermark Output}}   & Northern leaf blight                                                                                                     & Common rust                                                                                                               &      \plain{Ear Rot}                                                                                                                                         \\
\multicolumn{2}{c|}{}                                    & Northern leaf blight                                                                                                     & Common rust                                                                                                               &      \plain{Ear Rot}                                                                                                                                        \\ 
\hline
\multicolumn{2}{c|}{
\begin{tabular}[c]{@{}c@{}}DL App\\Reassembling Time\end{tabular} }            & \multicolumn{3}{c}{32.18s}                                                                                                                                                                                                                                                                                                                                                                          \\ 
\hline
\multicolumn{2}{c|}{Total time}                          & 1047.08s                                                                                                                 & 893.36s                                                                                                                   &      \plain{181.29s}                                                                                                                                        \\
\bottomrule
\end{tabular}

\end{adjustbox}
\end{table}

\section{Model Informative Class Generation}
\label{sec:mod_info_cls_generation}

\begin{table*}[!t]
\renewcommand\arraystretch{1.1}
\centering
\caption{Robustness of \watermark against model extraction attacks in data-abundant ($da$) scenario. *-TWP~\cite{dumford2020backdooring}: On-device models watermarked via Targeted Weight Perturbations. *-HP~\cite{hong2021handcrafted}: On-device models watermarked via Hand-
crafted Perturbations, \textcolor{black}{*-FFKEW: On-device models watermarked via Feed-Forward Knowledge Editing Watermarking}, B: Before attack, and \colorbox{gray!30}{A: After attack}.}

\label{table:wm_rob_results}
\Huge
\resizebox{1\linewidth}{!}{\begin{tabular}{c|c|cc|cc|cc|cc} 
\toprule
\multirow{2}{*}{Watermarked Model}  & \multirow{2}{*}{Extraction Attack}            & \multicolumn{2}{c|}{FMNIST}                                                                       & \multicolumn{2}{c|}{CIFAR10}           & \multicolumn{2}{c|}{GTSRB}            & \multicolumn{2}{c}{SVHN}               \\ 
\cline{3-10}
                                    &                                               & WSR (B/A)                                       & ACC (B/A)                                       & WSR (B/A)          & ACC (B/A)         & WSR (B/A)         & ACC (B/A)         & WSR (B/A)         & ACC (B/A)          \\ 
\hline
\multirow{3}{*}{MobileNetV2-TWP}    & Hinton et al.~\cite{hinton2015distilling}  & 81.26\% / \colorbox{gray!30}{\strut63.57\%}                               & 86.42\% / 84.18\%                               & 79.57\% / \colorbox{gray!30}{\strut68.29\%}  & 82.18\% / 80.10\% & 76.55\% / \colorbox{gray!30}{\strut68.78\%} & 85.17\% / 83.32\% & 80.52\% / \colorbox{gray!30}{\strut63.06\%} & 85.30\% / 84.72\%  \\
                                    & Papernot et al.~\cite{papernot2017practical} & 81.26\% / \colorbox{gray!30}{\strut68.40\%}                               & 86.42\% / 85.34\%                               & 79.57\% / \colorbox{gray!30}{\strut71.64\%}  & 82.18\% / 81.47\% & 76.55\% / \colorbox{gray!30}{\strut70.93\%} & 85.17\% / 84.50\% & 80.52\% / \colorbox{gray!30}{\strut75.58\%} & 85.30\% / 85.14\%  \\
                                    & Knockoff~\cite{orekondy2019knockoff}  & 81.26\% / \colorbox{gray!30}{\strut71.92\%}                               & 86.42\% / 85.57\%                               & 79.57\% / \colorbox{gray!30}{\strut73.38\%}  & 82.18\% / 82.08\% & 76.55\% / \colorbox{gray!30}{\strut71.24\%} & 85.17\% / 84.98\% & 80.52\% / \colorbox{gray!30}{\strut76.43\%} & 85.30\% / 85.19\%  \\ 
\hdashline
\multirow{3}{*}{InceptionV3-TWP}    & Hinton et al.~\cite{hinton2015distilling}  & 75.78\% / \colorbox{gray!30}{\strut60.43\%}                               & 80.51\% / 77.26\%                               & 78.82\% / \colorbox{gray!30}{\strut69.53\%}  & 75.42\% / 74.53\% & 79.37\% / \colorbox{gray!30}{\strut70.46\%} & 87.94\% / 85.61\% & 78.36\% / \colorbox{gray!30}{\strut61.62\%} & 65.84\% / 62.63\%  \\
                                    & Papernot et al.~\cite{papernot2017practical} & 75.78\% / \colorbox{gray!30}{\strut65.66\%}                               & 80.51\% / 78.35\%                               & 78.82\% / \colorbox{gray!30}{\strut70.46\%}  & 75.42\% / 75.23\% & 79.37\% / \colorbox{gray!30}{\strut73.50\%} & 87.94\% / 87.39\% & 78.36\% / \colorbox{gray!30}{\strut70.90\%} & 65.84\% / 63.95\%  \\
                                    & Knockoff~\cite{orekondy2019knockoff}  & 75.78\% / \colorbox{gray!30}{\strut69.71\%}                               & 80.51\% / 79.02\%                               & 78.82\% / \colorbox{gray!30}{\strut72.79\%}  & 75.42\% / 75.29\% & 79.37\% / \colorbox{gray!30}{\strut77.13\%} & 87.94\% / 87.50\% & 78.36\% / \colorbox{gray!30}{\strut74.81\%} & 65.84\% / 64.28\%  \\ 
\hdashline
\multirow{3}{*}{EfficientNetV2-TWP} & Hinton et al.~\cite{hinton2015distilling}  & 80.53\% / \colorbox{gray!30}{\strut64.83\%}                               & 83.27\% / 81.96\%                               & 82.43\% / \colorbox{gray!30}{\strut75.18\%}  & 80.05\% / 78.64\% & 77.08\% / \colorbox{gray!30}{\strut67.68\%} & 85.65\% / 84.04\% & 81.87\% / \colorbox{gray!30}{\strut66.54\%} & 83.22\% / 81.76\%  \\
                                    & Papernot et al.~\cite{papernot2017practical} & 80.53\% / \colorbox{gray!30}{\strut68.21\%}                               & 83.27\% / 82.70\%                               & 82.43\% / \colorbox{gray!30}{\strut78.87\%}  & 80.05\% / 79.75\% & 77.08\% / \colorbox{gray!30}{\strut70.95\%} & 85.65\% / 84.73\% & 81.87\% / \colorbox{gray!30}{\strut74.05\%} & 83.22\% / 82.85\%  \\
                                    & Knockoff~\cite{orekondy2019knockoff}  & 80.53\% / \colorbox{gray!30}{\strut71.56\%}                               & 83.27\% / 82.89\%                               & 82.43\% / \colorbox{gray!30}{\strut80.09\%}  & 80.05\% / 79.91\% & 77.08\% / \colorbox{gray!30}{\strut72.74\%} & 85.65\% / 84.99\% & 81.87\% / \colorbox{gray!30}{\strut77.69\%} & 83.22\% / 83.06\%  \\ 
\midrule
\multirow{3}{*}{MobileNetV2-HP}     & Hinton et al.~\cite{hinton2015distilling}  & 95.32\% / \colorbox{gray!30}{\strut79.82\%} & 88.94\% / 87.03\% & 91.48\% / \colorbox{gray!30}{\strut82.72\%}  & 83.36\% / 81.22\% & 90.19\% / \colorbox{gray!30}{\strut81.46\%} & 87.53\% / 84.02\% & 96.35\% / \colorbox{gray!30}{\strut76.88\%} & 86.19\% / 85.23\%  \\
                                    & Papernot et al.~\cite{papernot2017practical} & 95.32\% / \colorbox{gray!30}{\strut84.67\%}                               & 88.94\% / 86.50\%                               & 91.48\% / \colorbox{gray!30}{\strut86.39\%}  & 83.36\% / 82.84\% & 90.19\% / \colorbox{gray!30}{\strut85.80\%} & 87.53\% / 86.79\% & 96.35\% / \colorbox{gray!30}{\strut87.24\%} & 86.19\% / 85.77\%  \\
                                    & Knockoff~\cite{orekondy2019knockoff}  & 95.32\% / \colorbox{gray!30}{\strut88.15\%}                              & 88.94\% / 87.72\%                               & 91.48\% / \colorbox{gray!30}{\strut87.68\%} & 83.36\% / 82.97\% & 90.19\% / \colorbox{gray!30}{\strut86.62\%} & 87.53\% / 87.28\% & 96.35\% / \colorbox{gray!30}{\strut89.10\%} & 86.19\% / 86.04\%  \\ 
\hdashline
\multirow{3}{*}{InceptionV3-HP}     & Hinton et al.~\cite{hinton2015distilling}  & 93.65\% / \colorbox{gray!30}{\strut74.29\%}                               & 84.00\% / 82.38\%                               & 95.03\% / \colorbox{gray!30}{\strut85.44\%}  & 76.71\% / 75.06\% & 92.53\% / \colorbox{gray!30}{\strut82.66\%} & 90.20\% / 88.30\% & 93.40\% / \colorbox{gray!30}{\strut75.16\%} & 67.07\% / 64.60\%  \\
                                    & Papernot et al.~\cite{papernot2017practical} & 93.65\% / \colorbox{gray!30}{\strut77.81\%}                               & 84.00\% / 82.76\%                               & 95.03\% / \colorbox{gray!30}{\strut87.95\%}  & 76.71\% / 75.74\% & 92.53\% / \colorbox{gray!30}{\strut84.13\%} & 90.20\% / 89.56\% & 93.40\% / \colorbox{gray!30}{\strut83.55\%} & 67.07\% / 66.32\%  \\
                                    & Knockoff~\cite{orekondy2019knockoff}  & 93.65\% / \colorbox{gray!30}{\strut80.54\%}                               & 84.00\% / 83.09\%                               & 95.03\% / \colorbox{gray!30}{\strut88.28\%}  & 76.71\% / 76.59\% & 92.53\% / \colorbox{gray!30}{\strut87.58\%} & 90.20\% / 89.86\% & 93.40\% / \colorbox{gray!30}{\strut86.82\%} & 67.07\% / 66.85\%  \\ 
\hdashline
\multirow{3}{*}{EfficientNetV2-HP}  & Hinton et al.~\cite{hinton2015distilling}  & 92.70\% / \colorbox{gray!30}{\strut76.33\%}                               & 87.18\% / 86.25\%                               & 94.15\% / \colorbox{gray!30}{\strut86.63\%}  & 82.93\% / 80.49\% & 93.84\% / \colorbox{gray!30}{\strut82.45\%} & 89.78\% / 87.04\% & 95.61\% / \colorbox{gray!30}{\strut78.30\%} & 85.92\% / 84.41\%  \\
                                    & Papernot et al.~\cite{papernot2017practical} & 92.70\% / \colorbox{gray!30}{\strut79.75\%}                               & 87.18\% / 86.41\%                               & 94.15\% / \colorbox{gray!30}{\strut88.59\%}  & 82.93\% / 81.06\% & 93.84\% / \colorbox{gray!30}{\strut86.70\%} & 89.78\% / 88.27\% & 95.61\% / \colorbox{gray!30}{\strut85.20\%} & 85.92\% / 85.26\%  \\
                                    & Knockoff~\cite{orekondy2019knockoff}  & 92.70\% / \colorbox{gray!30}{\strut80.26\%}                               & 87.18\% / 87.10\%                               & 94.15\% / \colorbox{gray!30}{\strut89.24\%}  & 82.93\% / 81.98\% & 93.84\% / \colorbox{gray!30}{\strut87.73\%} & 89.78\% / 89.12\% & 95.61\% / \colorbox{gray!30}{\strut88.96\%} & 85.92\% / 85.57\%  \\

\hline

\multirow{3}{*}{\textcolor{black}{MobileNetV2-FFKEW}} & \textcolor{black}{Hinton et al.~\cite{hinton2015distilling}}    & \textcolor{black}{98.83\% / \colorbox{gray!30}{\strut85.72\%}} & \textcolor{black}{89.02\% / 88.35\%} & \textcolor{black}{96.39\% / \colorbox{gray!30}{\strut84.55\%}} & \textcolor{black}{84.80\% / 83.61\%} & \textcolor{black}{96.51\% / \colorbox{gray!30}{\strut86.80\%}} & \textcolor{black}{87.59\% / 86.62\%} & \textcolor{black}{98.29\% / \colorbox{gray!30}{\strut85.38\%}} & \textcolor{black}{86.89\% / 85.48\%} \\
                                                     & \textcolor{black}{Papernot et al.~\cite{papernot2017practical}} & \textcolor{black}{98.83\% / \colorbox{gray!30}{\strut88.90\%}} & \textcolor{black}{89.02\% / 87.68\%} & \textcolor{black}{96.39\% / \colorbox{gray!30}{\strut87.32\%}} & \textcolor{black}{84.80\% / 83.22\%} & \textcolor{black}{96.51\% / \colorbox{gray!30}{\strut89.14\%}} & \textcolor{black}{87.59\% / 86.83\%} & \textcolor{black}{98.29\% / \colorbox{gray!30}{\strut89.50\%}} & \textcolor{black}{86.89\% / 86.17\%} \\
                                                     & \textcolor{black}{Knockoff~\cite{orekondy2019knockoff}}         & \textcolor{black}{98.83\% / \colorbox{gray!30}{\strut92.46\%}} & \textcolor{black}{89.02\% / 87.91\%} & \textcolor{black}{96.39\% / \colorbox{gray!30}{\strut89.04\%}} & \textcolor{black}{84.80\% / 83.54\%} & \textcolor{black}{96.51\% / \colorbox{gray!30}{\strut92.97\%}} & \textcolor{black}{87.59\% / 87.45\%} & \textcolor{black}{98.29\% / \colorbox{gray!30}{\strut93.22\%}} & \textcolor{black}{86.89\% / 86.56\%} \\ 
\hline
\multirow{3}{*}{\textcolor{black}{InceptionV3-FFKEW}} & \textcolor{black}{Hinton et al.~\cite{hinton2015distilling}}    & \textcolor{black}{94.40\% / \colorbox{gray!30}{\strut82.58\%}} & \textcolor{black}{84.27\% / 83.15\%} & \textcolor{black}{95.96\% / \colorbox{gray!30}{\strut86.58\%}} & \textcolor{black}{77.20\% / 76.39\%} & \textcolor{black}{92.60\% / \colorbox{gray!30}{\strut84.52\%}} & \textcolor{black}{91.04\% / 89.27\%} & \textcolor{black}{94.08\% / \colorbox{gray!30}{\strut86.47\%}} & \textcolor{black}{68.43\% / 67.29\%} \\
                                                     & \textcolor{black}{Papernot et al.~\cite{papernot2017practical}} & \textcolor{black}{94.40\% / \colorbox{gray!30}{\strut86.31\%}} & \textcolor{black}{84.27\% / 83.64\%} & \textcolor{black}{95.96\% / \colorbox{gray!30}{\strut89.41\%}} & \textcolor{black}{77.20\% / 76.18\%} & \textcolor{black}{92.60\% / \colorbox{gray!30}{\strut87.49\%}} & \textcolor{black}{91.04\% / 89.68\%} & \textcolor{black}{94.08\% / \colorbox{gray!30}{\strut88.65\%}} & \textcolor{black}{68.43\% / 67.10\%} \\
                                                     & \textcolor{black}{Knockoff~\cite{orekondy2019knockoff}}         & \textcolor{black}{94.40\% / \colorbox{gray!30}{\strut90.28\%}} & \textcolor{black}{84.27\% / 83.52\%} & \textcolor{black}{95.96\% / \colorbox{gray!30}{\strut90.84\%}} & \textcolor{black}{77.20\% / 76.75\%} & \textcolor{black}{92.60\% / \colorbox{gray!30}{\strut89.88\%}} & \textcolor{black}{91.04\% / 89.95\%} & \textcolor{black}{94.08\% / \colorbox{gray!30}{\strut91.73\%}} & \textcolor{black}{68.43\% / 67.83\%} \\ 
\hline
\multirow{3}{*}{\textcolor{black}{EfficientNetV2-FFKEW}} & \textcolor{black}{Hinton et al.~\cite{hinton2015distilling}}    & \textcolor{black}{98.19\% / \colorbox{gray!30}{\strut85.87\%}} & \textcolor{black}{88.75\% / 87.30\%} & \textcolor{black}{98.04\% / \colorbox{gray!30}{\strut88.69\%}} & \textcolor{black}{83.99\% / 81.45\%} & \textcolor{black}{97.21\% / \colorbox{gray!30}{\strut86.23\%}} & \textcolor{black}{90.45\% / 88.36\%} & \textcolor{black}{98.41\% / \colorbox{gray!30}{\strut85.58\%}} & \textcolor{black}{85.98\% / 84.70\%} \\
                                                     & \textcolor{black}{Papernot et al.~\cite{papernot2017practical}} & \textcolor{black}{98.19\% / \colorbox{gray!30}{\strut89.42\%}} & \textcolor{black}{88.75\% / 87.69\%} & \textcolor{black}{98.04\% / \colorbox{gray!30}{\strut89.92\%}} & \textcolor{black}{83.99\% / 82.52\%} & \textcolor{black}{97.21\% / \colorbox{gray!30}{\strut88.93\%}} & \textcolor{black}{90.45\% / 89.11\%} & \textcolor{black}{98.41\% / \colorbox{gray!30}{\strut90.15\%}} & \textcolor{black}{85.98\% / 85.34\%} \\
                                                     & \textcolor{black}{Knockoff~\cite{orekondy2019knockoff}}         & \textcolor{black}{98.19\% / \colorbox{gray!30}{\strut93.66\%}} & \textcolor{black}{88.75\% / 87.86\%} & \textcolor{black}{98.04\% / \colorbox{gray!30}{\strut91.13\%}} & \textcolor{black}{83.99\% / 82.37\%} & \textcolor{black}{97.21\% / \colorbox{gray!30}{\strut90.44\%}} & \textcolor{black}{90.45\% / 89.76\%} & \textcolor{black}{98.41\% / \colorbox{gray!30}{\strut92.94\%}} & \textcolor{black}{85.98\% / 85.63\%} \\

\bottomrule
\end{tabular}}
\end{table*}
Figure~\ref{fig:model_info_classes} shows an example of generating model informative classes for a 2D convolution operator and model parameters.
As seen, a FlatBuffers table ("Conv2DOptions" or "Buffer") and its field ("padding" or "data") are mapped to a Python class that has the same name as the table and a method matching the field.
The method (\texttt{\small Padding(self)} or \texttt{\small Data(self, j)}) is the core of the class as it allows us to access the serialized data (padding value and parameters) stored in a model via a specific offset that indicates the data location.
Note that we only present one matching method in one generated Python class for the demonstration purpose.
If there are multiple fields declared in a table, all fields are mapped accordingly.
Moreover, all tables (e.g., "RNNOptions", "Pool2DOptions" and "FullyConnectedOptions") in the schema are mapped through the same manner as shown in the figure.
For other data structures, such as enum and union, we directly replicate their constants into corresponding model informative classes.


\section{Utility Class Implementation}
\label{sec:util_cls_implementation}

For the sake of consistent presentation, we use the \texttt{\small Conv2DOptions} and \texttt{\small Buffer} model informative classes in Figure~\ref{fig:model_info_classes} as an example.
Their utility classes are depicted in Figure~\ref{fig:utility_classes}.
The \texttt{\small Conv2DOptionsT} class (Figure~\ref{subfig:utility_class_Conv2DOptionsT}) contains four methods that are self describing: 
(1) \texttt{\small\underline{\hspace{0.25cm}}init\underline{\hspace{0.25cm}}(self)} defines the attributes corresponding to the field matching methods in the \texttt{\small Conv2DOptions} class.
(2) \texttt{\small InitFromBuf(cls, buf, pos)} initializes an instance of the \texttt{\small Conv2DOptions} class with data from a FlatBuffers buffer at a given position.
(3) \texttt{\small InitFromObj(cls, conv2DOptions)} initializes an instance of the \texttt{\small Conv2DOptionsT} class with data from the \texttt{\small Conv2DOptions} instance.
(4) \texttt{\small\underline{\hspace{0.125cm}}UnPack(self, conv2DOptions)} sets the attributes of the \texttt{\small Conv2DOptionsT} instance to the values returned by the corresponding field matching methods in the \texttt{\small Conv2DOptions} instance.
The idea behind these methods is to extract the values (e.g., padding value) of a particular data structure (e.g., Conv2DOptions) based on its position in a FlatBuffers buffer (the victim model), and organize the extracted values into a corresponding instance (e.g., \texttt{\small x}) that
enables us to modify later.
Similarly, the \texttt{\small BufferT} class (Figure~\ref{subfig:utility_class_BufferT}) also has such methods, with the major difference in the  \texttt{\small\underline{\hspace{0.125cm}}UnPack(self, buffer)}, where the data for a particular operator (e.g., parameters of a Conv2D layer) can be extracted in accordance with its index in a FlatBuffers buffer.



\section{Data Serialization Function For Utility Class}
\label{sec:data_ser_function}
Following the same illustrative example in Figure~\ref{fig:utility_classes}, the data serialization functions for \texttt{\small Conv2DOptionsT} and \texttt{\small BufferT} classes are depicted in Listing~\ref{list:utility_class_data_serialization}.
We start with the methods added to the \texttt{\small Conv2DOptionsT} class. 
In lines 1-4, \texttt{\small Conv2DOptionsStart(builder)} initializes bookkeeping for writing a new \texttt{\small Conv2DOptions} instance, where the positional argument \texttt{\small builder} accepts an instance of the \texttt{\small Builder}\footnote{https://github.com/google/flatbuffers/tree/master/python} class used for constructing a FlatBuffers buffer.
Within the method, \texttt{\small builder.StartObject(6)} indicates the number of attributes to be written.
Here we have six attributes that align to the FlatBuffers table "Conv2DOptions" in Figure~\ref{fig:model_info_classes}.
Next, \texttt{\small Conv2DOptionsAddPadding(builder, padding)} adds a passed in padding value into the \texttt{\small builder} used by the \texttt{\small Conv2DOptionsStart(builder)}.
Note that we only present one attribute-related method for ease of presentation, other methods can refer to our detailed implementation\footnote{\href{https://github.com/Jinxhy/THEMIS}{https://github.com/Jinxhy/THEMIS}}.
Finally, \texttt{\small Conv2DOptionsEnd(builder)} serializes the data contained in the \texttt{\small Conv2DOptions} instance into a FlatBuffers buffer using the same \texttt{\small builder} as before.
To make it easy to use, we integrate all aforementioned methods into the \texttt{\small Pack(self, builder)}, as shown in lines 6-11.
For the \texttt{\small BufferT} class, it has five methods from lines 14 to 32, namely \texttt{\small BufferStart(builder)}, \texttt{\small BufferAddData(builder, data)}, \texttt{\small BufferStartDataVector(builder, numElems)}, \texttt{\small BufferEnd(builder)}, and \texttt{\small Pack(self, builder)}.
Among these methods, the \texttt{\small BufferStartDataVector(builder, numElems)} is designed specifically for writing model parameters as a deep learning model stores its parameters in matrix (vectors) form.
The remaining methods are similar to those in the \texttt{\small Conv2DOptionsT} class. 

\section{Additional Experimental Results}
\label{sec:add_exp_results}
More experimental results are shown in Table~\ref{table:text_classification_results}, Table~\ref{table:torch_mobile_result}, and Table~\ref{table:wm_rob_results}.

\end{document}